\documentclass[journal]{new-aiaa}

\usepackage{epsfig}
\usepackage{subfig}

\usepackage{graphicx}

\title{Performance of wall-modeled LES with boundary-layer-conforming
  grids for external aerodynamics }

\author{Adri\'an Lozano-Dur\'an\footnote{Draper Assistant Professor, Department of Aeronautics and Astronautics, Massachusetts Institute of Technology, AIAA Senior Member (Corresponding Author).}}
\affil{Massachusetts Institute of Technology, Massachusetts 02139, USA}
\author{Sanjeeb T.~Bose\footnote{Chief Technology Officer of Cascade Technologies, Inc. and Center for Turbulence Research, Stanford University. Member AIAA.}}
\affil{Stanford University, Stanford, California 94305, USA}
\affil{Cascade Technologies, Inc., Palo Alto, California 94303, USA}
\author{Parviz Moin\footnote{Franklin P. and Caroline M. Johnson Professor, Center for Turbulence Research. Fellow AIAA.}}
\affil{Stanford University, Stanford, California 94305, USA}

\begin{document}

\maketitle

\begin{abstract}
We investigate the error scaling and computational cost of
wall-modeled large-eddy simulation (WMLES) for external aerodynamic
applications. The NASA Juncture Flow is used as representative of an
aircraft with trailing-edge smooth-body separation. Two gridding
strategies are examined: i) constant-size grid, in which the near-wall
grid size has a constant value and ii) boundary-layer-conforming grid
(BL-conforming grid), in which the grid size varies to accommodate the
growth of the boundary-layer thickness.  Our results are accompanied
by a theoretical analysis of the cost and expected error scaling for
the mean pressure coefficient ($C_p$) and mean velocity profiles. The
prediction of $C_p$ is within less than 5\% error for all the grids
studied, even when the boundary layers are marginally resolved.  The
high accuracy in the prediction of $C_p$ is attributed to the
outer-layer nature of the mean pressure in attached flows.  The errors
in the predicted mean velocity profiles exhibit a large variability
depending on the location considered, namely, fuselage, wing-body
juncture, or separated trailing-edge. WMLES performs as expected in
regions where the flow resembles a zero-pressure-gradient turbulent
boundary layer such as the fuselage ($<5\%$ error). However, there is
a decline in accuracy of WMLES predictions of mean velocities in the
vicinity of wing-body junctions and, more acutely, in separated zones.
The impact of the propagation of errors from the underresolved wing
leading-edge is also investigated.  It is shown that BL-conforming
grids enable a higher accuracy in wing-body junctions and separated
regions due to the more effective distribution of grid points, which
in turn diminishes the streamwise propagation of errors.
\end{abstract}

\newpage

\section{Introduction}

\lettrine{T}{he} use of computational fluid dynamics (CFD) for
external aerodynamic applications has been a key tool for aircraft
design in the modern aerospace industry~\citep{Johnson2005, Abbas2011,
  Spalart2016}. CFD methodologies with increasing functionality and
performance have greatly improved our understanding and predictive
capabilities of complex flows.  These improvements suggest that
Certification by Analysis (CbA) --prediction of the aerodynamic
quantities of interest by numerical simulations~\citep{Clark2020} may
soon be a reality.  CbA is expected to reduce the number of wind
tunnel tests, reducing both the turnover time and cost of the
design cycle~\citep{Practices2021}. However, flow predictions from the
state-of-the-art CFD solvers are still unable to comply with the
stringent accuracy requirements and computational efficiency demanded
by the industry~\citep{Slotnick2014}. These limitations are imposed,
largely, by the defiant ubiquity of turbulence.  In the present work,
we investigate the cost and performance of wall-modeled large-eddy
simulation (WMLES) to predict quantities of interest in the NASA
Juncture Flow Experiment~\citep{Rumsey2019}.

One of the major challenges in CFD is the prediction of corner flow
and smooth-body separation~\citep{Witherden2017, Cho2020}.  In the
latter, the loss of momentum across the boundary layer eventually
leads to flow detachment, which can significantly affect the
performance of an aircraft wing.  In wing-fuselage junctures, the flow
is often observed to separate in the corner region near the wing
trailing-edge. This is also the case at the angles of attack (AoA)
typically encountered during take-off and
landing~\citep{Rumsey2018_HL}. Current turbulence models, such as
those utilized in Reynolds-averaged Navier-Stokes CFD (RANS), have
performed poorly in predicting the onset and extent of the
three-dimensional separated flow in wing-fuselage
junctions~\citep{Rumsey2002}. These deficiencies have been exposed in
previous AIAA Drag Prediction Workshops~\citep{Vassberg2008}, where
large variations in the prediction of separation, skin friction, and
pressure were documented in the corner-flow region near the wing
trailing-edge.

To advance the state-of-the-art of CFD in realistic separated flows,
NASA has developed a validation experiment for a generic full-span
wing-fuselage junction model at subsonic conditions: the NASA Juncture
Flow Experiment. The reader is referred to \citet{Rumsey2019} for a
summary of the history and goals of the experiment~\citep[see
  also][]{Rumsey2016a, Rumsey2016b}. The geometry and flow conditions
are designed to trigger flow separation in the trailing edge corner of
the wing, with recirculation bubbles varying in size with the AoA. The
model is a full-span wing-fuselage body that was configured with
truncated DLR-F6 wings, both with and without leading-edge horn at the
wing root. The model has been tested at a chord Reynolds number of 2.4
million, and AoAs ranging from -10 degrees to +10 degrees in the
Langley 14- by 22-foot Subsonic Tunnel. An overview of the
experimental measurements can be found in \citet{Kegerise2019}. The
main aspects of the planning and execution of the project are
discussed by \citet{Rumsey2018}, along with details about the CFD and
experimental teams.

To date, most CFD efforts on the NASA Juncture Flow Experiment have
been conducted using RANS or hybrid-RANS solvers.  \citet{Lee2017}
performed the first CFD analysis to aid the NASA Juncture Flow
committee in selecting the wing configuration for the final
experiment. \citet{Lee2018} presented a preliminary CFD study of the
near wing-body juncture region to evaluate the best practices in
simulating wind tunnel effects.  \citet{Rumsey2019} used NASA's
FUN3D\citep{Anderson1994, Anderson1996, Anderson1999} to investigate
the ability of RANS-based CFD solvers to predict the flow details
leading up to separation. The study comprised different RANS
turbulence models such as a linear eddy viscosity one-equation model,
a nonlinear version of the same model, and a full second-moment
seven-equation model. \citet{Rumsey2019} also performed a grid
sensitivity analysis and CFD uncertainty quantification. Comparisons
between CFD simulations and the wind tunnel experimental results have
been recently documented by \citet{Lee2019}.

WMLES of the NASA Juncture Flow has been less thoroughly investigated,
despite NASA's recognition of WMLES as a critical pacing item for
``developing a visionary CFD capability required by the notional year
2030''. According to NASA's recent CFD Vision 2030 report
\citep{Slotnick2014}, hybrid RANS/LES \citep{Spalart1997, Spalart2009}
and WMLES \citep{Bose2018} are identified as the most viable
approaches for predicting realistic flows at high Reynolds numbers in
external aerodynamics. Previous attempts of WMLES of the NASA Juncture
Flow include the works by \citet{Iyer2020}, \citet{Ghate2020}, and
\citet{Lozano_AIAA_2020, Lozano_brief_2020a}.  These
studies highlighted the capabilities of WMLES to predict wall
pressure, velocity and Reynolds stresses, especially compared with
RANS-based methodologies. Nonetheless, it was noted that WMLES is
still far from providing the robustness and stringent accuracy
required for CbA, especially in the separated regions and
wing-fuselage juncture.

The goal of this study is to systematically quantify the errors in the
mean quantities of interest in the NASA Juncture Flow using WMLES.
Several strategies have been proposed to model the near-wall region in
WMLES, and comprehensive reviews can be found in \citet{Piomelli2002},
\citet{Cabot2000}, \citet{Spalart2009}, \citet{Larsson2015}, and the
most recent review by \citet{Bose2018}. We follow the wall-flux
modeling approach (or approximate boundary conditions modeling), where
the no-slip and thermal wall boundary conditions are replaced with
stress and heat flux boundary conditions provided by the wall
model. This category of wall models utilizes the LES solution at a
given location in the LES domain as input, and returns the wall-fluxes
needed by the LES solver. Examples of the most popular and well-known
approaches are those computing the wall stress using either the law of
the wall \citep{Deardorff1970, Schumann1975, Piomelli1989} or the
full/simplified RANS equations \citep{Balaras1996, Wang2002,
  Chung2009, Bodart2011, Kawai2012, Kawai2013, Bermejo-Moreno2014, Park2014,
  Yang2015}.  We will not attempt here to devise improvements that
alleviate current modeling limitations. The interested reader may
refer to \citet{Lozano_brief_2020b} for new modeling venues in WMLES
applied to the NASA Juncture Flow.

This work is organized as follows. The flow setup, mathematical
modeling, and numerical approach are presented in Section
\ref{sec:numerical}. The strategies for grid generation are discussed
in Section \ref{sec:gridding} along with the computational cost of
WMLES. In Section \ref{sec:errors}, we introduce the theoretical
analysis for the error scaling. WMLES of the NASA Juncture Flow is
presented in Section \ref{sec:JFE}, which includes predictions for
mean velocity profiles and Reynolds stresses for three different
locations on the aircraft: the upstream region of the fuselage, the
wing-body juncture, and the wing-body juncture close to the trailing
edge. We also discuss the prediction of the mean surface pressure
coefficient at different spanwise locations over the wing. Finally,
conclusions are offered in Section \ref{sec:conclusions}.

\section{Numerical Methods}\label{sec:numerical}

\subsection{Flow conditions and computational setup}\label{sec:setup}

\begin{figure}
\centering
\includegraphics[width=0.8\textwidth]{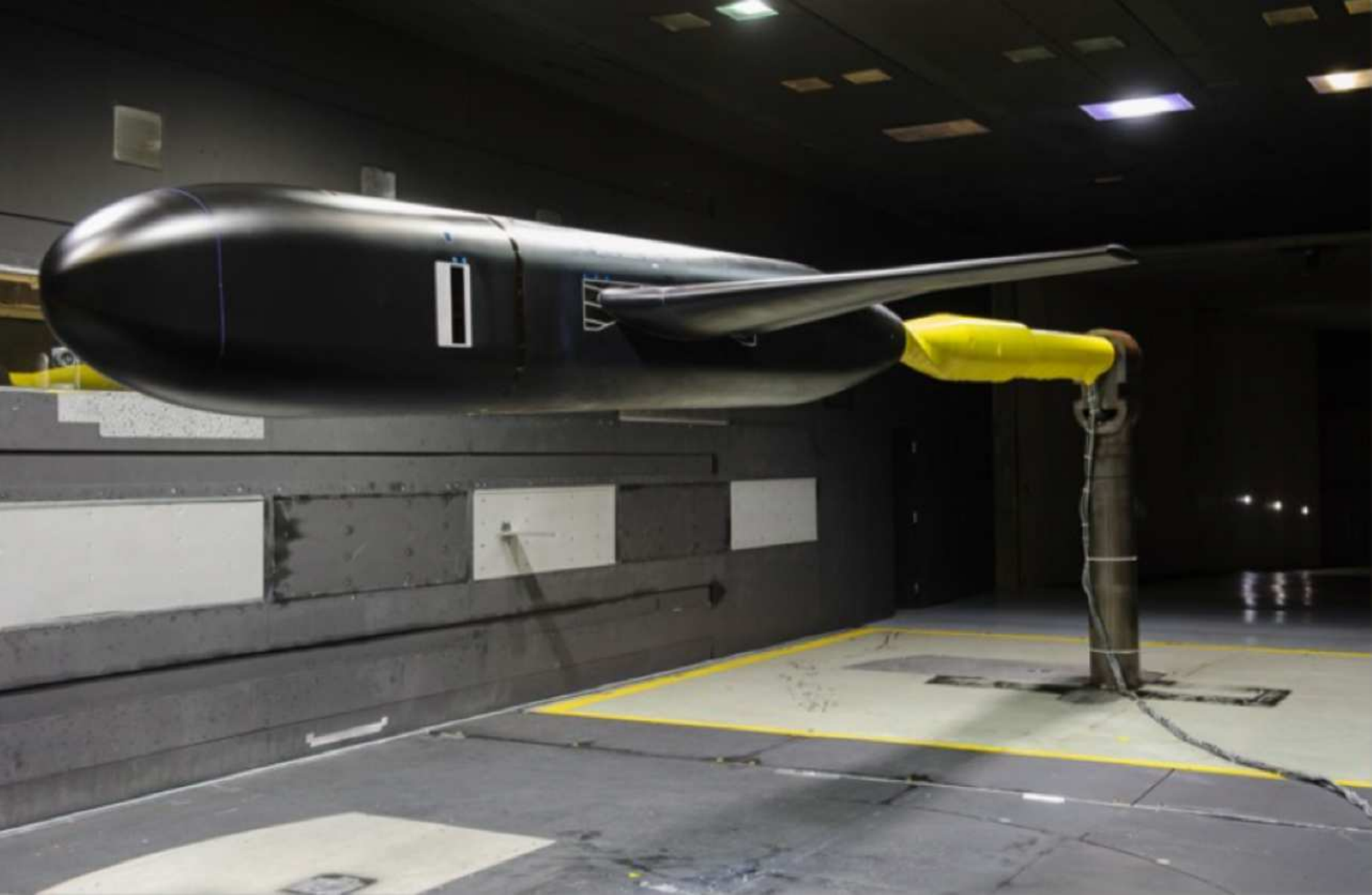}
\caption{Experimental setup of the NASA Juncture Flow at NASA Langley
 14- by 22-Foot Subsonic Wind Tunnel.\label{fig:experiment}}
\end{figure}

We use the NASA Juncture Flow geometry with a wing based on the DLR-F6
and a leading-edge horn to mitigate the effect of the horseshoe vortex
over the wing-fuselage juncture (figure \ref{fig:experiment}).  The
model wingspan is nominally 3397.2~mm, the fuselage length is
4839.2~mm, and the crank chord (chord length at the Yehudi break) is
$L=557.1$~mm. The frame of reference is such that the fuselage nose is
located at $x = 0$, the $x$-axis is aligned with the fuselage
centerline, the $y$-axis denotes spanwise direction, and the $z$-axis
is the vertical direction (away from the fuselage upstream of the
wing).  The associated instantaneous velocities are denoted by $u$,
$v$, and $w$, and occasionally by $u_1$, $u_2$, and $u_3$.  The wing
leading-edge horn meets the fuselage at $x = 1925$ mm, and the wing
root trailing-edge is located at $x = 2961.9$ mm.

In the experiment, the model was tripped near the front of the
fuselage and on the upper and lower surfaces of both wings.  In our
case, preliminary calculations showed that tripping was also necessary
to trigger the transition to turbulence over the wing. Hence, the
geometry of the wing was modified by displacing in the $z$ direction a
line of surface mesh points close to the leading edge by 1~mm along the
suction side of the wing, and by -1~mm along the pressure side. The
tripping lines follow approximately the location of the tripping dots
used in the experimental setup for the left wing (lower surface $x =
(4144-y)/2.082$; upper surface $x = (3775-y)/1.975$ for $y<-362$ and
$x = (2847-y)/1.532$ for $y>-362$). Tripping using dots mimicking the
experimental setup was also tested.  It was found that the results
over the wing-body juncture show little sensitivity to the tripping
due to the presence of the incoming boundary layer from the
fuselage. No tripping was needed on the fuselage, which naturally
transitioned from laminar to turbulence.

\begin{figure}
\centering
\includegraphics[width=1\textwidth]{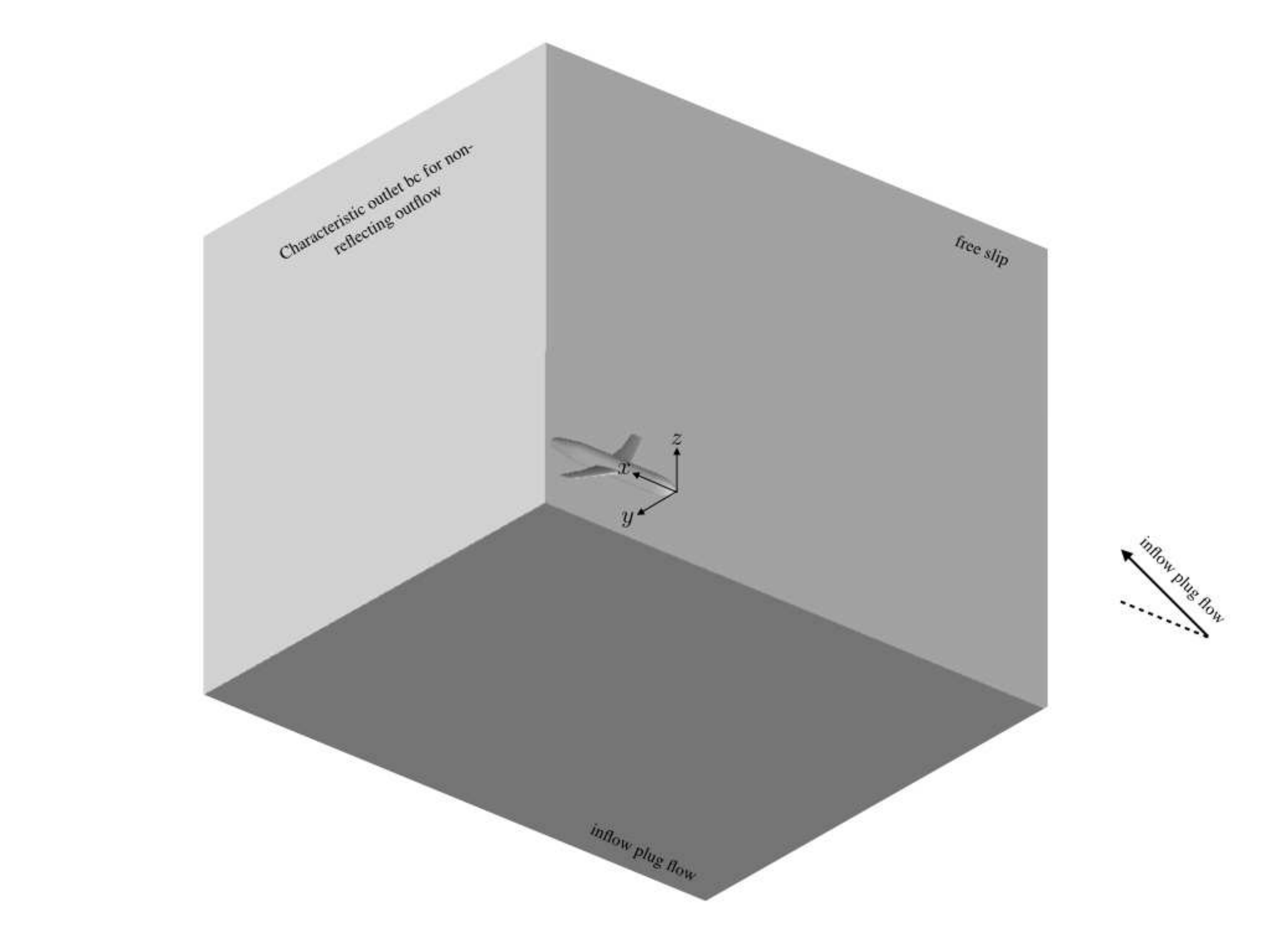}
\caption{Computational domain and NASA Juncture Flow
 model.\label{fig:setup}}
\end{figure}

In the wind tunnel, the model was mounted on a sting aligned with the
fuselage axis.  The sting was attached to a mast that emerged from the
wind tunnel floor. Here, all calculations are performed in free air
conditions, and the sting and mast are ignored. The computational
setup is such that the dimensions of the domain are about five times
the length of the fuselage in the three directions (figure
\ref{fig:setup}). The Reynolds number is $Re = L U_\infty/\nu=2.4$
million based on the crank chord length $L$, freestream velocity
$U_\infty$, and the kinematic viscosity $\nu$. The freestream Mach
number is Ma $= 0.189$, the freestream temperature is $T = 288.84$ K,
and the dynamic pressure is 2476 Pa.  We impose a uniform plug flow as
inflow boundary condition in the front and bottom boundaries of the
domain. The Navier--Stokes characteristic boundary condition for
subsonic non-reflecting outflow is imposed at the outflow and top
boundaries~\citep{Poinsot1992} and free-slip is used at the lateral
boundaries.  At the aircraft wall, we impose Neumann boundary
condition with the shear stress provided by the wall model as
described in Section \ref{sec:models}.
%

\subsection{Numerical methods, subgrid-scale model and wall model}\label{sec:models}

The simulations are conducted with the high-fidelity solver charLES
developed by Cascade Technologies, Inc~\citep{Bres2018, Fu2021}. The
code integrates the compressible LES equations using a kinetic-energy
conserving, second-order accurate, finite volume method.  The
numerical discretization relies on a flux formulation which is
approximately entropy preserving in the inviscid limit, thereby
limiting the amount of numerical dissipation added into the
calculation. The time integration is performed with a third-order
Runge-Kutta explicit method. The SGS model is the dynamic Smagorinsky
model \citep{Moin1991} with the modification by \citet{Lilly1992}.

We utilize a wall model to overcome the restrictive grid-resolution
requirements imposed by the small-scale motions in the vicinity of the
walls. The no-slip boundary condition at the walls is replaced by a
wall-stress boundary condition.  The walls are assumed adiabatic and
the wall stress is obtained by an algebraic equilibrium wall model
derived from the integration of the one-dimensional stress model along
the wall-normal direction~\citep{Deardorff1970, Piomelli1989},
\begin{equation}
  u_{||}^+(y_\bot^+) =
  \begin{cases}
    y_\bot^+ + a_1 (y_\bot^{+})^2 \text{\, \, \, for $y_\bot^+ < 23$}, \\
    \frac{1}{\kappa}\ln{y_\bot^+} + B \text{\, \, \, \, otherwise}
  \end{cases}
  \label{eq:charles_algwm}
\end{equation}
where $u_{||}$ is the model wall-parallel velocity, $y_\bot$ is the
wall-normal direction to the aircraft surface, $\kappa=0.41$ is the
K\'arm\'an constant, $B = 5.2$ is the intercept constant, and $a_1$ is
computed to ensure $C^1$ continuity. The superscript $+$ denotes inner
units defined in terms of wall friction velocity ($u_\tau$) and
$\nu$. The matching location for the wall model is the first off-wall
cell center of the LES grid, denoted by $h_w$, at which $u_{||}$
equals the wall-parallel velocity of the LES solution. No temporal
filtering or other treatments were used for the LES velocity at the
matching location.

\section{Grid strategies and computational cost}\label{sec:gridding}

\subsection{Grid generation: constant-size grid vs. boundary-layer-conforming grid}
\label{subsec:tbl}

The mesh generation is based on a Voronoi hexagonal close-packed
point-seeding method. We examine two strategies to
distribute the cell centroids of the control volumes as illustrated in
Figure~\ref{fig:grid_types}:
%
\begin{figure}
\begin{center}
\includegraphics[width=1\textwidth]{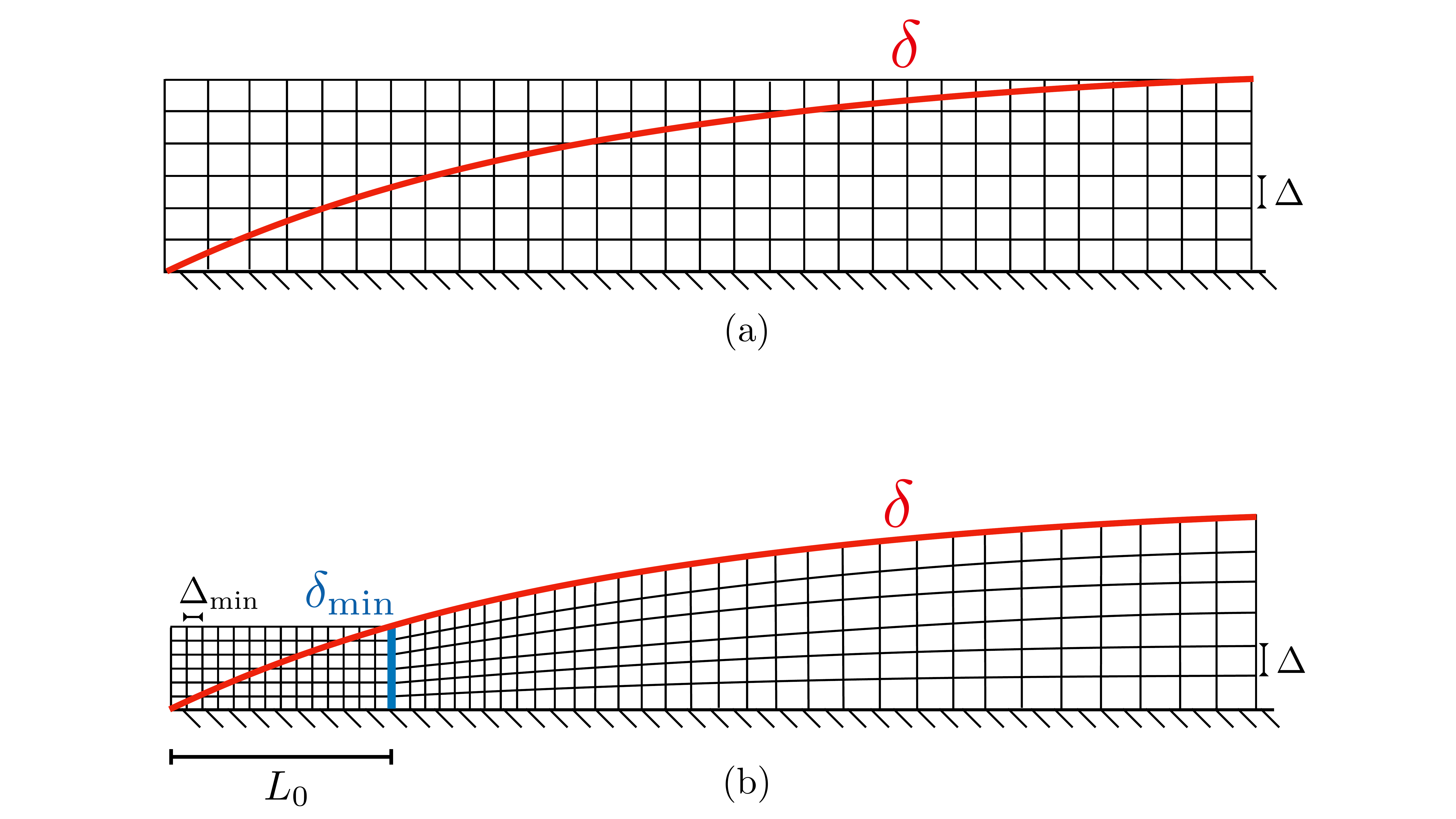}
\end{center}
\caption{ Schematic of (a) constant-size grid and (b) BL-conforming
  grid for a ZPGTBL. \label{fig:grid_types}}
\end{figure}
%
\begin{itemize}
\item[(i)] \underline{Constant-size grid}. In the first approach, we
  set the grid size in the vicinity of the aircraft surface to be
  roughly constant and isotropic $\Delta\approx \Delta_x \approx
  \Delta_y \approx \Delta_z$, where $\Delta_x$, $\Delta_y$ and
  $\Delta_z$ are the characteristic grid sizes in $x$, $y$, and $z$
  directions, respectively. Starting from the wall and building up the
  grid, the number of cell layers with size $\Delta$ is specified to
  be five. We set the farfield grid resolution,
  $\Delta_\mathrm{far}\gg\Delta$, and create additional layers with
  varying grid size to blend the near-wall grid with the farfield
  grid. The meshes are constructed using a Voronoi diagram and ten
  iterations of Lloyd's algorithm to smooth the transition between
  layers with different grid resolutions. The concept is illustrated
  in Figure~\ref{fig:grid_types}(a) for a flat plate, while Figure
  \ref{fig:tbl_grids}(a) shows the actual grid structure in the NASA
  Juncture Flow for $\Delta=2$~mm and $\Delta_\mathrm{far}=200$~mm.
  This grid-generation approach is algorithmically simple and
  efficient. However, it is agnostic to details of the actual flow
  such as wake/shear regions and boundary-layer growth. This implies
  that flow regions close to the fuselage nose and wing leading-edge
  are underresolved (less than one point per boundary-layer
  thickness), whereas the wing trailing edge and the
  downstream-fuselage regions are seeded with up to hundreds of points
  per boundary-layer thickness. The gridding strategy (ii) aims at
  providing a more equitable distribution of grid points.
%
\begin{figure}
  \begin{center}
    \includegraphics[width=1.0\textwidth]{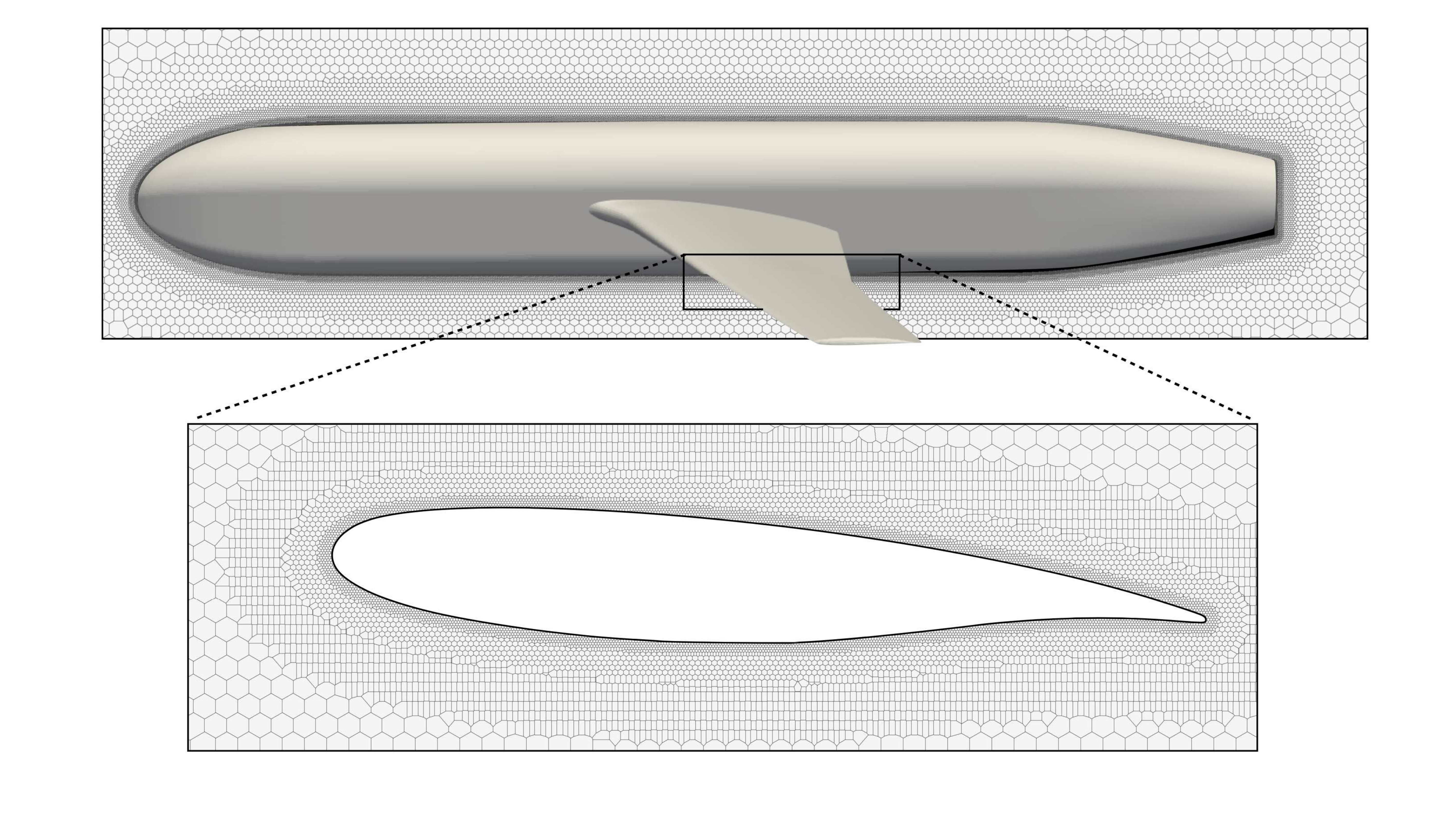}
  \end{center}
    \begin{center}
  \includegraphics[width=1.0\textwidth]{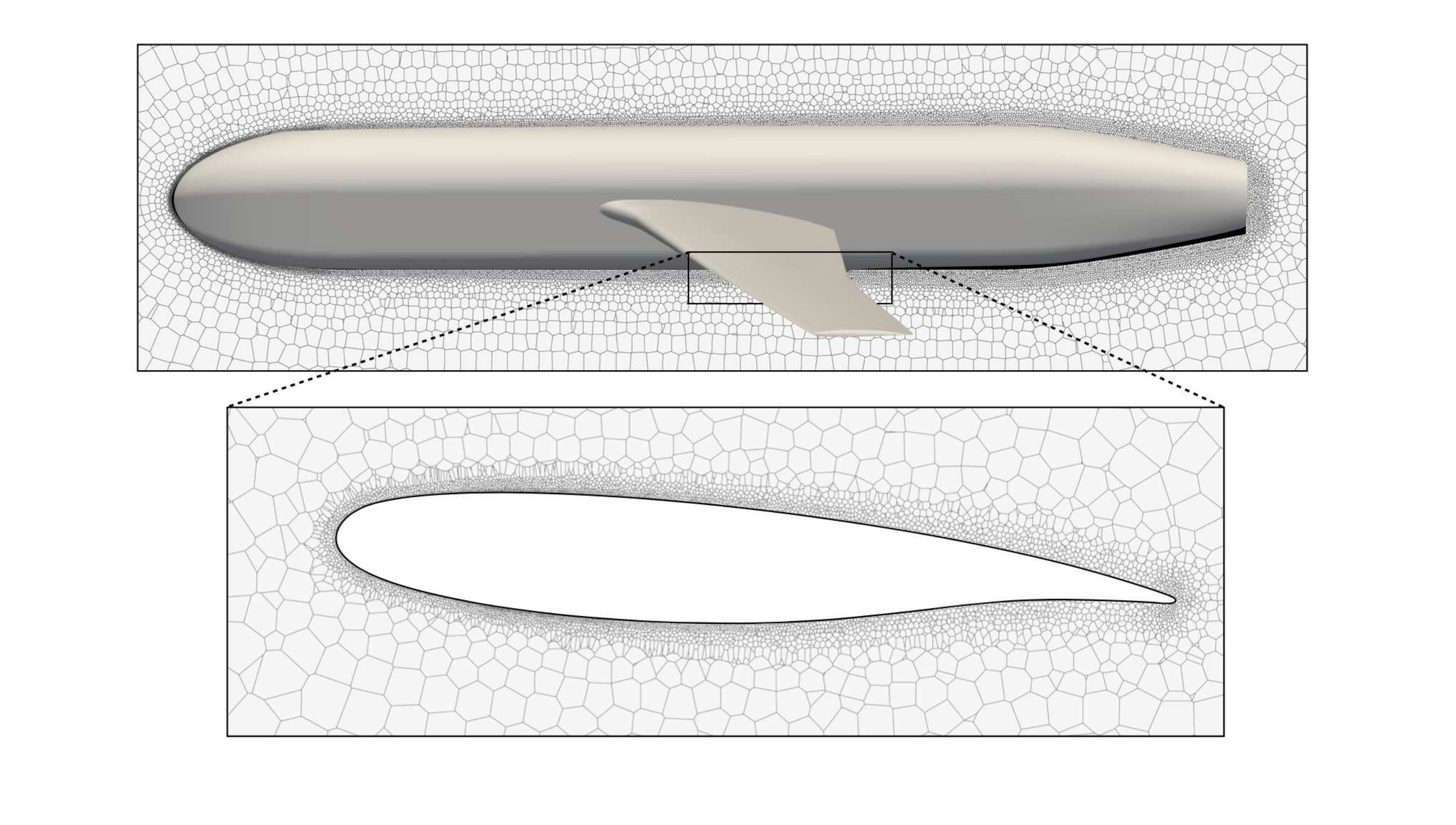}
  \end{center}
  \caption{Visualization of Voronoi control volumes for (top)
    constant-size grid following strategy i) with $\Delta=2$~mm and
    $\Delta_\mathrm{far}=200$~mm and (bottom) boundary-layer-conforming grid
    following strategy ii) with $N_\mathrm{bl} = 5$ and
    $Re_\Delta^\mathrm{min}=2.8 \times 10^3$.\label{fig:tbl_grids}}
\end{figure}
%
\item[(ii)] \underline{Boundary-layer-conforming grid}. In the second
  gridding strategy, we account for the actual growth of the turbulent
  boundary layers, denoted by $\delta$, by seeding the control volumes
  consistently with its growth. We refer to this approach as
  boundary-layer-conforming grid (BL-conforming grid). The method
  necessitates two parameters. The first one is the number of points
  per boundary-layer thickness, $N_\mathrm{bl}$, such that $\Delta_x
  \approx \Delta_y \approx \Delta_z \approx \Delta \approx
  \delta/N_\mathrm{bl}$ is a function of space.  The second parameter
  is less often discussed in the literature and is the minimum local
  Reynolds number that we are willing to marginally resolve in the
  flow, $Re_{\Delta}^\mathrm{min} \equiv \Delta_\mathrm{min}
  U_\infty/\nu$, where $\Delta_\mathrm{min}$ is the smallest grid
  resolution permitted. This is a necessary constraint as $\delta
  \rightarrow 0$ at the leading edge of the body, which would impose a
  large burden on the number of points required to cover this region.
  Hence, the grid resolution is kept constant and equal to
  $\Delta_\mathrm{min}$ at those regions where the boundary-layer
  thickness is below $\delta_\mathrm{min}=N_\mathrm{bl}
  \Delta_\mathrm{min}$ (see Figure~\ref{fig:grid_types}(b)). We also
  impose a geometric constraint on the grid size such that $\Delta$
  must be smaller than the local radius of curvature $R$ of the
  surface. The grid is then constructed by seeding points within the
  boundary layer with space-varying grid size
  \begin{equation}
    \Delta(x,y,z) \approx \mathrm{min}\left[ \mathrm{max}\left( 
    \frac{\gamma\delta}{N_\mathrm{bl}}, \frac{Re_{\Delta}^\mathrm{min} \nu}{U_\infty}\right), \beta R \right],
  \end{equation}
  where $\gamma=1.2$ is a correction factor for $\delta$ to ensure
  that the near-wall grid contains the instantaneous boundary layer,
  and $\beta=1/2$. Note that the grid is still locally isotropic and
  the characteristic size of the control volumes is $\Delta \approx
  \delta/N_\mathrm{bl}$ in the three spatial directions. Figure
  \ref{fig:tbl_grids}(b) shows the structure of a BL-conforming grid
  in the NASA Juncture Flow with $N_\mathrm{bl} = 5$ and
  $Re_\Delta^\mathrm{min}=2.8 \times 10^3$. Additional control volumes
  of increasing size are created to blend the near-wall grid with the
  farfield grid of size $\Delta_{\mathrm{far}}=200$ mm.
\end{itemize}

\begin{figure}
\begin{center}
  \subfloat[]{\includegraphics[width=0.47\textwidth]{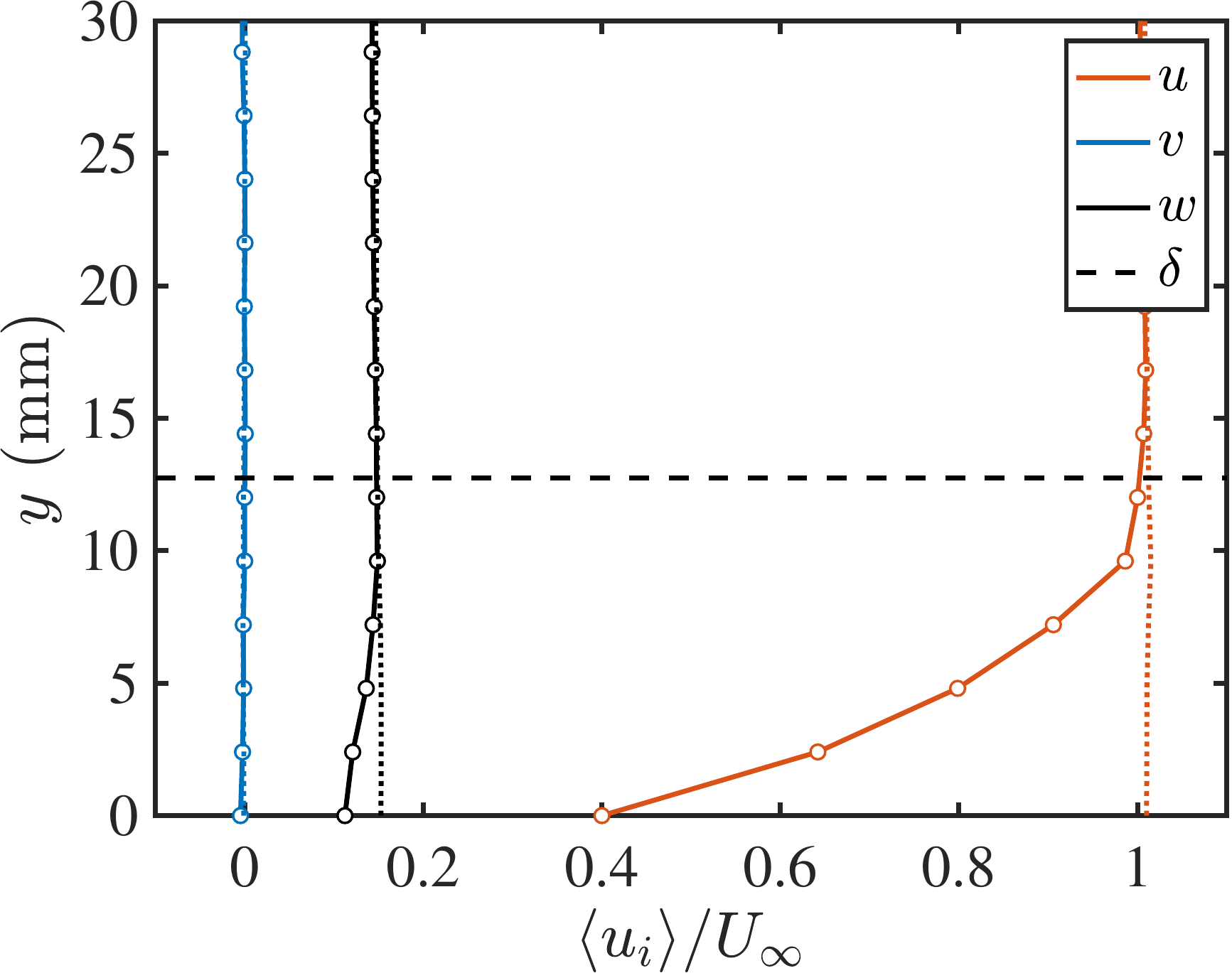}}
  \hspace{0.1cm}
  \subfloat[]{\includegraphics[width=0.47\textwidth]{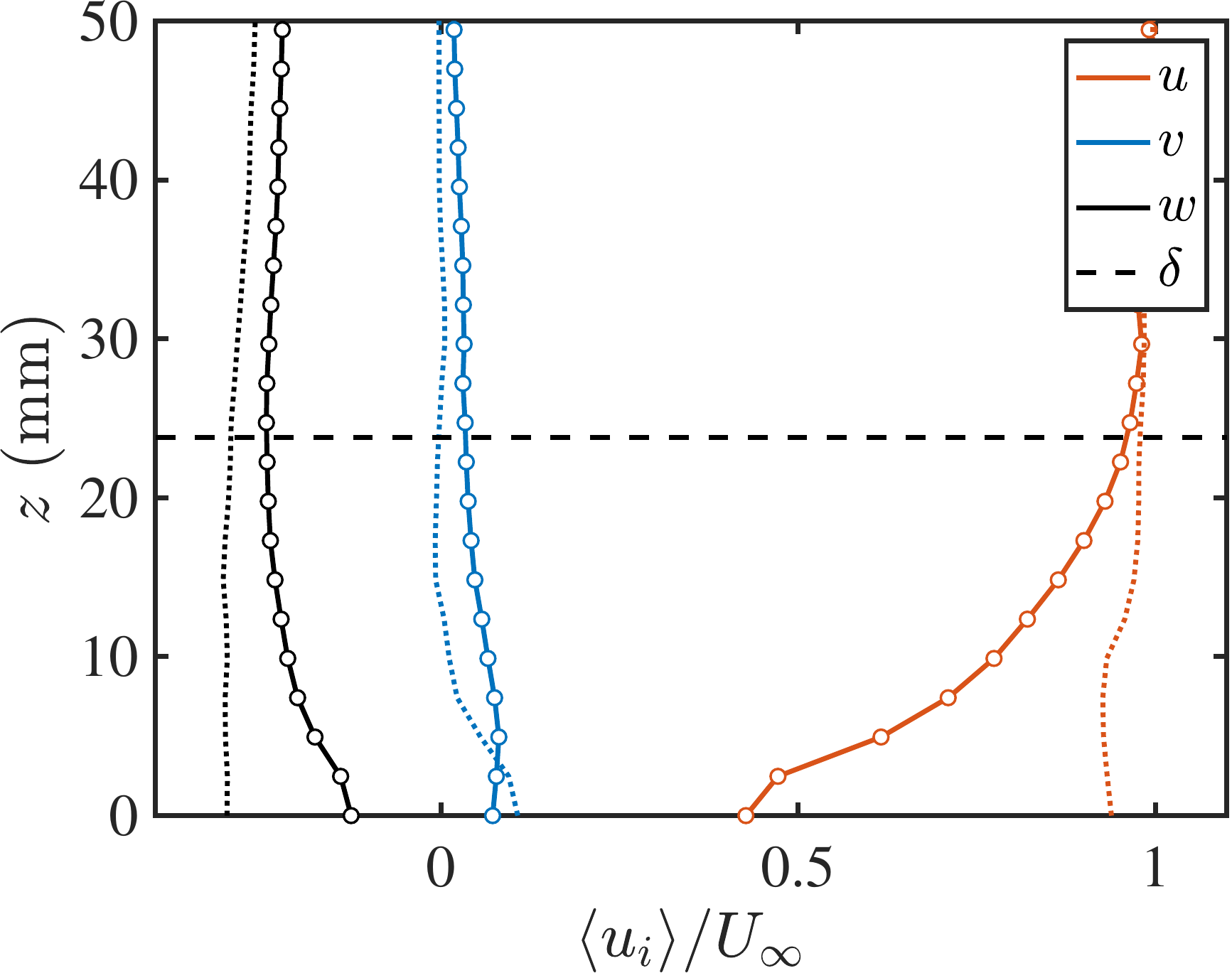}}
\end{center}
\begin{center}
  \includegraphics[width=0.8\textwidth]{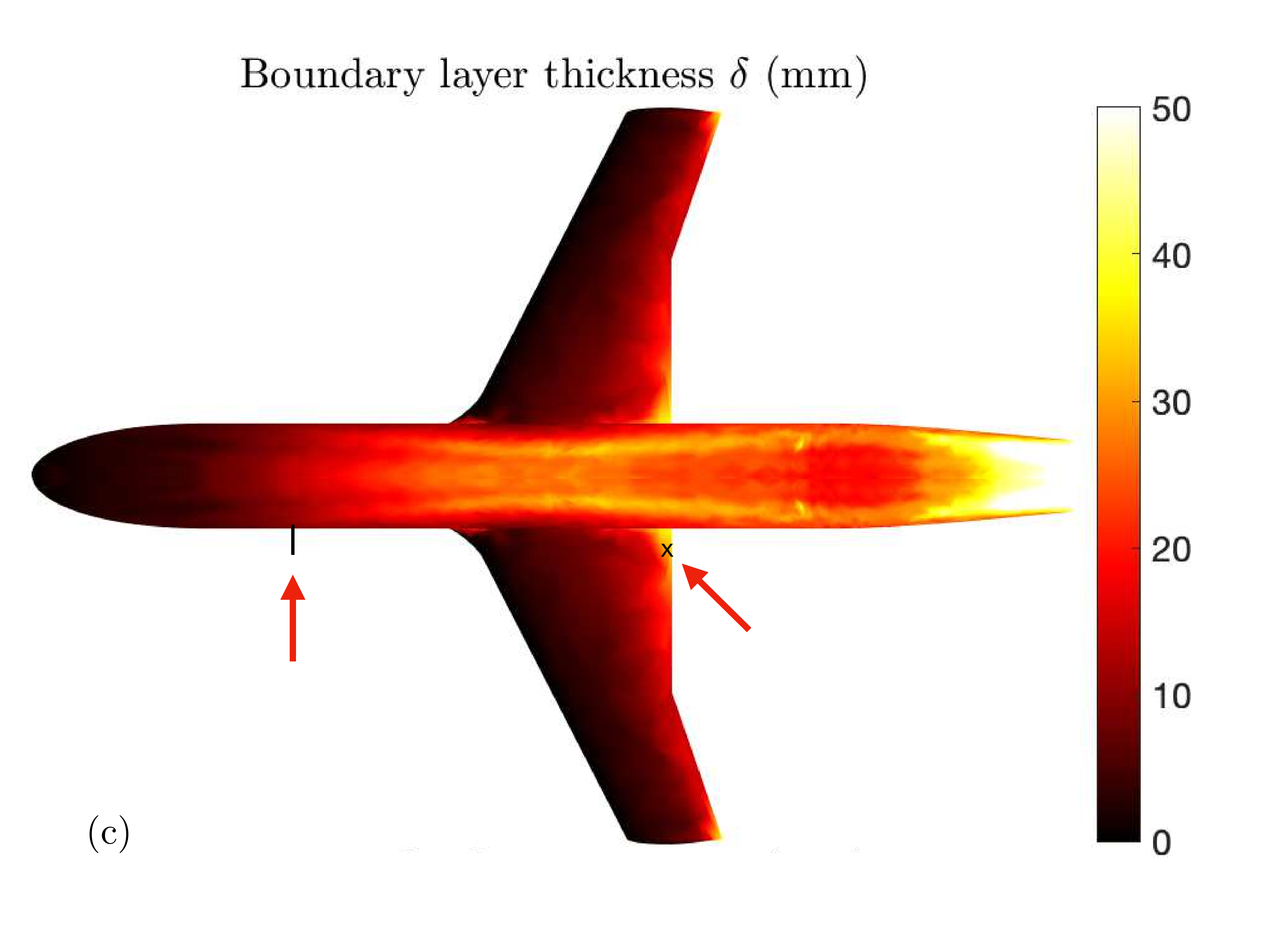}
\end{center}
\caption{The three mean velocity components for $\langle
  \boldsymbol{u}\rangle$ (lines with symbols) and $\langle
  \boldsymbol{u}_I\rangle$ (dotted lines), and boundary layer height
  (dashed). The locations of the mean profiles are indicated in panel
  (c) by the solid line for panel (a) and the cross for panel (b). (c)
  Boundary-layer thickness (in millimeters) for the NASA Juncture Flow
  at AoA = 5 degrees and $Re= 2.4 \times
  10^6$. \label{fig:tbl_thickness}}
\end{figure}

The gridding approach above requires an estimation of the
boundary-layer thickness at each location of the aircraft surface.
Given that boundary layers originate from viscous effects, the method
proposed here is based on measuring the deviation of the viscous
solution from a reference inviscid flow. This is achieved by
conducting two simulations: one WMLES, whose velocity is denoted as
$\boldsymbol{u}$, and one inviscid simulation (no SGS model and
free-slip at the wall), with velocity denoted by
$\boldsymbol{u}_I$. The grid generation for the two simulations
follows strategy (i) with $\Delta =2$~mm. Boundary layers at the
leading edge with thickness below $2$~mm are estimated by
extrapolating the solution using a power law.  Two examples of mean
velocity profiles for $\boldsymbol{u}$ and $\boldsymbol{u}_I$ are
shown in Figures \ref{fig:tbl_thickness}(a) and (b). The
three-dimensional surface representing the boundary layer edge
$S_\mathrm{bl}$ is identified as the loci of
  \begin{equation}\label{eq:Stbl}
    S_\mathrm{bl} \equiv \left\{ (x,y,z)  : \frac{||\langle \boldsymbol{u}_I(x,y,z) \rangle
      - \langle \boldsymbol{u}(x,y,z) \rangle||}{|| \langle \boldsymbol{u}_I(x,y,z) \rangle||} = 0.01 \right\},
  \end{equation}
where $\langle \cdot \rangle$ denotes time-average. Finally, at each
point of the aircraft surface $(x_a,y_a,z_a)$, the boundary-layer
thickness $\delta$ is defined as the minimum spherical distance
between $\boldsymbol{x}_a = (x_a,y_a,z_a)$ and $\boldsymbol{x} =
(x,y,z) \in S_\mathrm{bl}$,
  \begin{equation}\label{eq:delta_def}
    \delta(\boldsymbol{x}_a) \equiv ||\boldsymbol{x}_a - \boldsymbol{x}||_\mathrm{min},
    \ \forall \boldsymbol{x} \in S_\mathrm{bl}.
  \end{equation}
The boundary-layer thickness for the NASA Juncture Flow at $Re=2.4$M
and AoA=5 is shown in Figure \ref{fig:tbl_thickness}(c). The values of
$\delta$ range from $\sim$$0$~mm at the leading edge of the wing to
$\sim$30~mm at the trailing edge of the wing. Thicker boundary layers
above $50$~mm are found in the downstream region of the
fuselage. Equation (\ref{eq:Stbl}) might be interpreted as the
definition of a turbulent/nonturbulent interface, although it also
applies to laminar regions. Other approaches for defining
$S_{\mathrm{bl}}$, such as isosurfaces of Q-criterion, were also
explored and combined with Eq. (\ref{eq:delta_def}) to obtained a
boundary-layer interface. The results were similar to the ones
reported in Figure \ref{fig:tbl_thickness}(c), probably because the
present flow is dominated by attached boundary layers with very mild
separations. In these cases, Eq. (\ref{eq:Stbl}) stands as a
reasonable indicator of the regions influenced by viscous effects for
the purpose of generating BL-conforming grids. It is worth mentioning
that other methods for estimating $\delta$ are available in the
literature~\citep[e.g.][]{Vinuesa2016, Asada2018, Coleman2018,
  Spalart1993, Uzun2020, Griffin2021}. They feature varying degrees of
complexity and generalizability and also may offer a sensible
estimation of $\delta$ when properly conditioned to complex
geometries.
%

\subsection{Required number of grid points and time steps}\label{subsec:number}

\begin{figure}
  \begin{center}
   \subfloat[]{\includegraphics[width=0.49\textwidth]{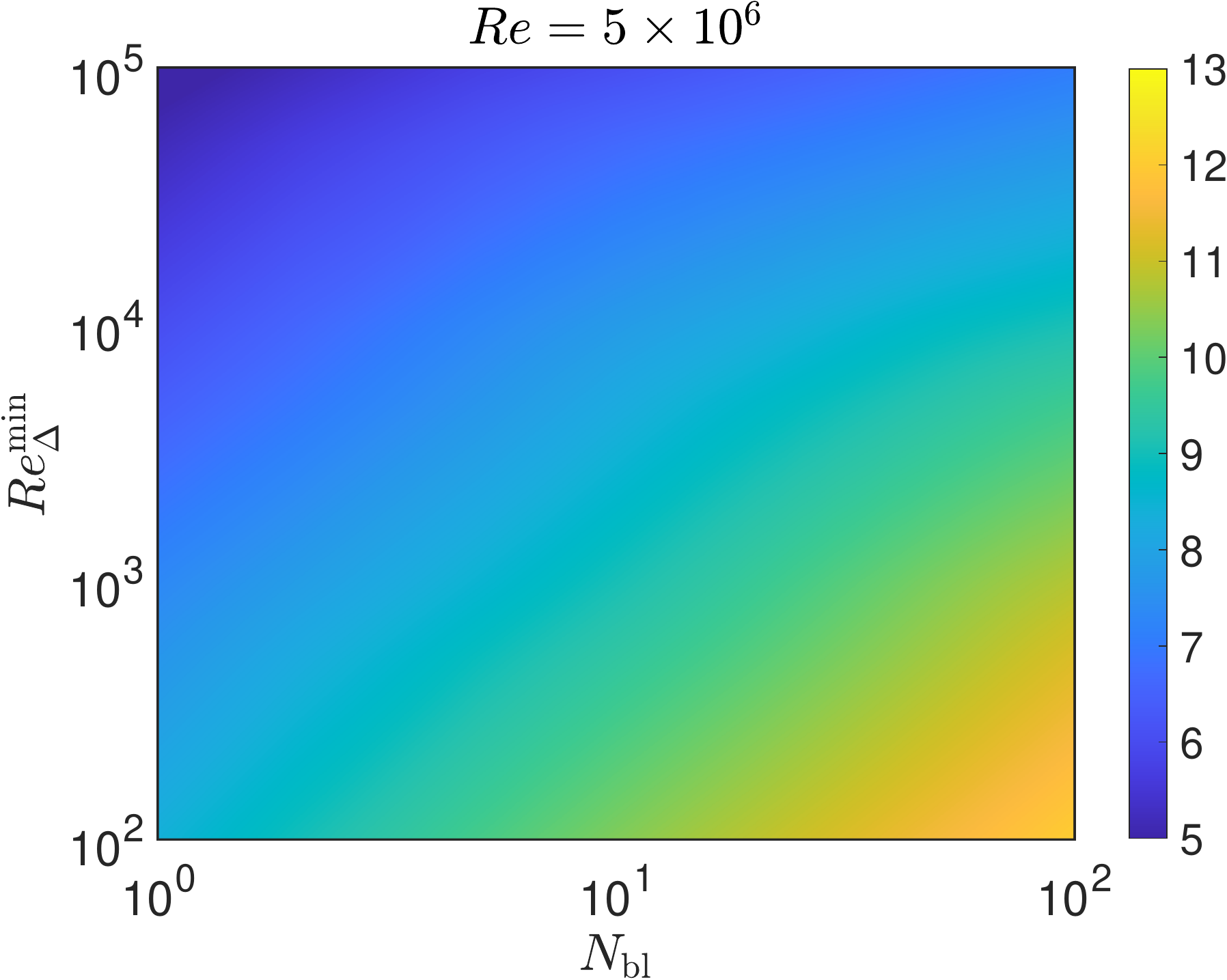}}
    \hspace{0.3cm}
   \subfloat[]{\includegraphics[width=0.44\textwidth]{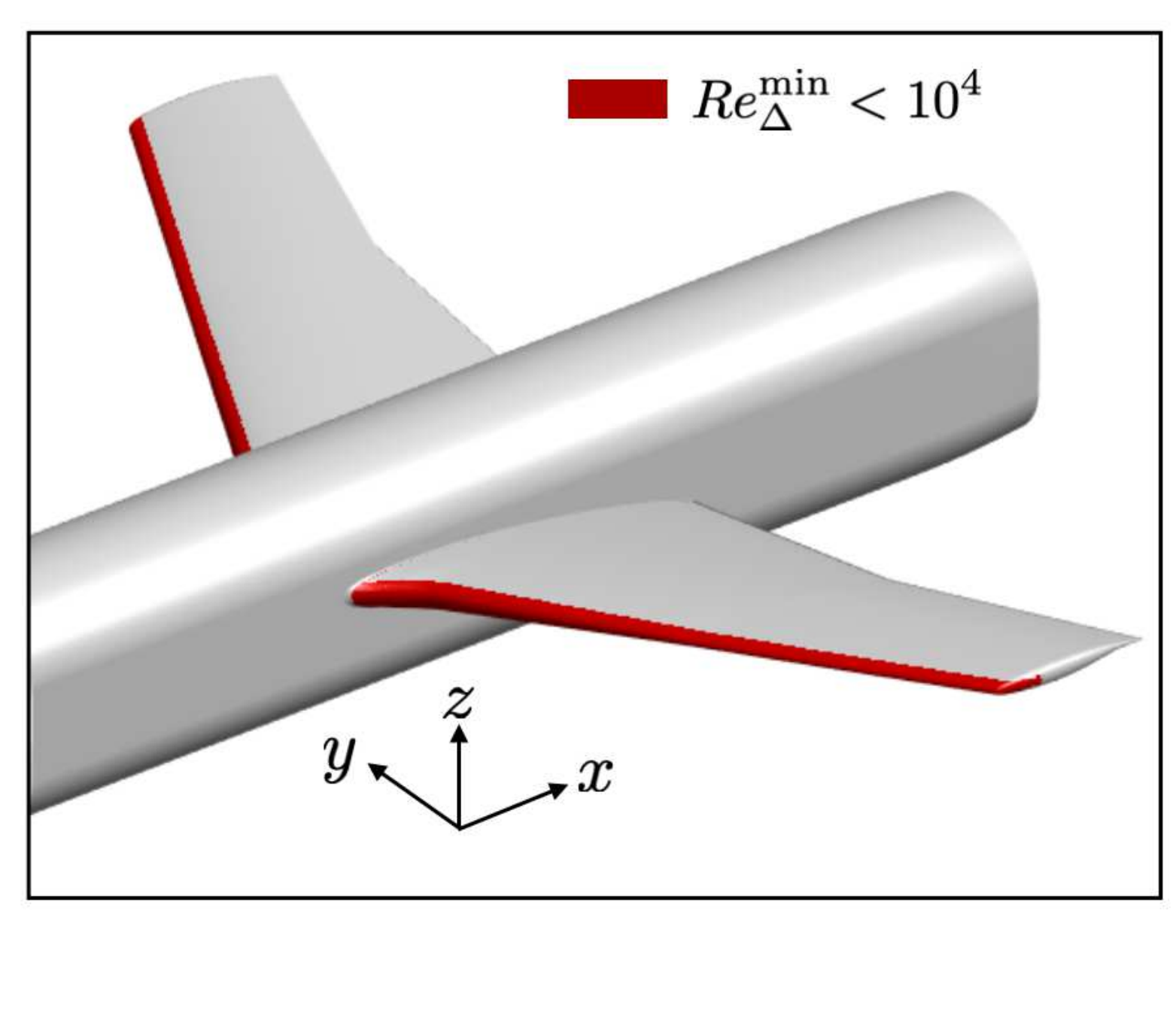}}
\end{center}
  \begin{center}
    \subfloat[]{\includegraphics[width=0.47\textwidth]{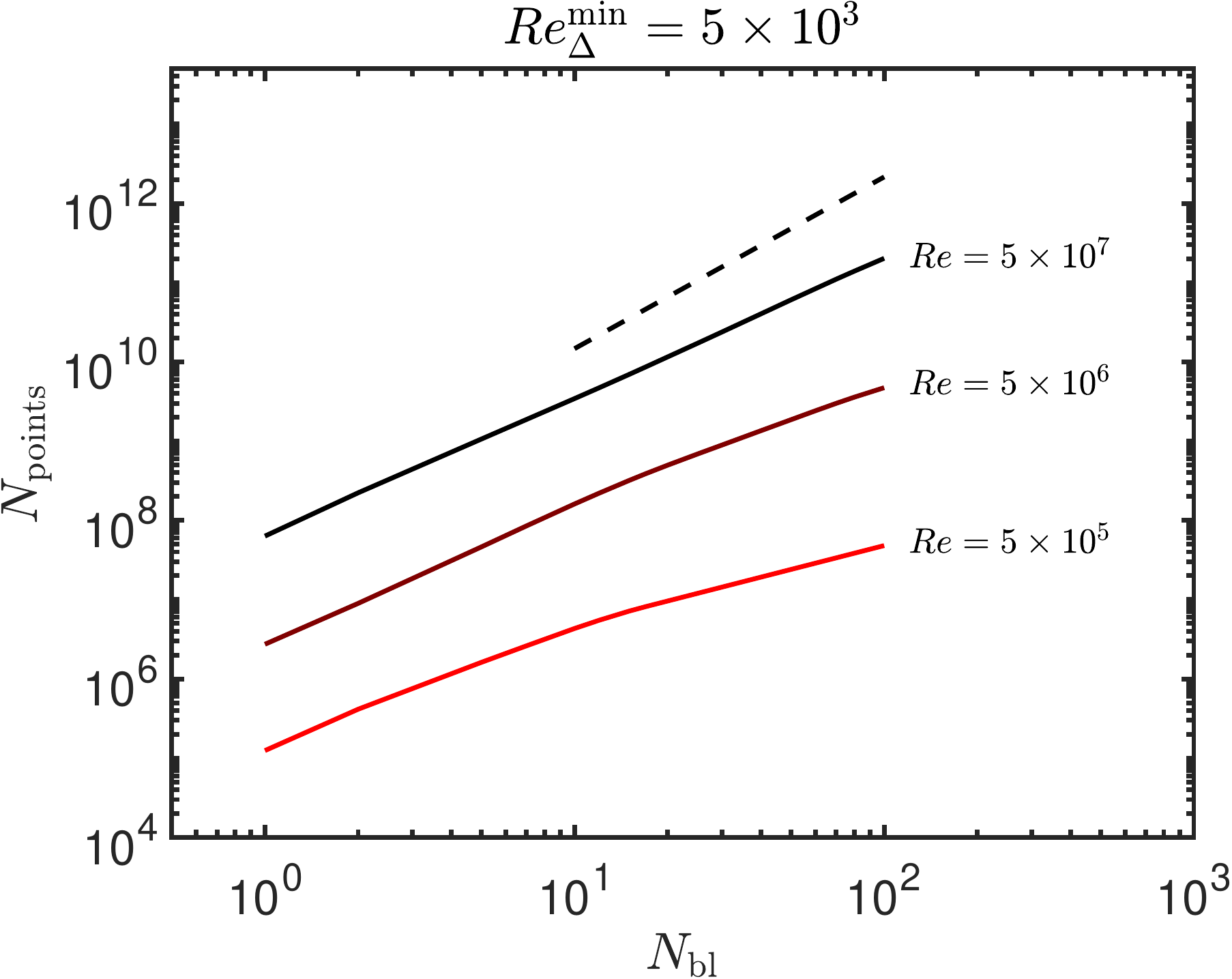}}
        \hspace{0.3cm}
  \subfloat[]{\includegraphics[width=0.47\textwidth]{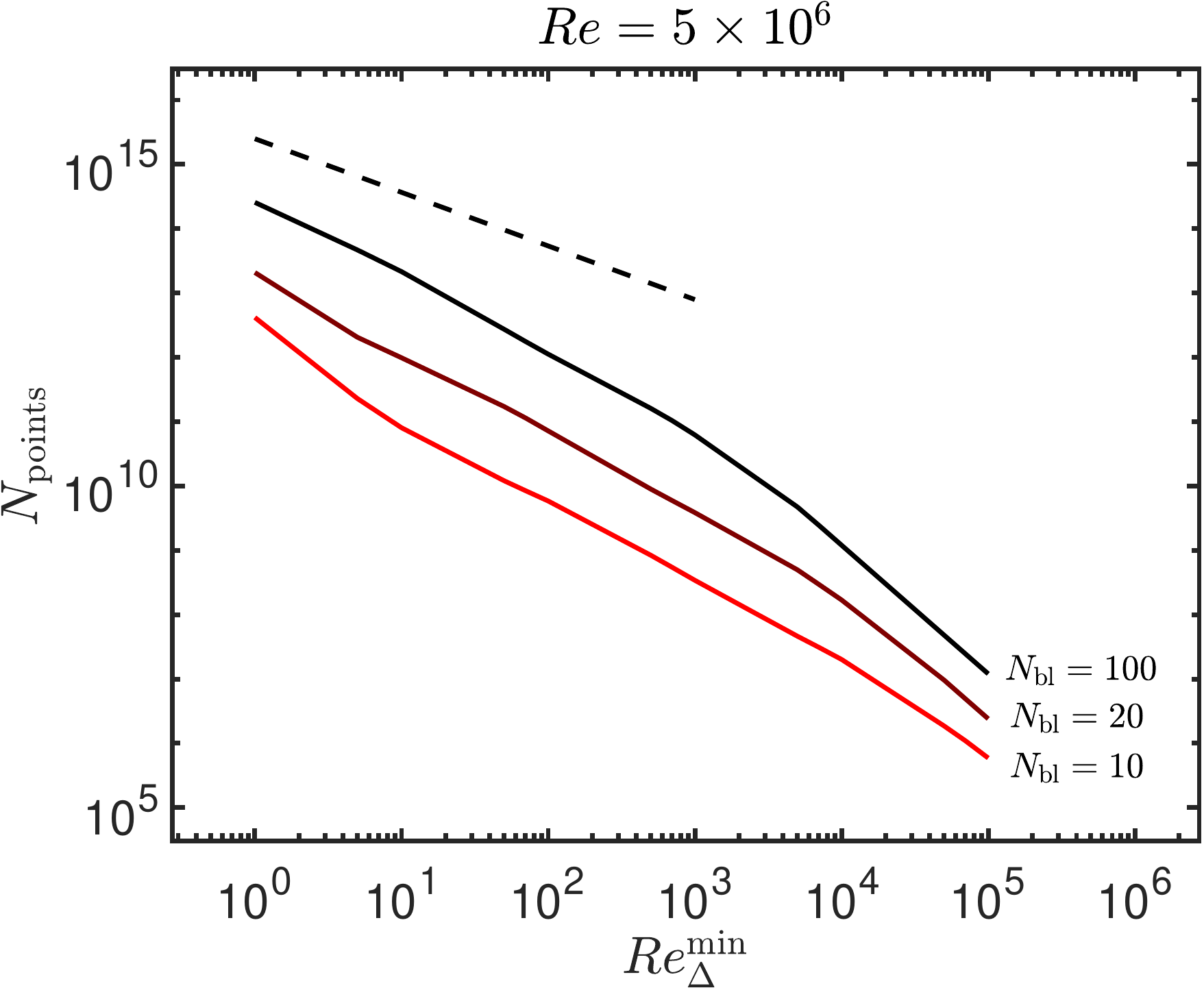}}
\end{center}
\caption{ (a) Logarithm of the number of points
  ($\log_{10}N_\mathrm{points}$) required for WMLES of the NASA
  Juncture Flow geometry as a function of the number of grid points
  per boundary-layer thickness ($N_\mathrm{bl}$) and minimum grid
  Reynolds number ($Re_\Delta^\mathrm{min}$) for $Re = 5 \times
  10^6$. (b) Subgrid boundary-layer region (in red) for
  $Re_\Delta^\mathrm{min} = 10^4$ at $Re = 5 \times 10^6$. Panels (c)
  and (d) are the number of grid points as a function of (c)
  $N_\mathrm{bl}$ and (d) $Re_\Delta^\mathrm{min}$. The dashed lines
  in panels (c) and (d) represent the power laws in
  Eq. (\ref{eq:growth}). \label{fig:cost}}
\end{figure}

We estimate the number of grid points (or control volumes) to conduct
WMLES of the NASA Juncture Flow as a function of the Reynolds number
($Re$), the number of points per boundary-layer thickness
($N_\mathrm{bl}$) and the minimum grid Reynolds number
($Re_{\Delta}^\mathrm{min}$). We assume the gridding strategy (ii) and
utilize the Juncture Flow geometry. The boundary-layer thickness was
obtained following the procedure in Section \ref{subsec:tbl}. The
total number of points, $N_\mathrm{points}$, to grid the boundary
layer spanning the surface area of the aircraft $S_a$ is
\begin{equation}\label{eq:points}
  N_\mathrm{points} = \int_{0}^{\delta} \iint_{S_a} \frac{1}{\Delta(x_{||},z_{||})^3} \mathrm{d}x_{||} \mathrm{d}z_{||} \mathrm{d}y_\bot 
  = \iint_{S_a} \frac{N_\mathrm{bl}}{\Delta(x_{\parallel},z_{\parallel})^2} \mathrm{d}x_{||} \mathrm{d}z_{||}, 
\end{equation}
where $x_{\parallel}$ and $z_{\parallel}$ are the local aircraft
wall-parallel directions, and $y_\bot$ is the local wall-normal
direction. Equation (\ref{eq:points}) only accounts for the points
within the boundary layer and neglects the points in the farfield and
the transition grid in between. However, the latter are usually
smaller than the number of points in the vicinity of the aircraft
wall. Note that all the cost estimates of WMLES are predicated upon
Eq. (\ref{eq:points}) with BL-conforming
grids~\citep[see][]{Chapman1979, Spalart1997, Choi2012, Yang2021};
yet, to the best of our knowledge WMLES of an aircraft geometry has
never been conducted using a true BL-conforming grids until the
present work. Equation (\ref{eq:points}) is integrated numerically and
the results are shown in Figure \ref{fig:cost}.  The cost map in
Figure \ref{fig:cost}(a) contains $\log_{10}(N_\mathrm{points})$ as a
function of $N_\mathrm{bl}$ and $Re_{\Delta}^\mathrm{min}$. The
accuracy of the solution is expected to improve for increasing values
of $N_\mathrm{bl}$ (i.e., higher energy content resolved by the LES
grid), and decrease with increasing $Re_{\Delta}^\mathrm{min}$ (i.e.,
worse leading-edge resolution). $Re_{\Delta}^\mathrm{min}$ can also be
understood as the largest subgrid boundary-layer that can be resolved
by the LES grid using $N_\mathrm{bl}$ points.  Figure
\ref{fig:cost}(b) provides an  illustration of the
subgrid-boundary-layer region for $Re_\Delta^\mathrm{min} < 10^4$,
which is confined to a small region (less than 10\% of the chord) at
the leading edge of the wing.
%

The scaling properties of Eq. (\ref{eq:points}) can be unfolded by
considering a wall-attached, zero-pressure gradient flat-plate
turbulent boundary layer (ZPGTBL) with $\delta \sim
(x-x_e)[(x-x_e)U_\infty/\nu]^{-m}$, where $x_e$ is the streamwise
distance to the leading edge and $m\approx-1/7$~\citep{Nagib2007}. If
we further assume that $Re \gg Re_\Delta^\mathrm{min}$, the number of
control volumes can be shown to scale as
\begin{equation}\label{eq:growth}
N_\mathrm{points} \sim N_\mathrm{bl}^{13/6} Re \left( Re_\Delta^\mathrm{min} \right)^{-5/6},
\end{equation}
which is also included in Figures \ref{fig:cost}(c) and (d) for
reference.  Equation (\ref{eq:growth}) is the generalization of the
WMLES cost scaling from previous works~\citep{Chapman1979,
  Spalart1997, Choi2012, Yang2021} that explicitly accounts for
$N_\mathrm{bl}$ and $Re_\Delta^\mathrm{min}$ (see appendix A for the
derivation). Additionally, the number of time steps to integrate the
equations of motion for a time $T$ is
\begin{equation}\label{eq:Nsteps}
N_\mathrm{steps} =
T/\Delta t \sim T U_\infty/\Delta_\mathrm{min} = \tilde{T} Re ( Re_\Delta^\mathrm{min})^{-1},
\end{equation}
where $\Delta t$ is the time step, $\tilde{T} = T U_\infty/L$ is the
flow-through time with respect to the crank chord, and we have assumed
that the simulation is constrained by the convective time-step $\Delta
t \sim \Delta_\mathrm{min}/U_\infty$. Equation (\ref{eq:Nsteps}) shows
that both $Re$ and $Re_\Delta^\mathrm{min}$ (i.e., leading-edge
resolution) are key contributors to the time-integration cost of
WMLES. The total, space-time computational cost is given by $N_c = C(
N_\mathrm{points} \cdot N_\mathrm{steps})$, where $C$ is a
machine/code-dependent function than converts $N_\mathrm{points} \cdot
N_\mathrm{steps}$ to core-hours.
 
The Reynolds number considered in Figure \ref{fig:cost} is $Re=5
\times 10^6$, which is representative of wind tunnel experiments. For
an actual aircraft in flight conditions, the typical Reynolds number
is $Re\approx5 \times 10^7$, that would increase $N_\mathrm{points}$
by roughly a factor of ten due to the thinning of the boundary layers
as seen in Eq. (\ref{eq:growth}).  More notably, Eq. (\ref{eq:growth})
shows that the cost of WMLES with $N_\mathrm{bl}$ scales roughly as
$N_\mathrm{bl}^{13/6} \approx N_\mathrm{bl}^2$, which becomes very
computationally demanding even for moderate values of
$N_\mathrm{bl}$. The dependence of $N_\mathrm{points}$ on
$Re_\Delta^\mathrm{min}$ is milder, but values of
$Re_\Delta^\mathrm{min} < 10^3$ become rapidly unattainable. This
reiterates the critical importance of establishing the range of
admissible $N_\mathrm{bl}$ and $Re_\Delta^\mathrm{min}$ to achieve the
desire accuracy for a given quantity of interest. Additionally,
different quantities of interest might not need to share the same grid
requirements. For example, if $N_\mathrm{bl}\approx 5$ and
$Re_\Delta^\mathrm{min} \approx 10^4$ suffice to attain the desired
accuracy in the quantities of interest at $Re=5 \times 10^6$, then the
required number of points is $\mathcal{O}(10)$ million, which can be
currently simulated in hours using $\mathcal{O}(1000)$ CPU
cores. However, if the desired accuracy for the quantities of interest
is such that $N_\mathrm{bl}\approx 20$ and $Re_\Delta^\mathrm{min}
\approx 10^3$, the number of grid points increases up to
$\mathcal{O}(1000)$ million, which renders WMLES unfeasible as a
routine tool in industry. Hence, the key to the success of WMLES as a
design tool resides in the accuracy of the solution achieved as a
function of $N_\mathrm{bl}$ and $Re_\Delta^\mathrm{min}$. This calls
for a systematic error characterization of the quantities of interest,
which is the objective of the following sections.

 \section{Error scaling of WMLES}
 \label{sec:errors}

\subsection{Error definition} 

The purpose of this section is to establish the expectations of the
error scaling of WMLES in ZPGTBL.  This error will serve as the
baseline to determine whether WMLES is over- or under-performing in
the NASA Juncture Flow calculations. The analysis is motivated by the
fact that the solutions from WMLES are intrinsically grid-dependent,
i.e. the grid size is an explicit variable of the governing
equations. As such, WMLES should be framed as a convergence study and
multiple computations are required in order to faithfully assess the
quality of the results.  This raises the fundamental question of what
is the expected WMLES error as a function of the flow parameters and
grid resolution. To tackle this problem, we follow the error-scaling
methodology from \citet{Lozano2019a}. Taking the experimental values
($\boldsymbol{q}^\mathrm{exp}$) as ground truth, the relative error in
a quantity of interest $\langle \boldsymbol{q} \rangle$ is defined as
\begin{equation}\label{eq:error_general}
  \varepsilon_q \equiv \frac{||\langle \boldsymbol{q}^\mathrm{exp}\rangle-\langle \boldsymbol{q} \rangle||_n}{||\langle \boldsymbol{q}^\mathrm{exp}\rangle||_n}
  = f\left( \frac{\Delta}{\delta}, Re, \mathrm{Ma},\mathrm{geometry},...\right),
\end{equation}
where $||\cdot||_n$ is the L$_2$-norm over the vector components and
spatial coordinates of $\langle\boldsymbol{q}\rangle$, and $f$ is the
error function that in general depends on the non-dimensional
parameters of the problem and the geometry.  For a given geometry and
flow regime, the error function in Eq. (\ref{eq:error_general}) in
conjunction with the cost map in Figure \ref{fig:cost}(a) determines
whether WMLES is a viable approach in terms of accuracy of the
quantity of interest and computational resources available. For the
case of NASA Juncture Flow considered here, the geometry, $Re$, and Ma
are set parameters. If we further assume that the error of a quantity
$\langle\boldsymbol{q}\rangle$ follows a power law (i.e., functional
invariance under grid-size rescaling), Eq. (\ref{eq:error_general})
can be simplified as
\begin{equation}\label{eq:error_simply}
\varepsilon_q = \beta_q \left( \frac{\Delta}{\delta} \right)^{\alpha_q},
\end{equation}
where $\beta_q$ and $\alpha_q$ are the error constant of
proportionality and error convergence rate, respectively, that depend
on the modeling approach (SGS model, wall model, numerical scheme,...)
and flow regime (i.e., laminar flow, fully turbulent flow, separated
flow,...).

We focus on the error scaling of the surface pressure coefficient
($\varepsilon_p$) and mean velocity profile in outer units
($\varepsilon_u$).  It is certain that, from an engineering viewpoint,
the lift and drag coefficients are the most pressing quantities of
interest in aerodynamic applications. However, both are integrated
quantities susceptible to error cancellation.  They also lack
information about the spatial structure of the flow, which makes more
challenging the detection of modeling errors. On the other hand, the
granularity provided by pointwise time-averaged quantities, such as
the mean pressure and velocity, greatly facilitates the identification
of modeling deficiencies.  Hence, we will exploit the errors in
pressure coefficient and mean velocity profile as a proxy to measure
the quality of the WMLES solution.

\subsection{Reference error scaling for mean velocity profile and pressure coefficient} 
\label{subsec:errors}

To aid the interpretation of the results, it is informative to derive
theoretical estimations for the error scaling of the pressure
coefficient, the mean velocity profile, and the wall stress in
simplified flow scenarios.  For wall-attached flows, errors in $C_p$
can be assumed to be dominated by inviscid effects. Under the thin
boundary-layer approximation, the wall-normal-integrated spanwise mean
momentum equation yields $ \overline{p} + \overline{\rho v^2} \approx
p_I \Rightarrow p_\mathrm{wall} = p_I$, where $p_I$ is the inviscid
farfield pressure and $\overline{(\cdot)}$ is the average in the
spanwise direction and time.  Thus, the pressure at the surface is
mostly controlled by the inviscid imprint of the outer flow and we can
expect
\begin{equation}\label{eq:error_p}
\varepsilon_p = \beta_p
\left(\frac{\Delta}{\delta}\right)^{\alpha_p} Re^0,
\mathrm{with} \ \beta_p\ll 1, \ {\alpha_q} \approx 0.
\end{equation}
Equation (\ref{eq:error_p}) implies that errors in $C_p$ should be small
even when the boundary layer is marginally resolved (i.e., $\Delta
\approx \delta)$, and should exhibit a weak dependence on the
grid resolution.

The error scaling of the mean velocity profile can be estimated by
assuming WMLES of a turbulent channel flow in which the kinetic energy
spectrum follows $E_k \sim 1/\Delta^a$ and the velocity gradients
scale as $\partial u_i/\partial x_j \sim
\sqrt{E_k}/\Delta$~\citep{Lozano2019a}. The exponent $a$ depends on
the regime the SGS models operates in: $a = -1$ for the
shear-dominated range~\citep{Perry1977, Jimenez2018} and $a= -5/3$ for
the inertial range~\citep{Kolmogorov1941}. Taking into account the
scaling above, the expected error of the mean velocity profile is
\begin{equation}
  \label{eq:error_mean}
\varepsilon_u = \beta_u \left( \frac{\Delta}{\delta}\right)^{\alpha_u} Re^0, \ \mathrm{with} \ \alpha_u=0 \ \mathrm{or} \ \alpha_u=1,
\end{equation}
where $\alpha_u=0$ for $\Delta/\delta$ lying on the shear-dominated
region, and $\alpha_u=1$ for $\Delta/\delta$ within the inertial
range~\citep{Lozano2019a}. The constant $\beta_u$ depends on the SGS
model, numerical schemes, and error propagation from the upstream
flow. In the case of turbulent channel flow (no upstream error
propagation) $\beta_u$ is bounded by the error of the inviscid
solution such that $\beta_u <\mathcal{O}(10^{-1})$ as discussed below.
The results from Eq. (\ref{eq:error_mean}) also indicate that no
improvement in the error is expected for grid resolutions comparable
to the scales in the shear-dominated region, whereas an approximately
linear scaling can be anticipated for finer grids with sizes
comparable to the scales in the inertial range. The conclusion is
consistent with the phenomenological argument that capturing the
energy injection mechanism from the mean shear is critical to achieve
accurate LES results.

\begin{figure}
\begin{center}
\subfloat[]{\includegraphics[width=0.47\textwidth]{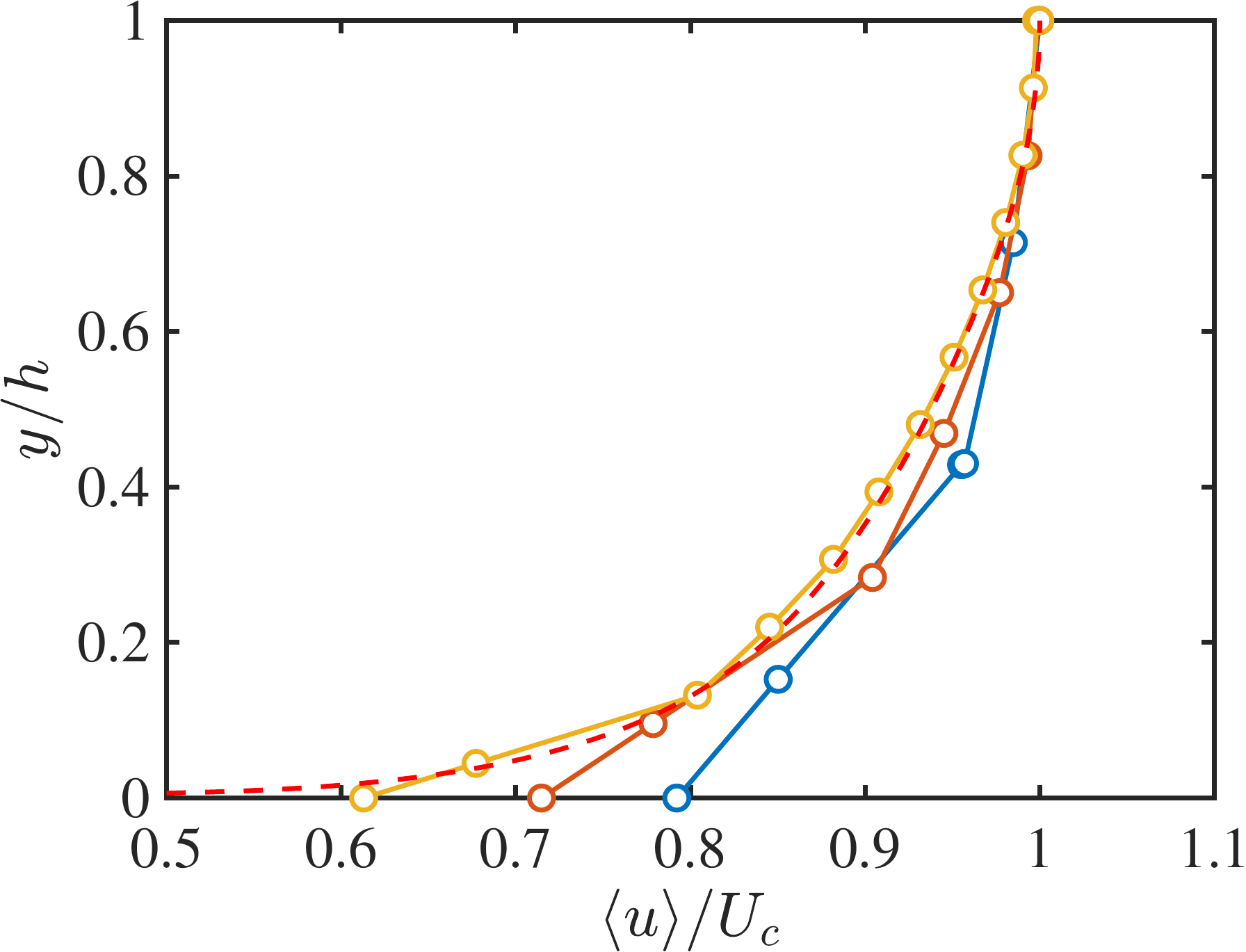}}
\hspace{0.5cm}
\subfloat[]{\includegraphics[width=0.47\textwidth]{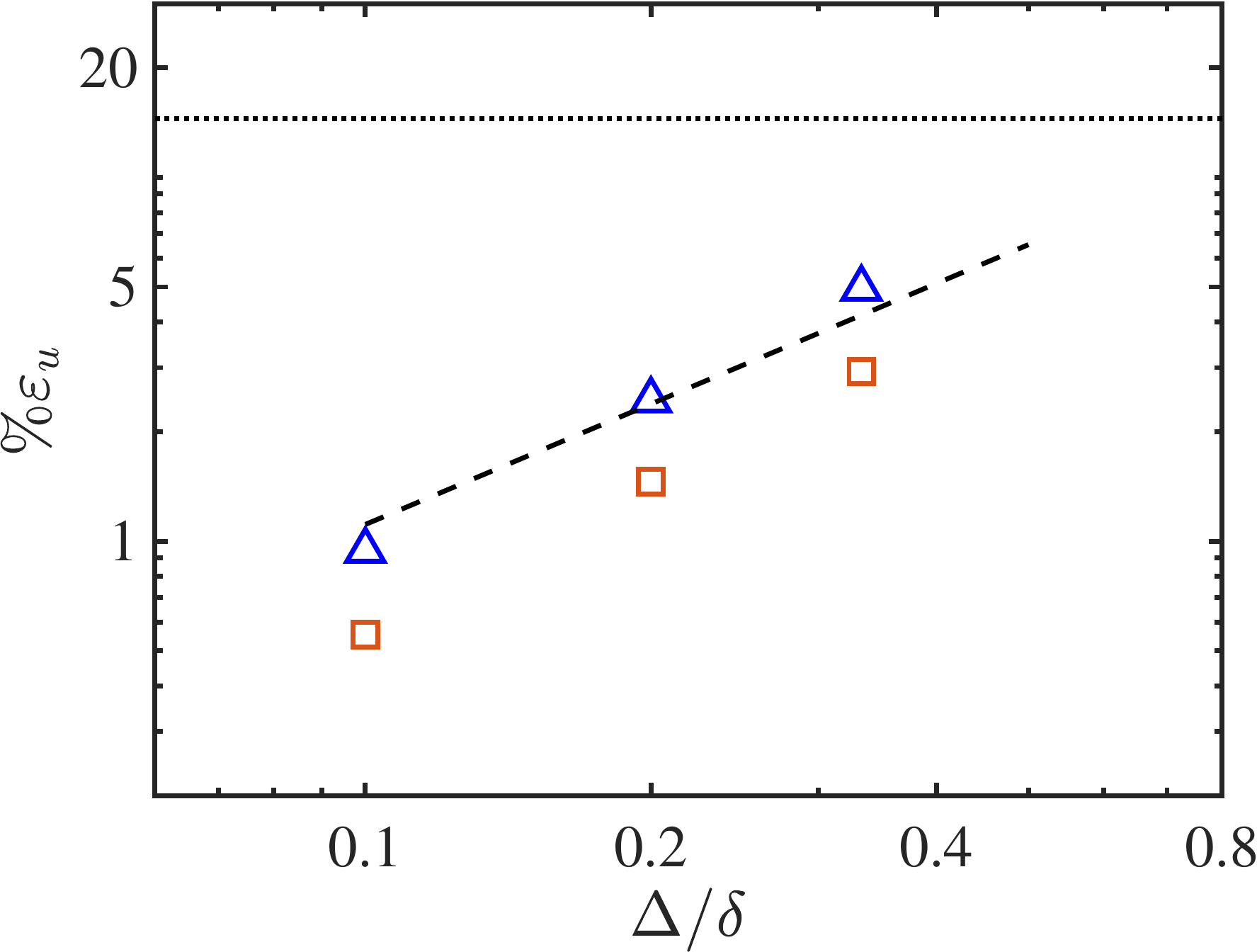}}
\end{center}
\caption{(a) The mean velocity profile for WMLES with charLES of a
  turbulent channel flow. The colors denote different grid resolutions
  $\Delta/\delta = 1/3, 1/5$, and $1/10$. (b) Error in the mean
  velocity profile $\varepsilon_u$ as a function of the grid
  resolution $\Delta$ for WMLES of turbulent channel flow.  The
  symbols denote simulations using charLES with Voronoi grids
  ($\square$) and finite-difference solver with staggered grids
  ($\triangle$). The dashed line is $\varepsilon_u \sim
  \Delta/\delta$. The horizontal dotted line is the error from the
  inviscid solution.  \label{fig:mean_error}}
\end{figure}

The error scaling from Eq. (\ref{eq:error_mean}) is validated for
WMLES of turbulent channel flows at $Re_\tau=4200$. The channels are
driven by setting the center line velocity ($U_c$) to a constant value
akin to $U_\infty$ in a turbulent boundary layer. The calculations are
conducted with charLES using dynamic Smagorinsky model and the
equilibrium wall model. We consider isotropic Voronoi grids with
characteristic sizes of $\Delta/\delta = 1/3, 1/5$, and $1/10$, which
are representative of the grid resolutions currently affordable for
WMLES of the NASA Juncture Flow Experiment.  Figure
\ref{fig:mean_error}(a) shows the mean velocity profiles for the three
grid resolutions and offers a visual reference of the convergence of
the mean velocity to the DNS solution from
\citet{Lozano-Duran2014}. The theoretical error scaling from
Eq. (\ref{eq:error_mean}) is tested in Figure
\ref{fig:mean_error}(b). The figure also compares the errors from
charLES (squares) with those obtained using a different numerical
scheme and grid strategy, namely, finite differences with staggered
grids (triangles) from \citet{Lozano2019a} (see appendix B). The
results show that $\varepsilon_u\sim (\Delta/\delta)$ stands as a
sensible approximation of the error for the two solvers considered,
providing confidence in the theoretical arguments involved in the
derivation of Eq. (\ref{eq:error_mean}). A useful approximation of the
error in charLES is $\varepsilon_u\approx 0.08 (\Delta/\delta)$, which
will be used later for comparison purposes with the NASA Juncture
Flow. As a reference, Figure \ref{fig:mean_error}(b) also includes the
largest error expected in a turbulent channel flow (dotted line)
defined as the difference between the mean velocity from DNS and the
inviscid solution, which gives an error of about $\sim$15\%. Thus,
even a low performance WMLES is bounded by a maximum error of
$\sim$15\% in canonical ZPGTBL (and channel flows) due to the
constraint imposed by the freestream. Note that the reference inviscid
error might be larger in more complex geometries.
%

Although not shown in Figure \ref{fig:mean_error}(b), the linear
convergence predicted from Eq. (\ref{eq:error_mean}) breaks down for
$\Delta \approx 0.03$--$0.05\delta$ due to the interplay between the
numerical errors and the SGS model~\citep{Lozano2019a}. This anomaly
does not manifest in our NASA Juncture Flow calculations because of
the choice of grid resolutions and will not be discussed here except
for the comments in appendix B. Nonetheless, this behavior is
responsible for the non-monotonic convergence of $\varepsilon_u$ with
$\Delta$, which is an outstanding problem of WMLES. Both charLES and
the finite-difference solver suffer from this non-monotonic behavior,
and the reader is referred to appendix B for more details.


The last important consideration is that Eq.  (\ref{eq:error_mean}) is
meant to be valid for wall turbulence over approximately flat
surfaces. In the case of flows in the vicinity of corners, corrections
are required to account for the errors due to the proximity of lateral
walls.  \citet{Lozano2019a} introduced a generalized version of
Eq. (\ref{eq:error_mean}) and showed that WMLES errors in the mean
velocity profile at a given wall-normal distance ($d$) are roughly
controlled by the local shear length-scale, $\varepsilon_u \sim
\Delta/L_s$, where $L_s \approx u_\tau/S$ and $S$ is the mean
shear at $d$. Under the assumption that $S \sim u_\tau /d$ (akin
to the log layer), the error follows $\varepsilon_u \sim \Delta/d$.
If we further assume that the error close to a corner is dominated by
the influence of the closest wall (located at a distance
$d_\mathrm{min}$), we can define the compensated error as $\varepsilon_u
d_\mathrm{min}/\delta$. The latter error will be instrumental to
assessing the accuracy of WMLES in corner regions with respect to the
reference error in turbulent channel flows.

\subsection{Wall-modeling errors} 

\begin{figure}
\begin{center}
\subfloat[]{\includegraphics[width=0.45\textwidth]{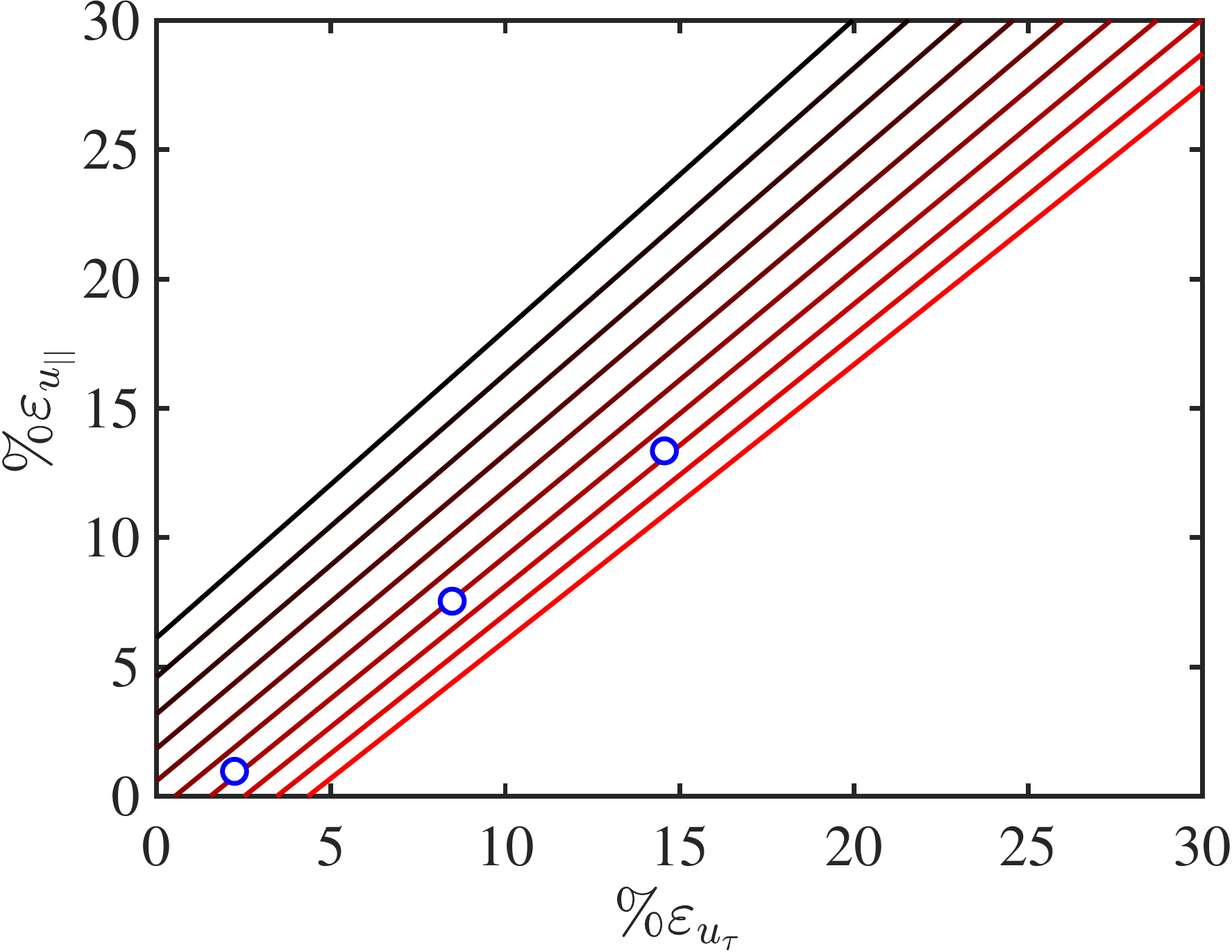}}
\hspace{0.5cm}
\subfloat[]{\includegraphics[width=0.45\textwidth]{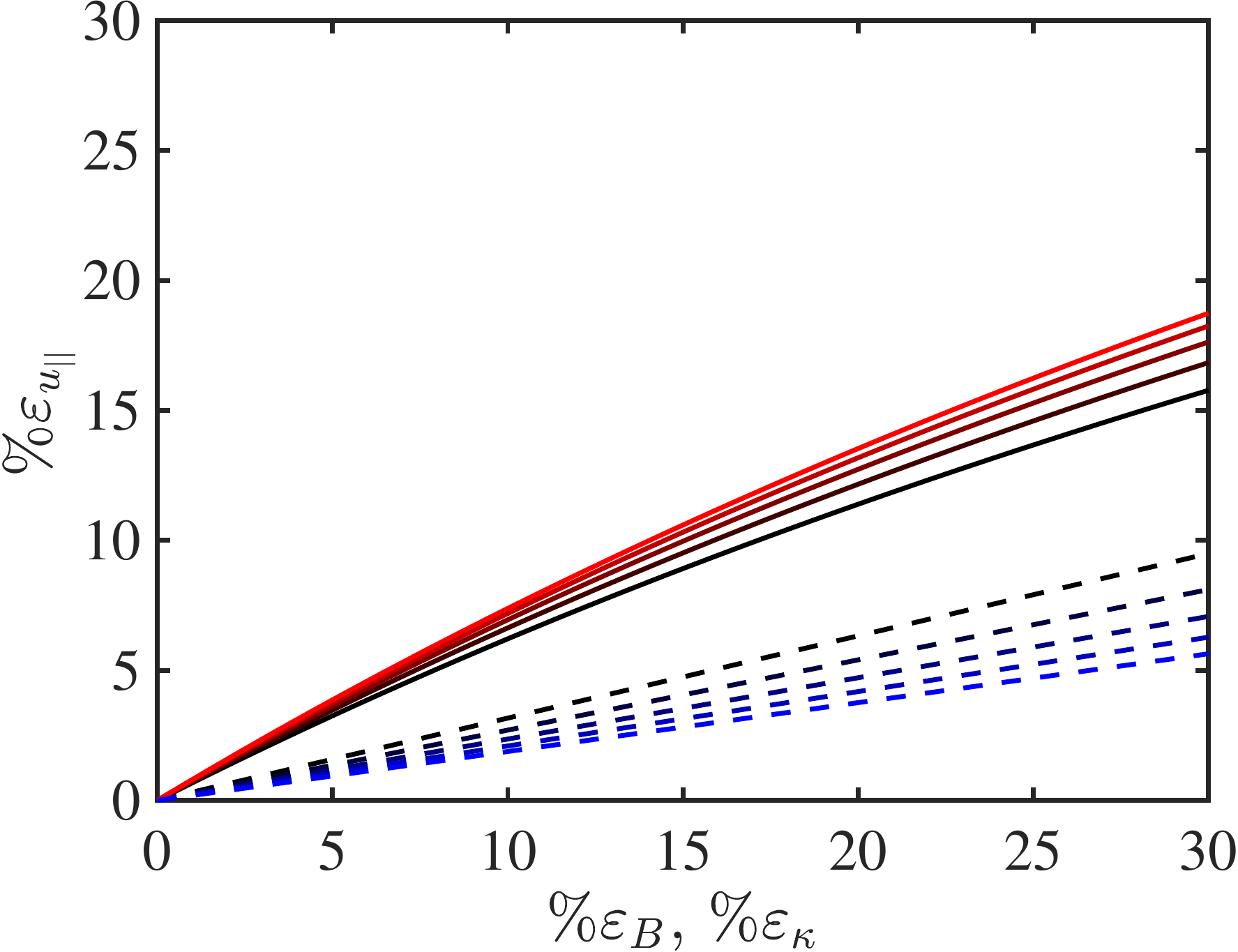}}
\end{center}
\caption{ Error propagation between LES and wall model from
  Eq. (\ref{eq:wm_error}). (a) Error in the LES wall-parallel velocity
  $u_{||}$ at the matching location as a function of the error in the
  wall-stress predicted by the equilibrium wall. The solid lines
  represent $\varepsilon_\kappa = \varepsilon_B = -10\%, -7.5\%,
  -3.5\%, -1\%, 1\%, 3.5\%, 7.5\%,10\%$ from black to red and
  $Re_{h_w} = 1000$. The circles are the actual errors measured by
  WMLES of turbulent channel flows with charLES for
  $\Delta/\delta=1/3, 1/6$ and $1/10$. (b) Error in the LES
  wall-parallel velocity $u_{||}$ at the matching location as a
  function of the error in the model parameters $\kappa$ (solid lines)
  and $B$ (dashed lines) for different values of matching location
  Reynolds number $Re_{h_w} = 100, 300, 1000, 3000$, and $10000$ from
  black to red/blue.\label{fig:WM_error}}
\end{figure}

We analyze the error propagation between the mean velocity profile and
the wall-stress predictions by the equilibrium wall model in
Eq. (\ref{eq:charles_algwm}). To that end, we express the quantity $q$
in terms of its exact value $\breve{q}$ and its relative error
$\varepsilon_q$ as $q = \breve{q}(1+\varepsilon_q)$. Assuming that the
matching location between LES and the wall model ($h_w$) lies in the
log layer, the error in the LES wall-parallel velocity ($u_{||}$) at
the matching location is given by
\begin{equation}
 \label{eq:wm_error}
  \varepsilon_{u_{||}} =
  \frac{1+\varepsilon_{u_\tau}}{1+\varepsilon_{\kappa}}
  \left(1+\frac{\ln(1+\varepsilon_{u_\tau})+\breve{\kappa} \breve{B}
    (\varepsilon_{\kappa}+\varepsilon_B+\varepsilon_{\kappa}
    \varepsilon_B)}{\breve{\kappa}\breve{\phi}}\right)-1,
\end{equation}
where $\breve{\phi} = 1/\breve{\kappa}\ln(Re_{h_w}) + \breve{B}$ is
the model reference solution, $Re_{h_w} = \breve{u}_\tau h_w/\nu$ is
the Reynolds number based on the matching location, and
$\varepsilon_{\kappa}$ and $\varepsilon_{B}$ are the relative errors
in the model constants $\kappa$ and $B$, respectively. Two sources of
errors can be identified in Eq. (\ref{eq:wm_error}): errors from the
LES mean velocity ($\varepsilon_{u_{||}}$) referred to as
\emph{external wall-modeling errors}, and errors from the model
parameters ($\varepsilon_{\kappa}$ and $\varepsilon_{B}$), referred to
as \emph{internal wall-modeling errors}. In the former, errors in the
LES mean velocity profile at the matching location propagate to the
value of $u_\tau$ predicted by the wall model. It can be shown that
this error is roughly linear, $\varepsilon_{u_\tau} \propto
\varepsilon_{u_{||}}$. This is seen in figure \ref{fig:WM_error}(a),
which features the error in $u_{||}$ as a function of the error in
$u_\tau$ evaluated from Eq. (\ref{eq:wm_error}) for different values
of $\varepsilon_{\kappa}$ and $\varepsilon_{B}$.  Figure
\ref{fig:WM_error}(a) also includes the actual errors (circles) for
WMLES of a turbulent channel flow using charLES with $\Delta/\delta =
1/3, 1/5$, and $1/10$, which follows the linear trend anticipated from
our analysis. Note that this error can be labeled as external to the
wall model inasmuch as it is present even if the wall model provides
an exact representation of the near-wall region (i.e.,
$\varepsilon_{\kappa} = \varepsilon_{B}=0$). Although not shown, the
sensitivity of $\varepsilon_{u_\tau}$ to $Re_{h_w}$ is weak and scales
as $\varepsilon_{u_{||}} \sim \ln(Re_{h_w})$.

The second source of errors represents the internal wall-model
limitations: even with a perfect prediction of $u_{||}$ (namely,
$\varepsilon_{u_{||}}=0$), the wall model might incur errors due to
i) uncertainties in the parameters $\kappa$ and $B$ embedded into the
model, ii) deviations of the actual flow from these parameters (e.g.,
pressure gradient effects or roughness changing $\kappa$ and $B$),
and/or iii) the fact that the log-layer law is no longer
representative of the mean velocity profile (e.g., as in separated
flows). A consequence of internal errors is that WMLES might not
converge to the DNS solution with grid refinements until $\Delta$ is
in the DNS-like regime and the contribution of the wall model is
negligible.  Figure \ref{fig:WM_error}(b) quantifies the internal
errors in the mean velocity profile as a function of
$\varepsilon_{\kappa}$ (solid lines) and $\varepsilon_{B}$ (dashed
lines). Overall, wall models are more resilient to internal errors
than to external errors. For example, purely internal errors of
$\varepsilon_{\kappa}=10\%$ and $\varepsilon_{B}=10\%$ yield
$\varepsilon_{u_\tau}\approx 6\%$ and $\varepsilon_{u_\tau}\approx
3\%$, respectively, both of them lower than the sole external error
$\varepsilon_{u_\tau}\approx \varepsilon_{u_{||}}=10\%$. Another
interesting observation from Figure \ref{fig:WM_error}(a) is that
certain combinations of $\varepsilon_{\kappa}$, $\varepsilon_{B}$ and
$\varepsilon_{u_{||}}$ are subject to error cancellation, resulting in
a fortuitous good prediction of $u_\tau$. Note that the expression in
Eq. (\ref{eq:wm_error}) might be used even when the near-wall flow
does not comply with the log-layer law; however in that case
$\varepsilon_{\kappa}$ and $\varepsilon_{B}$ become arbitrarily large
and Eq. (\ref{eq:wm_error}) ceases to be informative of the error
scaling.  
%

In summary, external errors will propagate linearly to the mean
velocity profile, increasing $\beta_u$ without affecting $\alpha_u$
appreciably, while internal errors will bound the maximum possible
accuracy of the WMLES solution. Both sources of error, external and
internal, will impact the predictions of the NASA Juncture Flow.

\section{WMLES of the NASA Juncture Flow Experiment}\label{sec:JFE}

\subsection{WMLES Cases and uncertainties}\label{Cases}

\begin{figure}
\begin{center}
\includegraphics[width=1\textwidth]{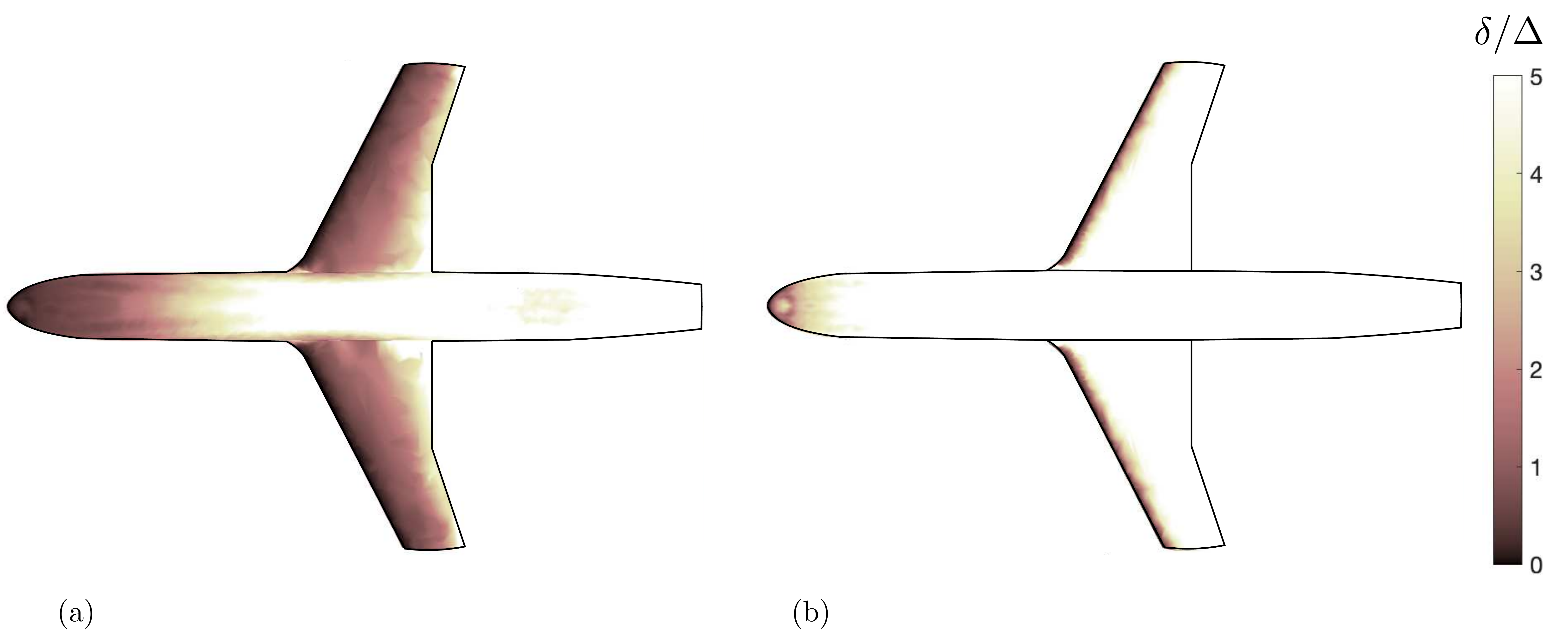}
\end{center}
\caption{Resolution map: points per boundary layer thickness
  ($\delta/\Delta$) for (a) constant-size grid for case C-D2 and (b)
  BL-conforming grid for case C-N5-R2e3. The colormap is clipped at
  $\delta/\Delta=5$ to facilitate the comparison between panel (a)
  and (b), but note that white areas might enclose regions with
  $\delta/\Delta>5$. \label{fig:res_maps}}
\end{figure}

We perform WMLES of the NASA Juncture Flow with a leading-edge horn at
$Re=2.4 \times 10^6$ and AoA=$5^\circ$.  Eight cases are
considered. In the first six cases, we employ grids generated using
strategy (i) with constant grid-size in millimeters similar to the
example offered in Figure \ref{fig:tbl_grids}(a). In this case, the
direct impact of $Re_\Delta^{\mathrm{min}}$ can be absorbed into
$\Delta/\delta$ as argued for Eq. (\ref{eq:error_simply}). The grid
sizes considered are $\Delta \approx 6.7, 4.3, 2.2, 1.1$, and $0.5$
mm, which are labeled as C-D7, C-D4, C-D2, C-D1, and C-D0.5,
respectively. Cases C-D7, C-D4, and C-D2 are obtained by refining the
grid across the entire aircraft surface. For cases C-D1 and C-D0.5,
the grid size is 1.1 and 0.5 mm, respectively, only within a box along
the fuselage and wing-body juncture defined by $x\in[1770,2970]$~mm,
$y\in[-300,-200]$~mm, and $z\in[-50,150]$~mm. The purpose of C-D1 and
C-D0.5 is to further examine the convergence of the solution in the
separated region, where we will show the performance of WMLES is the
poorest.  The refinement box was chosen to make the cases
computationally tractable. The computational cost $N_c$ of the
simulations was $31 \times 10^3$, $40 \times 10^3$, $75 \times 10^3$,
$116 \times 10^3$, $398 \times 10^3$ core-hours for C-D7, C-D4, C-D2,
C-D1, and C-D0.5, respectively, using Intel(R) Xeon(R) CPU E5-2670 0 @
2.60GHz (in the NASA Advanced Supercomputing facility Pleiades). The
total number of control volumes is 14, 17, 31, 50 and 174 million for
C-D7, C-D4, C-D2, C-D1, and C-D0.5, respectively.  Nonetheless, one
should be cautious in appraising the number of control volumes as an
indicator of the accuracy of the solution as discussed in Section
\ref{subsec:targeted}). A more sensible characterization of the grid
resolution is given by the \emph{resolution map}: distribution of
points per boundary layer thickness ($\delta/\Delta$) over the
aircraft surface.  Figure \ref{fig:res_maps}(a) features the
resolution map for case C-D4 and exposes the strong inhomogeneity in
$\delta/\Delta$ typical of constant-size grids.
%

\begin{figure}
\begin{center}
\includegraphics[width=0.7\textwidth]{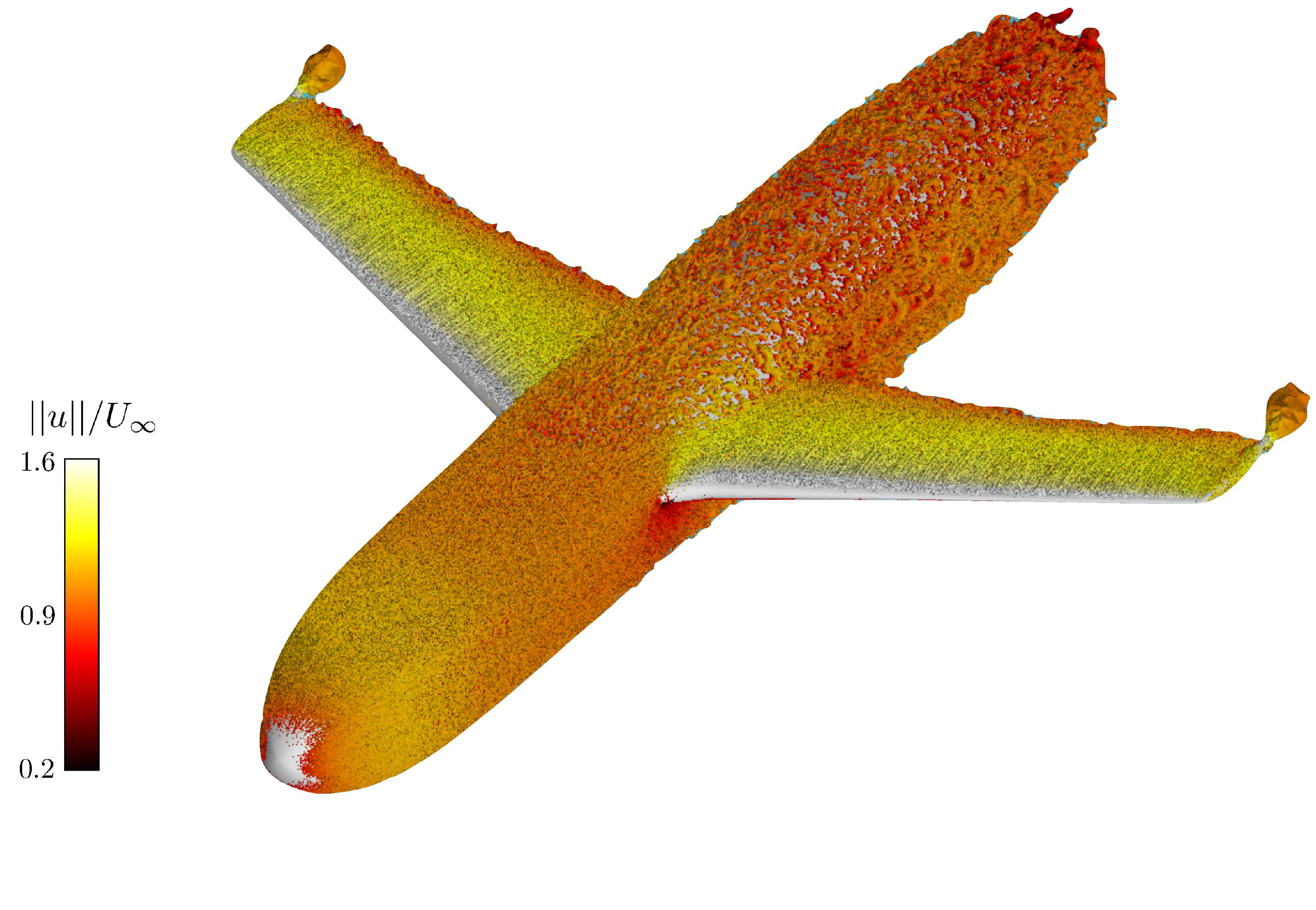}
\end{center}
\caption{Visualization of the instantaneous isosurfaces of the
  Q-criterion colored by the velocity magnitude. Data for
  C-N10-R2e3. \label{fig:q_criterion}}
\end{figure}

Three additional cases are considered to assess the impact of
BL-conforming grids on the accuracy of the predictions. The grids are
generated using strategy (ii) for $N_\mathrm{bl} = \delta/\Delta = 5$
and $Re_\Delta^\mathrm{min}=2.8\times 10^3$ (denoted as case
C-N5-R2e3); $N_\mathrm{bl} = \delta/\Delta = 10$ and
$Re_\Delta^\mathrm{min}=2.8\times 10^3$ (C-N10-R2e3); and
$N_\mathrm{bl} = \delta/\Delta = 5$ and
$Re_\Delta^\mathrm{min}=5.6\times 10^3$ (C-N5-R5e3). The first two
cases are intended to provide information about the effect of
$N_\mathrm{bl}$ on the accuracy of the results, whereas the third case
is used to evaluate the impact of the leading-edge resolution. The
grid Reynolds numbers considered set the minimum allowed grid
resolution to $\Delta_\mathrm{min}\approx0.6$ mm,
$\Delta_\mathrm{min}\approx0.3$ mm, and $\Delta_\mathrm{min}\approx
1.3$ mm for cases C-N5-R2e3, C-N10-R2e3, and C-N5-R5e3, respectively.
A cross section of the BL-conforming grid for C-N5-R2e3 is shown in
Figure \ref{fig:tbl_grids}(b). The computational cost $N_c$ of the
simulations was $27\times 10^3$, $90\times 10^3$, and $13\times 10^3$
core-hours for C-N5-R2e3, C-N10-R2e3, and C-N5-R5e3 respectively,
using Intel(R) Xeon(R) CPU E5-2670 0 @ 2.60GHz. The total number of
control volumes for C-N5-R2e3, C-N10-R2e3, and C-N5-R5e3 is 12 , 40,
and 6 million, respectively. Figure \ref{fig:res_maps}(b) shows the
resolution map for C-N5-R2e3. Compared to the constant-size grid
counterpart in Figure \ref{fig:res_maps}(a), BL-conforming grids
display an homogeneous grid resolution in terms of $\delta/\Delta$
except at the leading-edge region, where the limitation imposed by
$Re_\Delta^\mathrm{min}$ comes into effect. An instantaneous
visualization of the Q-criterion~\citep{Hunt1988} for case C-N5-R2e3
is shown in Figure \ref{fig:q_criterion} to provide a general overview
of turbulent flow around the Juncture Flow geometry.
%

In the following, $\langle q \rangle$ and $(q)'$ denote the
time-average and the fluctuating component of $q$, respectively. For
comparison purposes, the profiles from WMLES are interpolated to the
locations of the experiments. Statistical uncertainties owing to
limited samples in WMLES quantities are estimated assuming
uncorrelated and randomly distributed errors following a normal
distribution.  The uncertainty in $\langle q \rangle$ is then
estimated as $\langle q \rangle_u \equiv \sigma/\sqrt{N_s}$, where
$N_s$ is the number of samples in $\langle q \rangle$, and $\sigma$ is
the standard deviation of the samples.  The uncertainties were
computed for all the quantities of interest, but are only reported for
the tangential Reynolds stresses. The uncertainties for the mean
velocity profiles and pressure coefficient were found to be below 1\%
and are not included in the plots.

\subsection{Mean velocity profiles with constant-size grids}
\label{subsec:vel_stress}

\begin{figure}
\begin{center}
\includegraphics[width=0.6\textwidth]{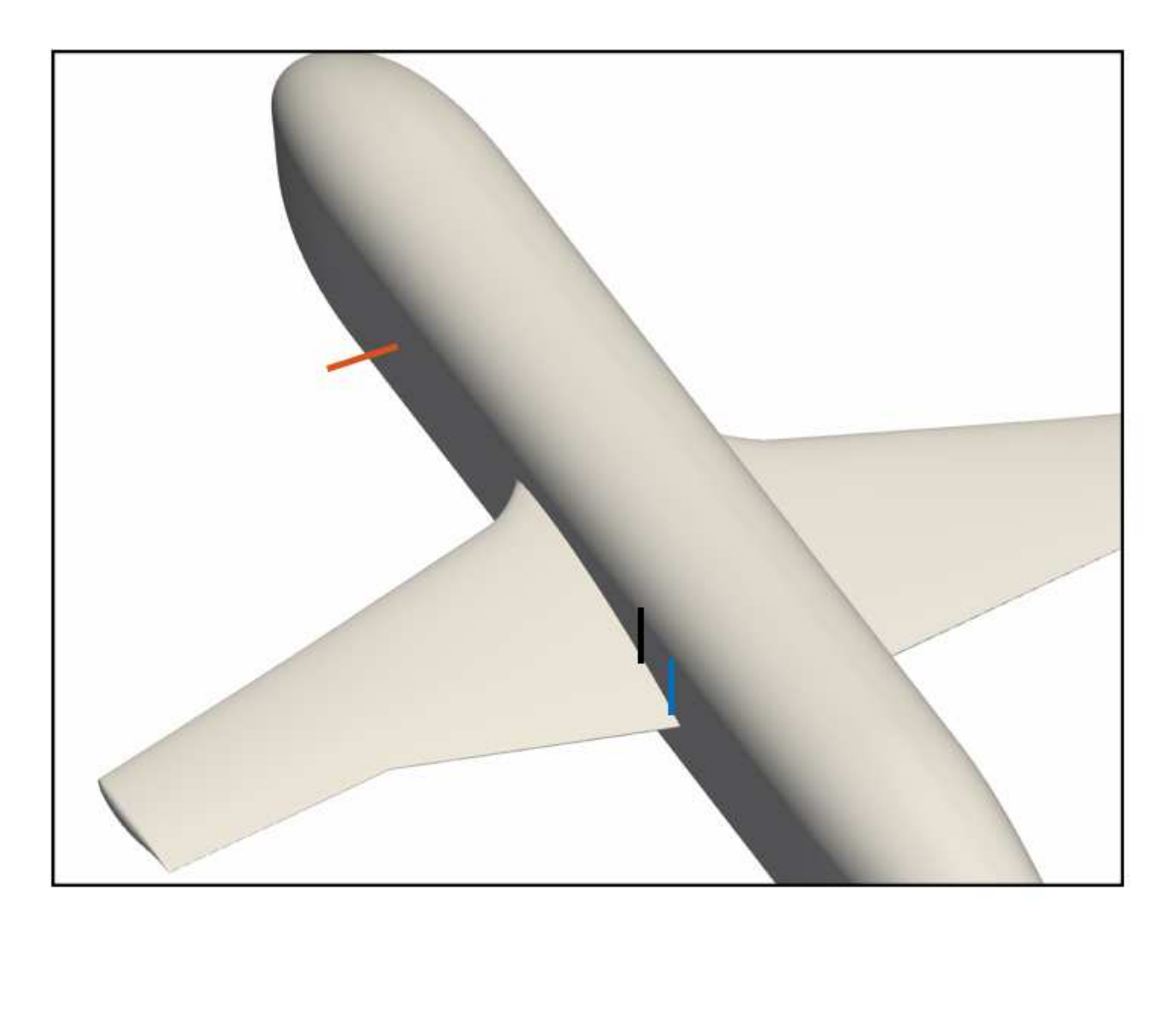}
\end{center}
\caption{ The three locations considered (solid lines) to investigate
  the error in the mean velocity profile: red (fuselage), black
  (juncture), and blue (separation). \label{fig:locations}}
\end{figure}

Three locations are considered to investigate the errors in the mean
velocity profile: (i) the upstream region of the fuselage
($x=1168.40$~mm, $z=0$~mm), (ii) the wing-body juncture
($x=2747.6$~mm, $y=239.1$~mm), and (iii) wing-body juncture with
separation close to the trailing edge ($x=2922.6$~mm,
$y=239.1$~mm). Figure~\ref{fig:locations} portrays the three locations
in the NASA Juncture Flow using solid lines of different
colors. Throughout the manuscript, we often refer to the three regions
simply as fuselage, juncture and separation/trailing-edge regions.  We
will assume that the mean velocity from WMLES is directly comparable
to unfiltered experimental data $\langle u_i^{\mathrm{exp}}
\rangle$, as it is case for quantities dominated by large-scale
motions~\citep{ Winckelmans2002, Lozano2019a}. The results for WMLES
with constant-size grids are presented next.
%

\begin{figure}
\begin{center}
\subfloat[]{\includegraphics[width=0.44\textwidth]{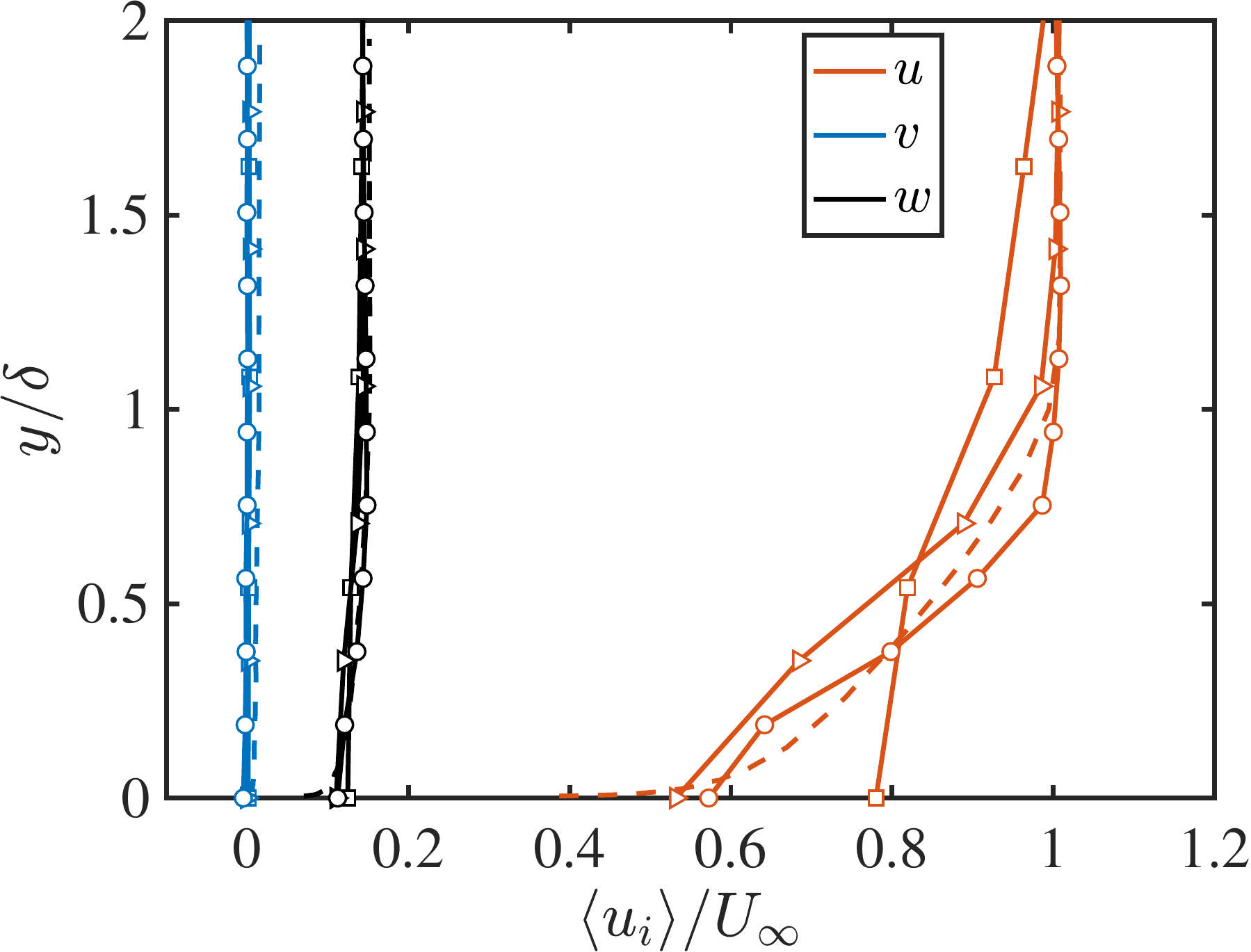}}
\hspace{0.5cm}
\subfloat[]{\includegraphics[width=0.47\textwidth]{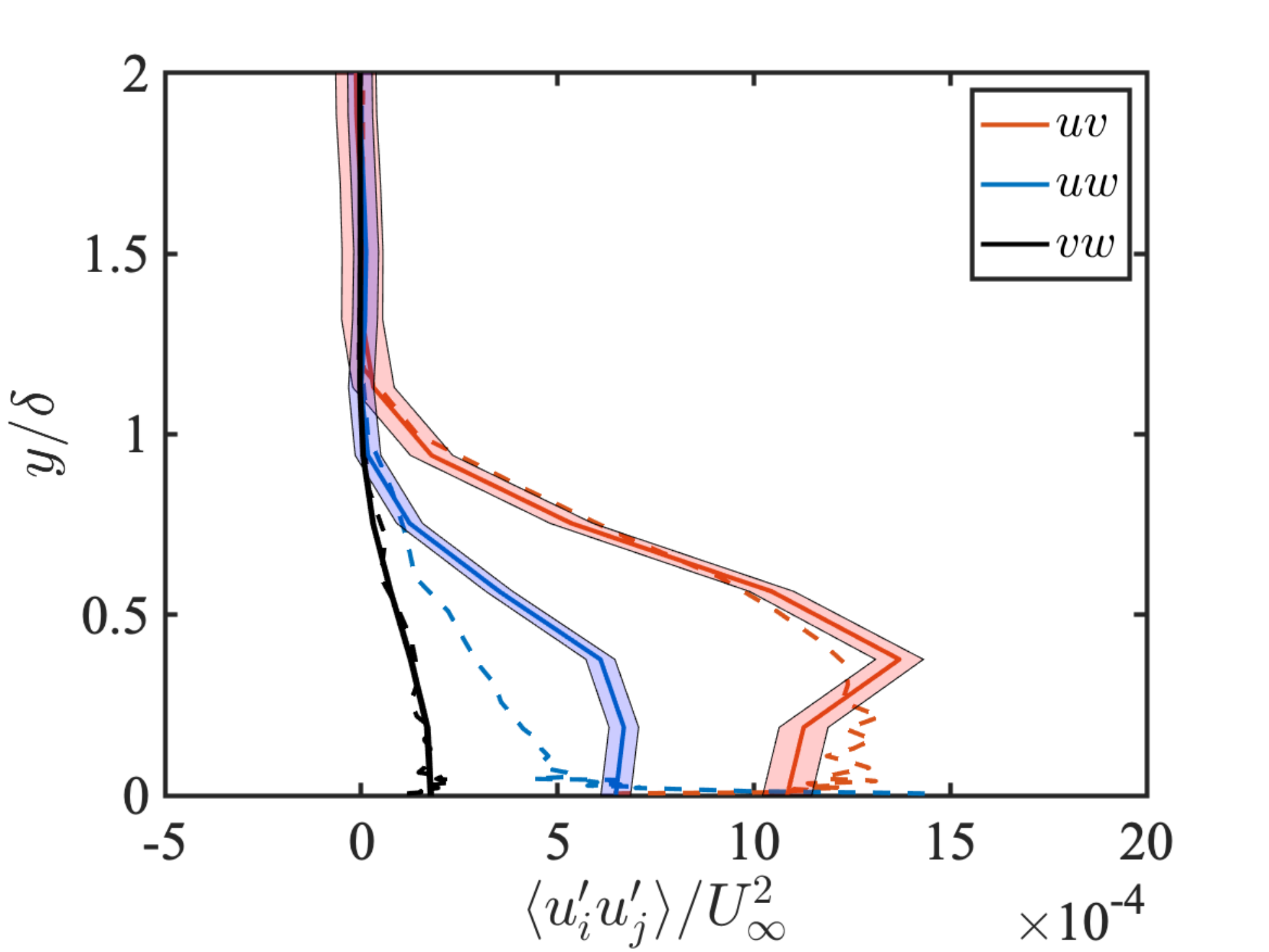}}
\end{center}
\caption{(a) Mean velocity profiles and (b) Reynolds shear stresses at
  location 1: upstream region of the fuselage $x=1168.4$~mm and
  $z=0$~mm (red line in Figure \ref{fig:locations}). Solid lines with
  symbols denote WMLES for cases C-D7 ($\square$), C-D4
  ($\triangleright$), and C-D2 ($\circ$). Colors denote different
  velocity components. Panel (b) only includes case C-D2 and the
  shaded area represents statistical uncertainty. Experiments are
  denoted by dashed lines. The distance $y$ is normalized by the local
  boundary-layer thickness $\delta$ at that
  location. \label{fig:fuselage}}
\end{figure}

\begin{figure}
\begin{center}
\subfloat[]{\includegraphics[width=0.44\textwidth]{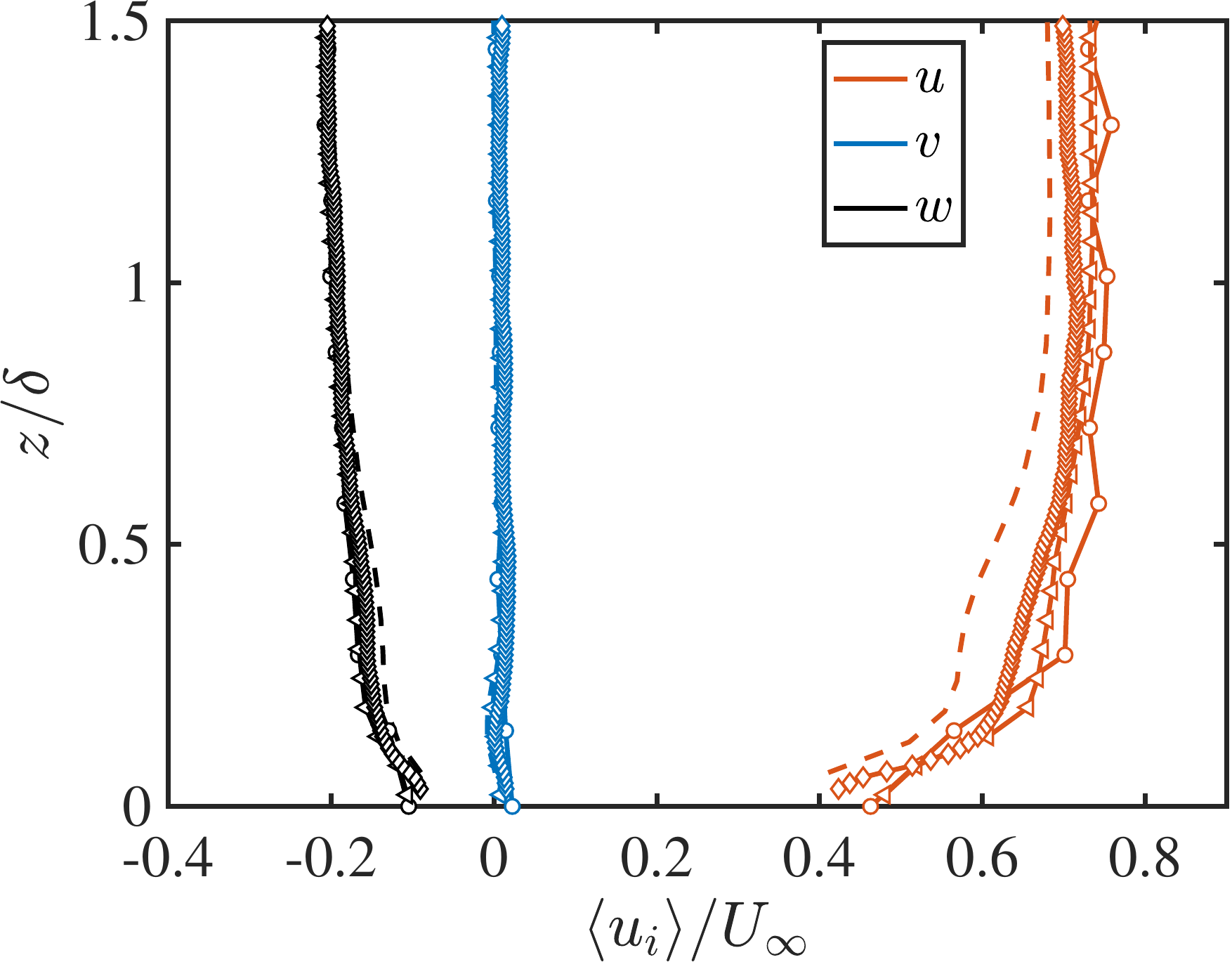}}
  \hspace{0.5cm}
\subfloat[]{\includegraphics[width=0.47\textwidth]{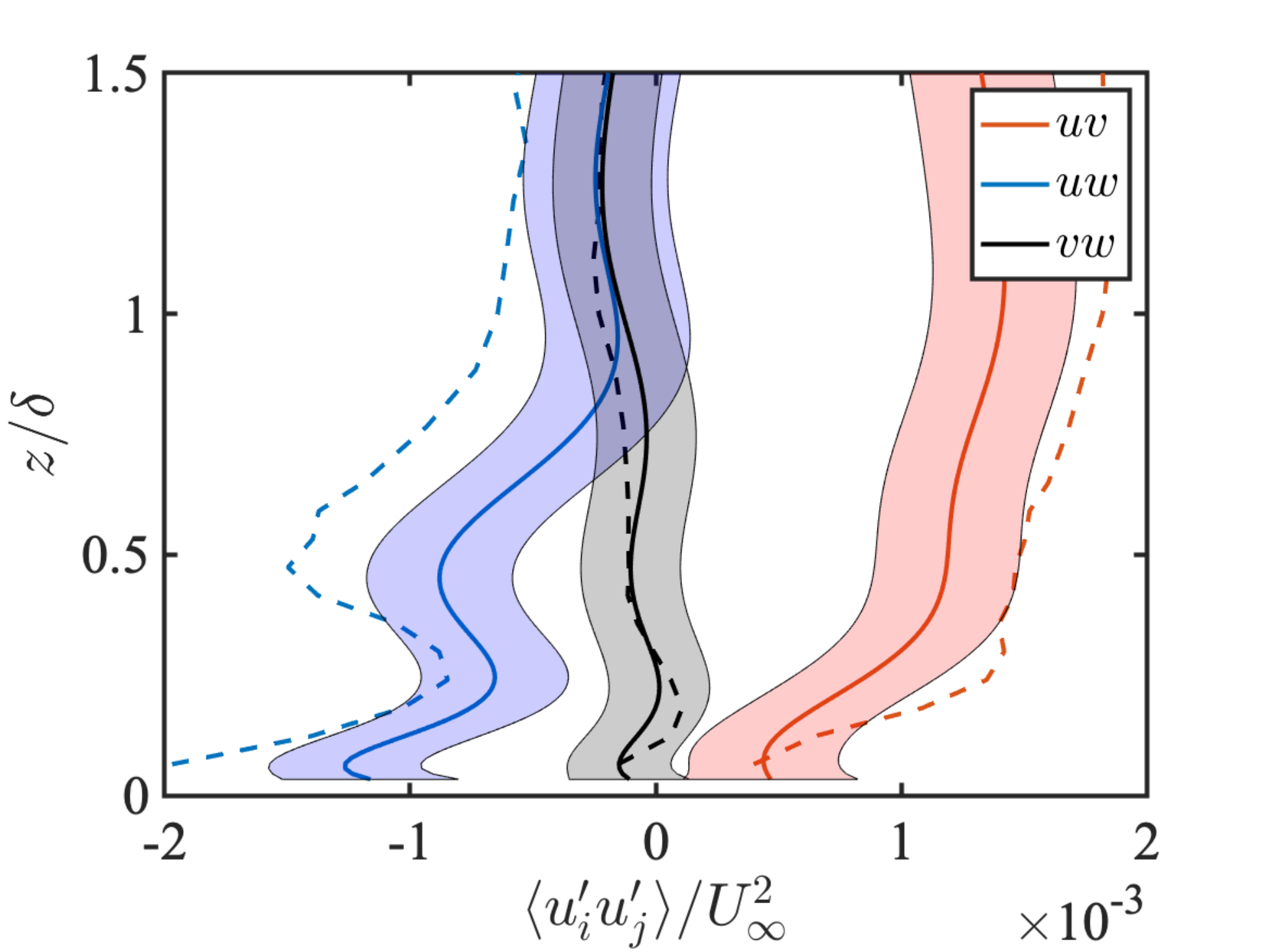}}
\end{center}
\caption{(a) Mean velocity profiles and (b) Reynolds shear stresses at
  location 2: wing-body juncture at $x=2747.6$~mm and $y=239.1$~mm
  (black line in Figure \ref{fig:locations}). In panel (a), lines with
  symbols are for cases C-D2 ($\circ$), C-D1 ($\triangleleft$), and
  C-D0.5 ($\diamond$). Colors denote different velocity
  components. Panel (b) only includes case C-D0.5 and the shaded area
  represents statistical uncertainty. Experiments are denoted by
  dashed lines. The distance $z$ is normalized by the local
  boundary-layer thickness $\delta$ at that
  location.  \label{fig:juncture}}
\end{figure}

\begin{figure}
\begin{center}
\subfloat[]{\includegraphics[width=0.44\textwidth]{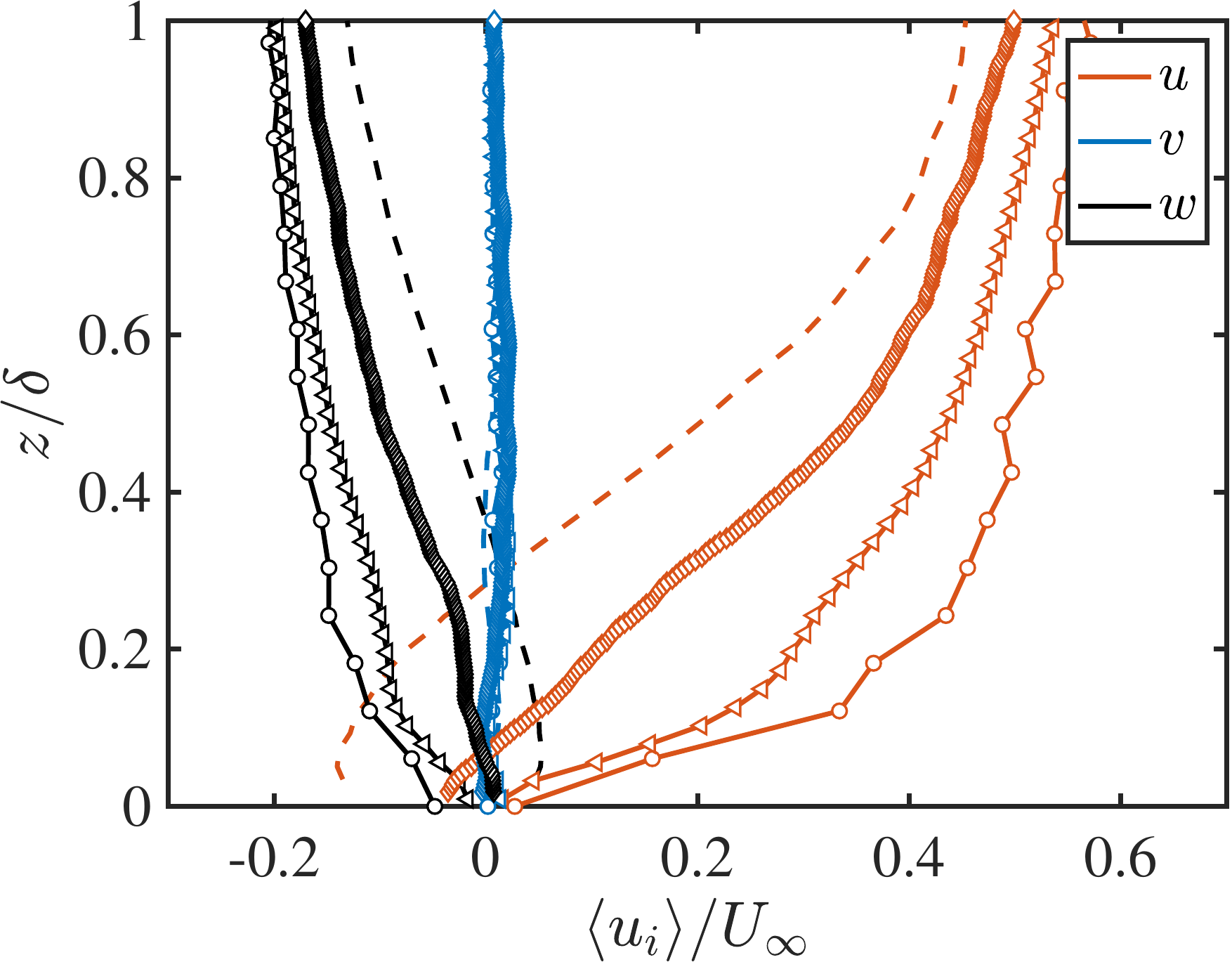}}
\hspace{0.5cm}
\subfloat[]{\includegraphics[width=0.47\textwidth]{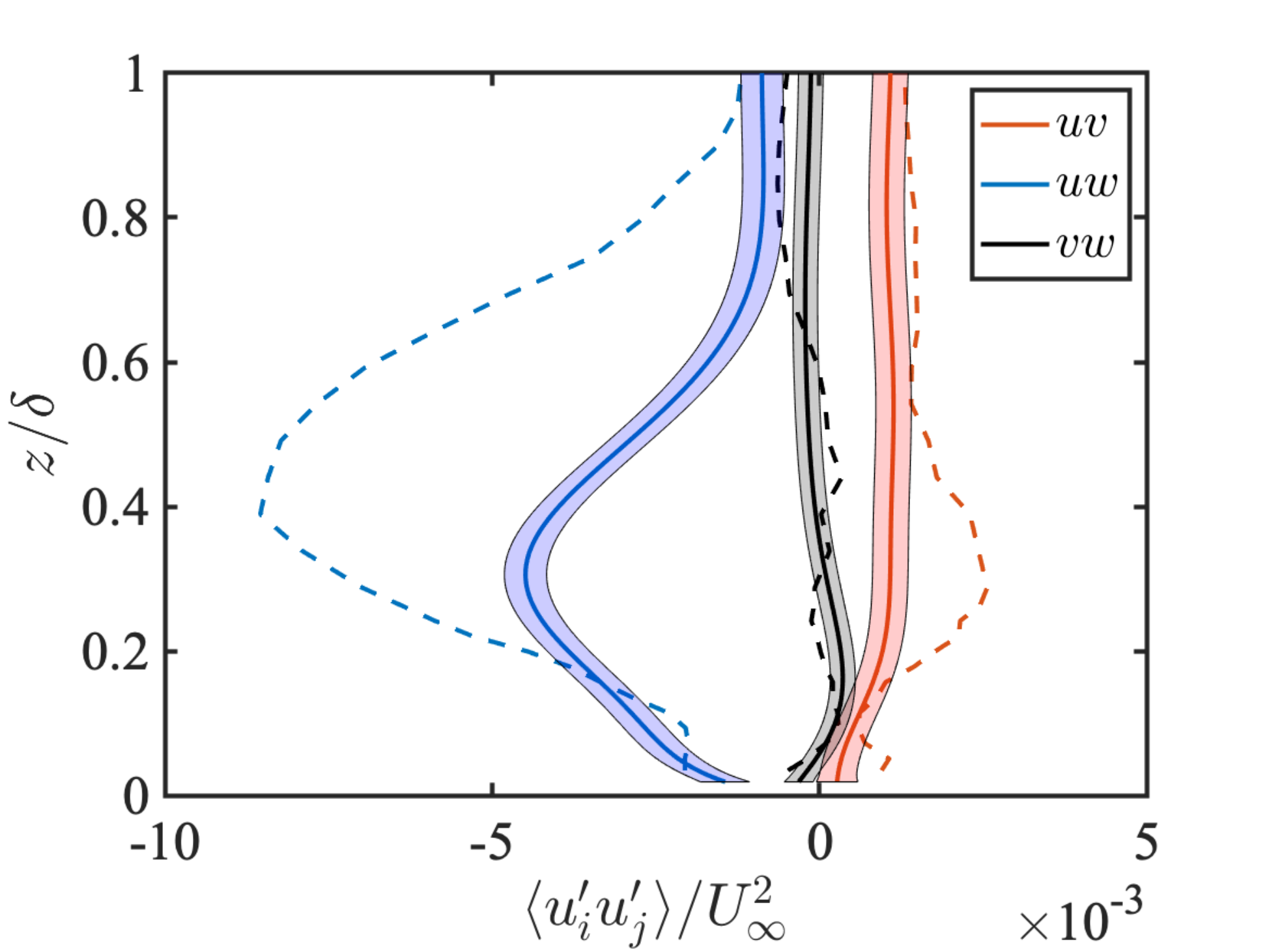}}
\end{center}
\caption{(a) Mean velocity profiles and (b) Reynolds shear stresses at
  location 3: wing-body juncture close to the trailing-edge at
  $x=2922.6$~mm and $y=239.1$~mm (blue line in Figure
  \ref{fig:locations}). In panel (a), lines with symbols are for cases
  C-D2 ($\circ$), C-D1 ($\triangleleft$), and C-D0.5
  ($\diamond$). Panel (b) only includes case C-D0.5 and the shaded
  area represents statistical uncertainty. Experiments are denoted by
  dashed lines. The distance $z$ is normalized by the local
  boundary-layer thickness $\delta$ at that
  location.\label{fig:separation}}
\end{figure}

The mean velocity profiles at the fuselage, juncture, and
trailing-edge region are shown in panel (a) of Figure
\ref{fig:fuselage}, \ref{fig:juncture}, and \ref{fig:separation},
respectively.  In the fuselage, the mean velocity profiles approach
the experimental results with grid refinement.  The turbulent boundary
layer over the fuselage is about $10$ to $20$~mm thick, which yields
roughly 3--6 points per boundary-layer thickness for the grid
resolutions considered.  The mean velocity profiles also get closer to
the experimental profile in the juncture region and trailing-edge
zone, as seen in Figures \ref{fig:juncture}(a) and
\ref{fig:separation}(a), but they do so at a slower pace.  The
accuracy of the predictions in the latter regions appears to be
inferior than those in the fuselage despite the fact that the local
$N_{\mathrm{bl}}$ in the juncture and trailing-edge is roughly 30--80
points per $\delta$, which is one order of magnitude larger than
$N_{\mathrm{bl}}$ in the fuselage.
%
%

\begin{figure}
\begin{center}
  \subfloat[]{\includegraphics[width=0.45\textwidth]{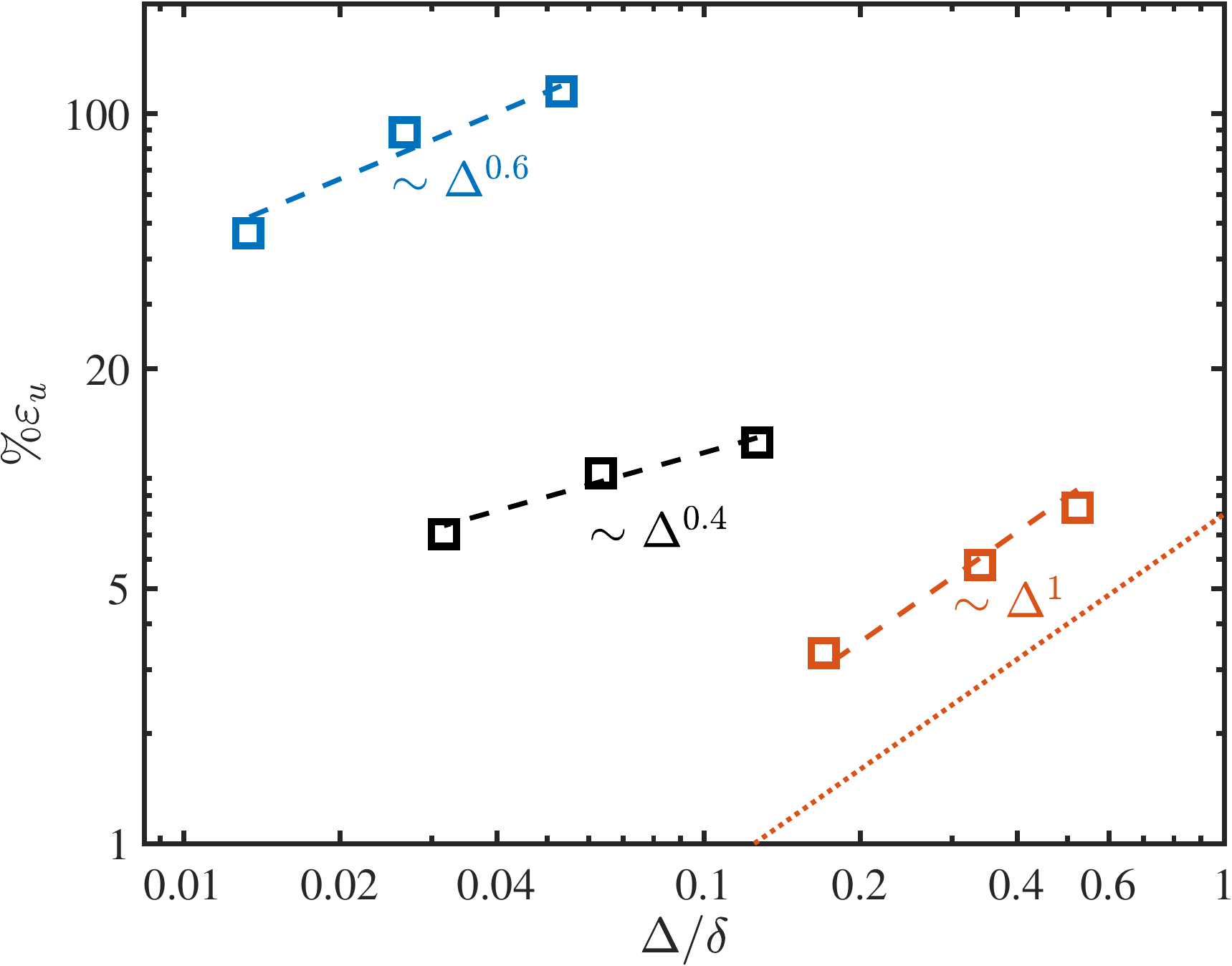}}
  \hspace{0.5cm}
  \subfloat[]{\includegraphics[width=0.45\textwidth]{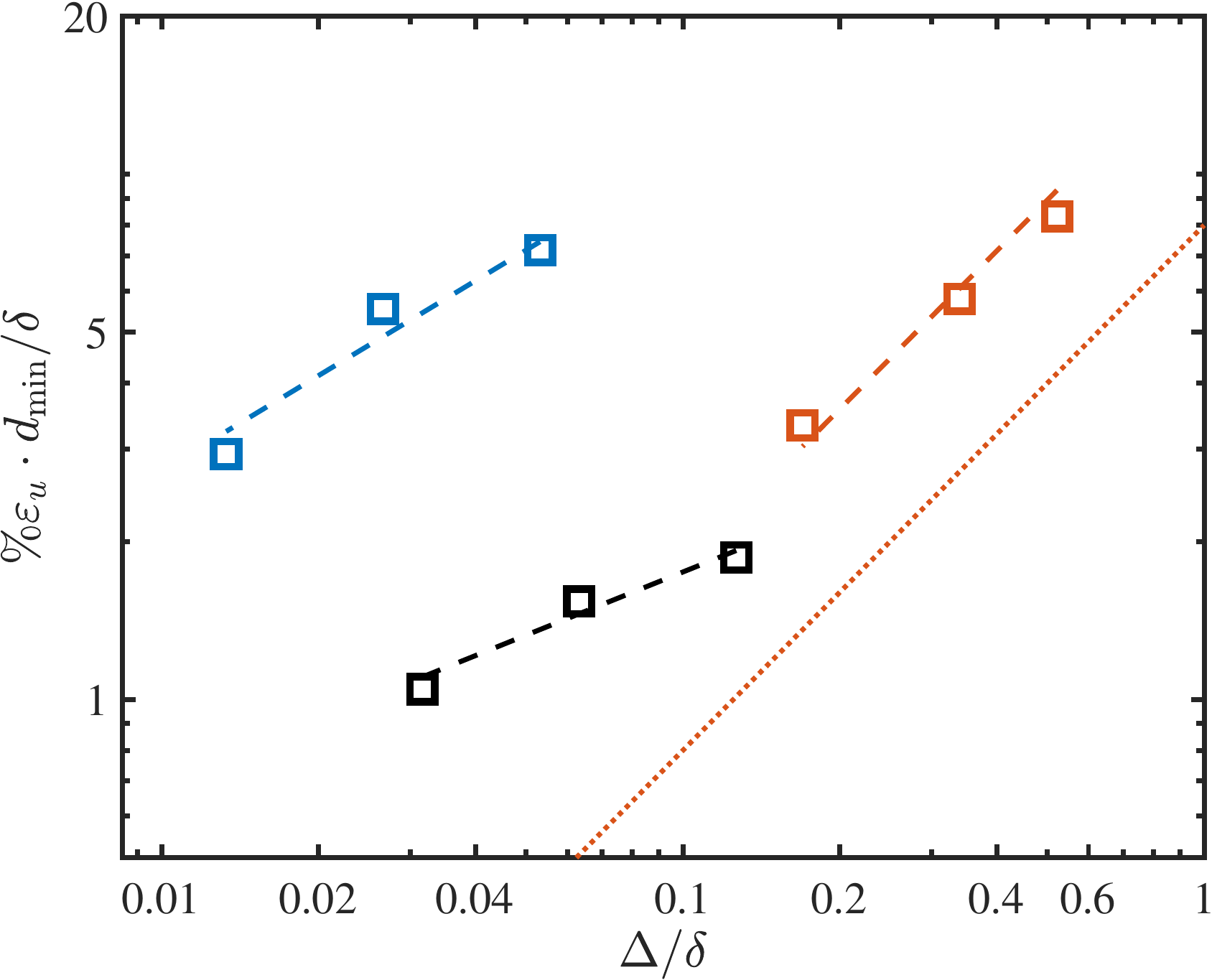}}  
\end{center}
\caption{ Errors in the mean velocity profile (symbols) at the three
  locations shown in Figure~\ref{fig:locations} using the same color
  code: red (fuselage), black (juncture), and blue (separation).  (a)
  Error in the mean velocity profile predicted by WMLES as a function
  of the grid resolution normalized with the local boundary-layer
  thickness. (b) Error in the mean velocity profile predicted by WMLES
  compensated by the closest distance to fuselage
  $d_\mathrm{min}/\delta$ as a function of the grid resolution. We
  have set $d_\mathrm{min} = \delta$ for the fuselage location. The
  red dotted line in panels is the reference error for a turbulent
  channel flow, $\varepsilon_u = 0.08
  \Delta/\delta$. \label{fig:Errors_all}}
\end{figure}

Figure \ref{fig:Errors_all} is the cornerstone of the present
study. It summarizes the relative errors in the prediction of the mean
velocity profile as a function of the local grid resolution. Figure
\ref{fig:Errors_all} also follows the same color code as Figure
\ref{fig:locations} and features the reference error for a turbulent
channel flow (dotted line) introduced in Section \ref{sec:errors}.  In
the fuselage, the errors are roughly 3\%--8\% and scale as
$\varepsilon_u \approx \beta_u (\Delta/\delta)$ (red symbols in Figure
\ref{fig:Errors_all}(a)) akin to the errors in a turbulent channel
flow. This suggest that the turbulent flow in the fuselage resembles a
ZPGTBL. As such, wall- and SGS models devised for, and validated in,
flat-plate turbulence perform as anticipated by the error analysis in
Section \ref{sec:errors}. The error constant in the fuselage,
$\beta_u\approx 0.14$, is larger than the value for turbulent channel
flows, $\beta_u\approx 0.08$ (see Section \ref{sec:errors}), and it
will be shown in \S \ref{subsec:targeted} that this is caused by the
propagation of errors from upstream of the fuselage.

In contrast, the wing-body juncture and trailing-edge region exhibit
larger errors than those reported in the fuselage. In the juncture
region, the errors in the mean velocity are about 15\% (black symbols
in Figure \ref{fig:Errors_all}a), whereas the errors in the
trailing-edge zone are about 100\% (blue symbols in Figure
\ref{fig:Errors_all}a). One of the reasons for the larger errors in
the juncture and trailing-edge is that the mean profiles are plotted
along a line parallel and close to the fuselage (see black and blue
lines in Figure \ref{fig:locations}). As argued in Section
\ref{sec:errors}, this would amplify the errors due to the additional
shear from the fuselage boundary-layer. To account for enhanced errors
in the proximity of a wall, Figure \ref{fig:Errors_all}(b) shows the
errors compensated by the closest distance to the fuselage
$d_\mathrm{min}/\delta$. The compensated errors in the juncture region
bear a closer resemble to the trends observed in the fuselage,
implying that most of the increase in error can be explained by the
proximity of the wall. Nonetheless, the convergence rate of the
solution at the juncture, $\alpha_u \approx 0.4$, is still slower than
the linear convergence expected from the channel reference error from
\S \ref{subsec:errors}.  The situation is less favorable for the
trailing-edge region: not only is the convergence rate slower
($\alpha_u \approx 0.6$), but its magnitude is still considerably
larger (i.e., higher $\beta_u$) even after error compensation. This
suggests that the errors in the separated region are not only the
consequence of the proximity to the fuselage but of something
else. The additional errors and slower convergence rate in the
wing-body juncture and trailing-edge may be attributed to two
causes. The first one might be the contamination of the solution from
upstream errors. The second option is that the presence of
three-dimensional boundary layers and/or flow separation renders
inaccurate the modeling assumptions predicated upon the local
similarity to a ZPGTBL~\citep{Lozano_2020_3D}. The first source of
error is examined in more detail in the following section using
BL-conforming grids.
%

For completeness, we close this section by briefly discussing the
errors in the prediction of the tangential Reynolds shear
stresses. Panel (b) of Figures \ref{fig:fuselage}, \ref{fig:juncture}
and \ref{fig:separation} contains the resolved portion of the
tangential Reynolds shear stresses, $\langle u_i' u_j'\rangle$.  The
profiles of $\langle u_i' u_j'\rangle$ capture the trends of the
experimental data at the stations investigated, although their
magnitudes are underpredicted for the juncture region and trailing
edge. \citet{Lozano2019a} showed that the values of the SGS tensor
($\tau_{ij}^\mathrm{SGS}$) required to predict the correct mean
velocity profile usually imply the underestimation of $\langle u_i'
u_j'\rangle$. Assuming that $\langle u_i^\mathrm{exp} \rangle \approx
\langle u_i\rangle$, then $\langle {u_i'}^\mathrm{exp}
        {u_j'}^\mathrm{exp}\rangle \approx \langle u_i' u_j'\rangle +
        \langle \tau_{ij}^\mathrm{SGS}\rangle$.  Thus, for the typical
        grid sizes in WMLES, we expect that $|\langle u_i'
        u_j'\rangle| < |\langle {u_i'}^\mathrm{exp}
        {u_j'}^\mathrm{exp}\rangle|$, i.e., underprediction of
        the tangential Reynolds stress.  Estimations from
        \citet{Lozano2019a} also showed that the error for the
        turbulence intensities in WMLES grids should scale as $\sim
        (\Delta/\delta)^{2/3}$ with values in the range of
        $10$--$30\%$, which is consistent with our observations.

\subsection{Mean velocity profiles with BL-conforming grids}
\label{subsec:targeted}

\begin{figure}
  \begin{center}
  \subfloat[]{\includegraphics[width=0.32\textwidth]{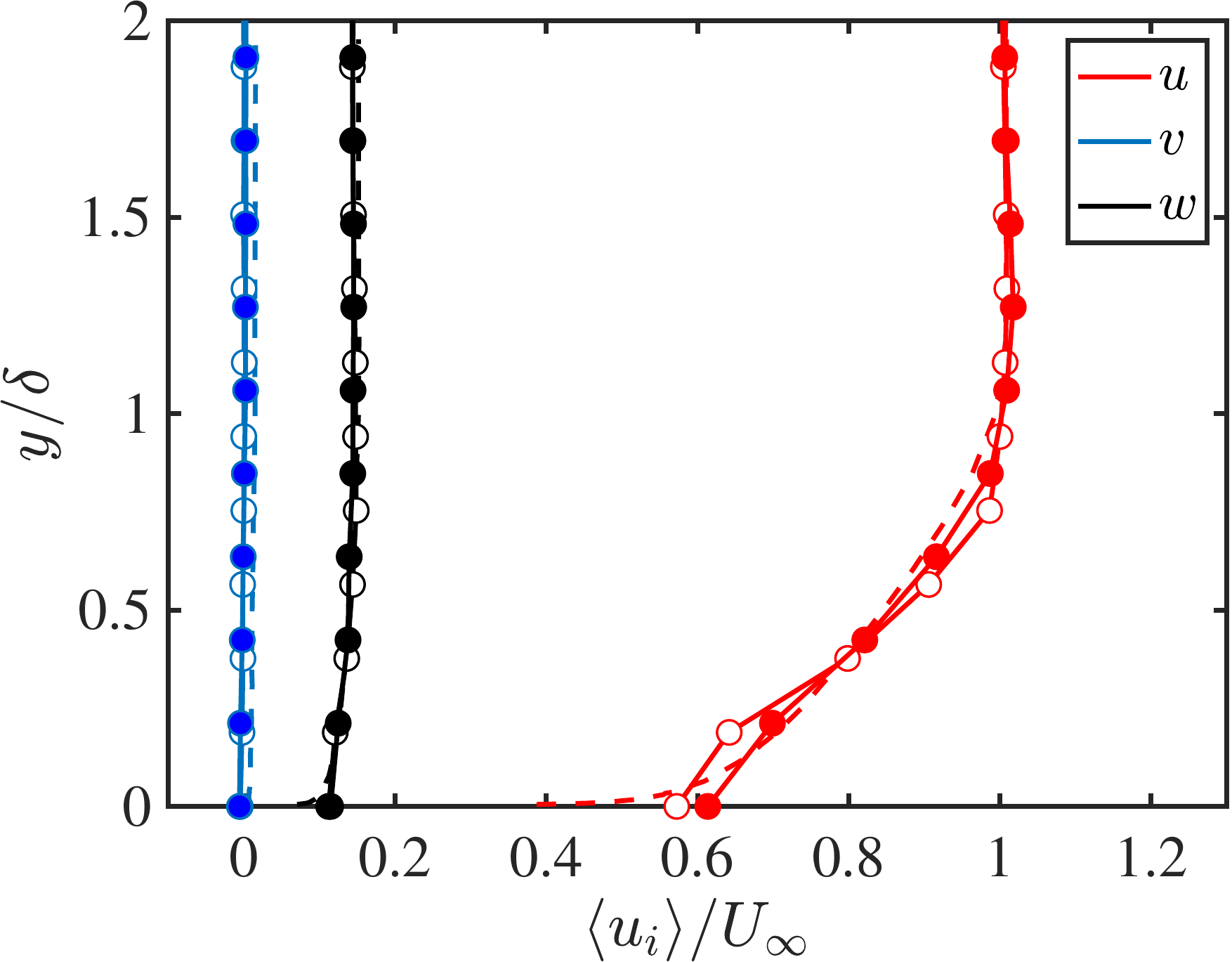}}
  \hspace{0.01cm}
  \subfloat[]{\includegraphics[width=0.32\textwidth]{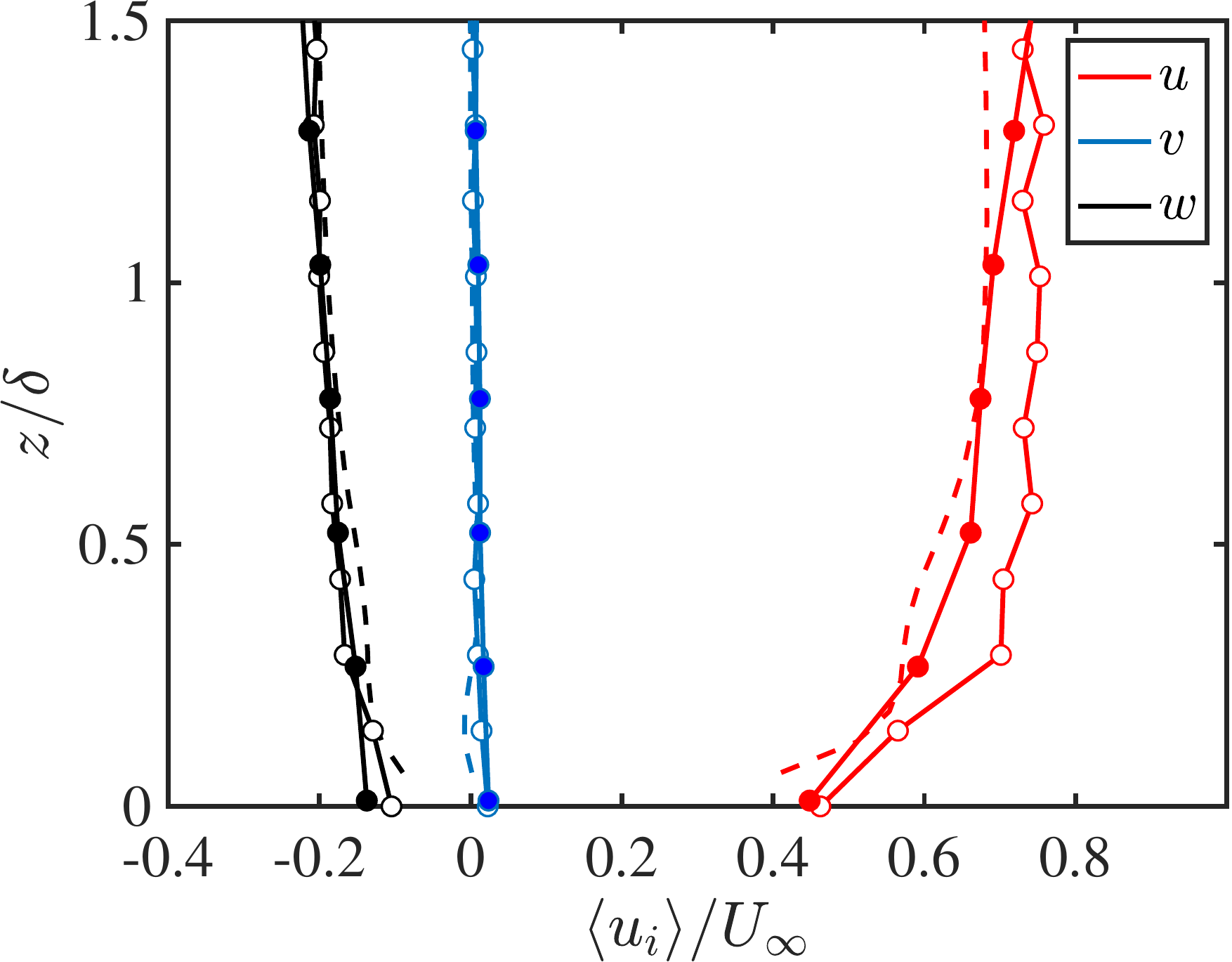}}
  \hspace{0.01cm}
  \subfloat[]{\includegraphics[width=0.32\textwidth]{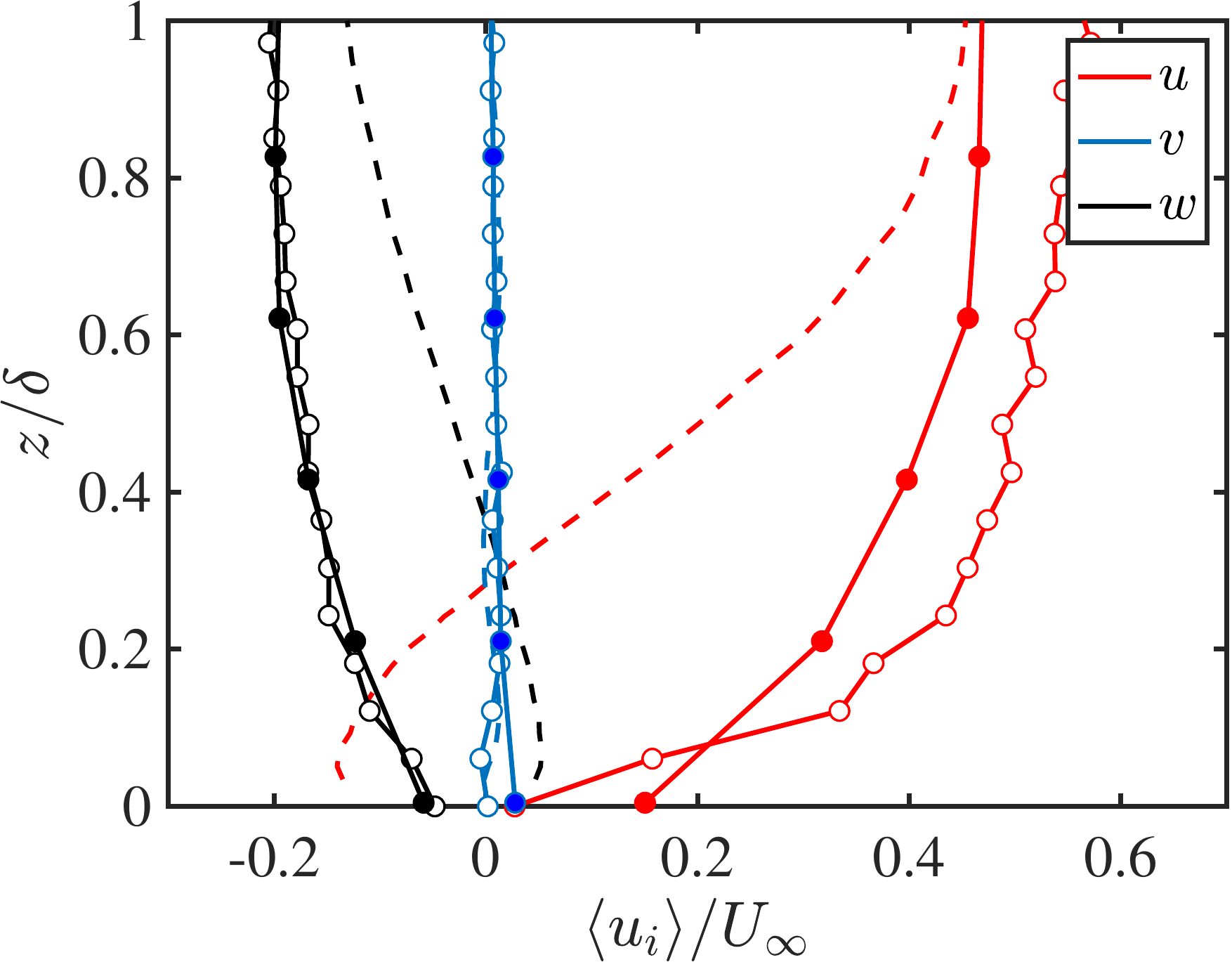}}
\end{center}
\caption{Mean velocity profiles for case C-N5-R2e3 (solid lines with
  $\bullet$) and C-D2 (solid lines with $\circ$) at (a) location 1:
  upstream region of the fuselage $x=1168.4$~mm and $z=0$~mm (b)
  location 2 wing-body juncture $x=2747.6$~mm and $y=239.1$~mm, and
  (c) location 3: wing-body juncture close to the trailing edge at
  $x=2922.6$~mm and $y=239.1$~mm.  Experiments are denoted by dashed
  lines. Colors denote different velocity components. The distance $y$
  is normalized by the local boundary-layer thickness $\delta$ at that
  location. (d) Error in the mean velocity profile prediction by WMLES
  as a function of the grid resolution. Open symbols represent
  constant-size grids (same as in figure \ref{fig:Errors_all}a) and
  closed symbols represent BL-conforming grids. The different colors
  denote the three locations as in Figure \ref{fig:Errors_all}(b)
  using the same color code. \label{fig:custom}}
\end{figure}

We evaluate the benefits of utilizing BL-conforming grids constructed
following the procedure described in Section \ref{subsec:tbl}. First,
we illustrate the improvements in the prediction of the mean velocity
by comparing the results for case C-D2 ($\Delta\approx 2$ mm) with
case C-N5-R2e3 ($N_\mathrm{bl}=5$ and $Re_\Delta^\mathrm{min}=2.8
\times 10^3$). The mean velocity profiles for both cases are shown in
Figure \ref{fig:custom} at the three locations under
consideration. Moderate improvements are attained for C-N5-R2e3 at the
fuselage, despite the fact C-N5-R2e3 and C-D2 share the same
$N_\mathrm{bl}$ at that location. The benefits enabled by
BL-conforming grids are accentuated at the juncture region and
trailing edge, at which C-N5-R2e3 outperforms C-D2 using only one
fourth of the number of grid points in each spatial direction (namely,
a factor of 64 less points in the local volume).  For reference, case
C-D2 has 31 million control volumes, whereas case C-N5-R2e3 has 12
million control volumes. The lower errors for case C-N5-R2e3 clearly
demonstrate that the sole use of the number of control volumes is a
misleading metric to quantify the quality of WMLES, and that the
spatial distribution of control volumes may have an appreciable impact
on the accuracy of the solution.

\begin{figure}
\begin{center}
  \subfloat[]{\includegraphics[width=0.45\textwidth]{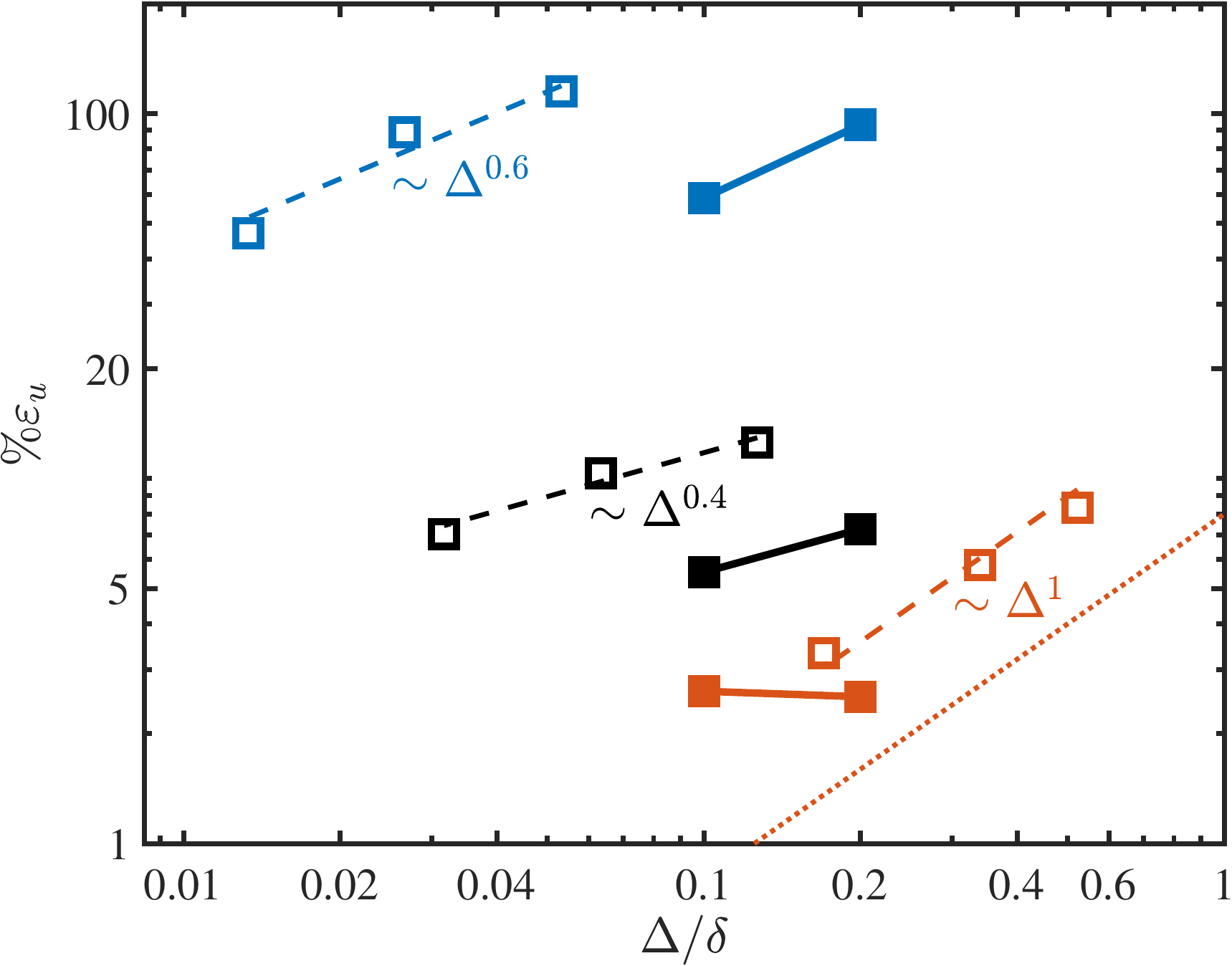}}
  \hspace{0.2cm}
  \subfloat[]{\includegraphics[width=0.45\textwidth]{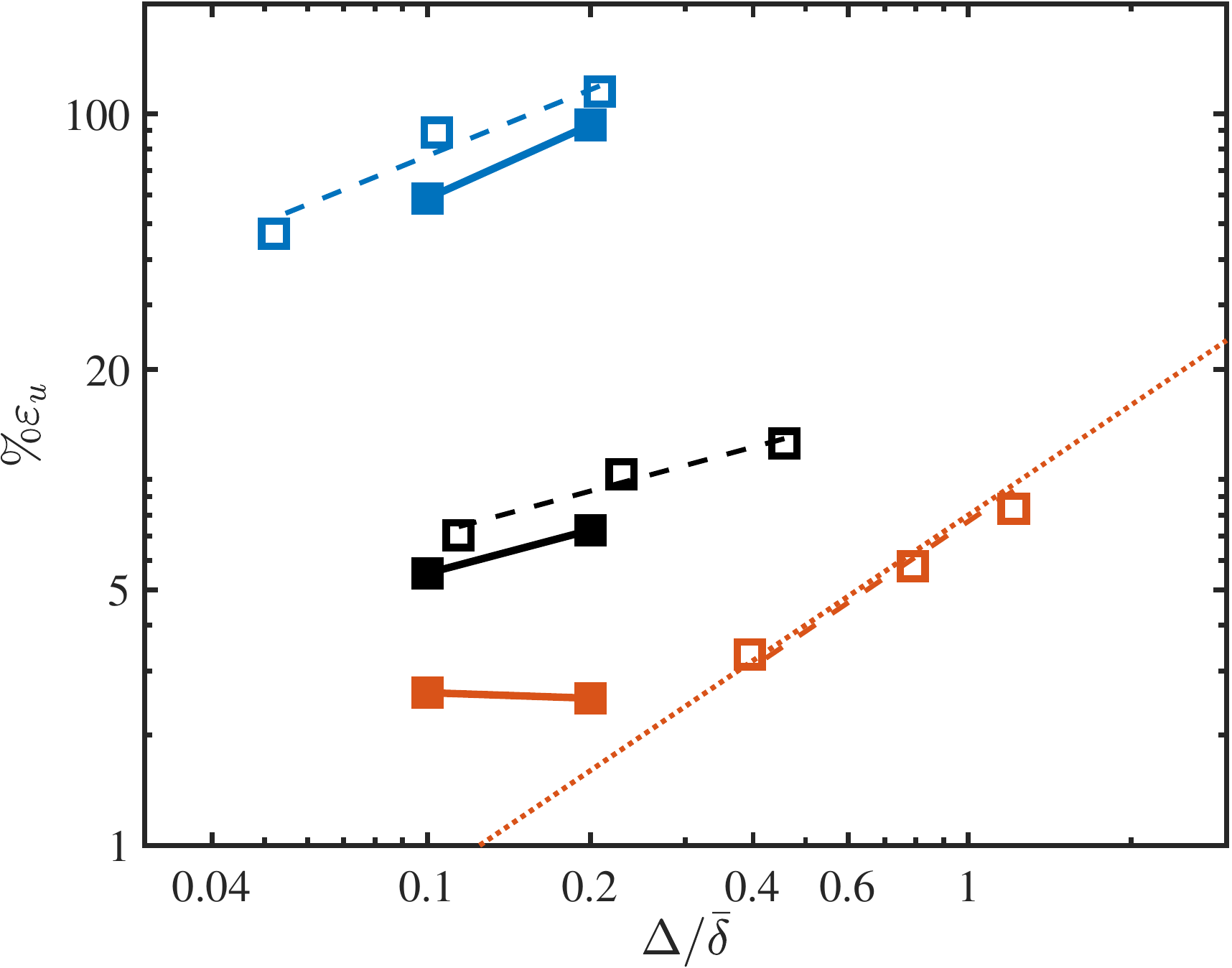}}  
\end{center}
\caption{ Errors in the mean velocity profile (symbols) at the three
  locations shown in Figure~\ref{fig:locations} using the same color
  code: red (fuselage), black (juncture), and blue (separation). (a)
  Error in the mean velocity profile predicted by WMLES as a function
  of the grid resolution normalized with the local boundary-layer
  thickness. (b) Error in the mean velocity profile predicted by WMLES
  as a function of the effective upstream-averaged grid resolution
  $\Delta/\bar{\delta}$. The value of $\bar{\delta}$ was calculated
  for $\Delta x_e=300\delta_0$. Open symbols denote constant-size
  grids and closed symbols denote BL-conforming grids. The red dotted
  line in panels is the reference error for a turbulent channel flow,
  $\varepsilon_u = 0.08 \Delta/\delta$.\label{fig:Errors_all_custom}}
\end{figure}

The systematic characterization of errors in the mean flow is shown in
Figure \ref{fig:Errors_all_custom}(a), which includes the errors for
constant-size grid (open symbols) and BL-conforming grids (closed
symbols).  The predictions obtained with BL-conforming grid are
consistently more accurate than those for constant-size grid at the
three locations considered. The improvements in the fuselage region are the
least pronounced, which might be an indication that the errors
saturate roughly at 2\%. This is consistent with the error analysis in
Section \ref{sec:errors}, where it was demonstrated that internal
wall-modeling errors caused by deviations of the actual flow from the
law-of-the-wall propagate to the mean velocity profile as dictated by
Eq.  (\ref{eq:wm_error}). These internal errors impose a lower limit
to the minimum error achievable by WMLES. For example, if the flow at
the fuselage differs from a ZPGTBL by 3\% ($\%\varepsilon_\kappa=3$),
the minimum error attainable by WMLES at that location would be 2\%,
which is consistent with our observations. Moreover, additional grid
refinements in the presence of internal errors are not expected to
improve the prediction of the mean velocity profile until the LES grid
resolution (and $h_w$) reaches the wall-resolved-LES regime, for which
internal errors would vanish. For the juncture and separation region,
the most noticeable improvement with BL-conforming grids is the
reduction in the error constant $\beta_u$, while the convergence rate
of the error $\alpha_u$ remains unchanged.  Thus, grids
designed to specifically target the boundary layer can improve the
overall accuracy of WMLES via a smaller value of $\beta_u$.
%
%

The improved outcome for BL-conforming grids might be justified by
considering the upstream history of the WMLES solution at a given
station. For the sake of simplicity, let us consider WMLES of a
flat-plate turbulent boundary layer evolving along the $x$ direction
using two grids as those shown in Figure~\ref{fig:grid_types}: a
constant-size grid and a BL-conforming grid. Given an $x$-location at
which both grids share the same $N_\mathrm{bl}$, the upstream flow for
the constant-size grid is underresolved compared to the BL-conforming
grid due to the thinning of $\delta$. Consequently, constant-size
grids display larger upstream errors that eventually propagate
downstream, contaminating the solution at $x$.  Even if at a given
$x$-location $N_\mathrm{bl}$ is larger for the constant-size grid than
for the BL-conforming grid, the solution for the former could still
exhibit poor accuracy because of the propagation of upstream errors.
Assuming that the lifetimes of the energy-containing eddies at $x_0<x$
are proportional to $\delta_0/u_\tau$, and that these eddies are
advected by $U_\infty$, the spatial extent for the downstream
propagation of errors, $\Delta x_e$, from a given location $x_0$ can be
estimated as $\Delta x_e \approx \Gamma U_\infty \delta_0/u_\tau$,
where $\Gamma$ is a flow-dependent constant. The distance $\Delta x_e$
is the streamwise recovery length for the energy-containing eddies to
forget their past history, and hence their errors. In ZPGTBL at high
$Re$, $\Delta x_e$ has been experimentally shown to reach values of
$\Delta x_e \approx 300\delta_0$~\citep{Sillero2013}. In our case,
this implies that errors from the underresolved leading edge
($\delta_0 \approx 1$~mm) are advected downstream for $\Delta x_e
\approx 300$~mm, which is $\sim$50\% of the crank chord.  On the other
hand, the BL-conforming grids maintain a constant grid resolution
scaled in $\delta$ units and effectively a higher resolution of the
upstream flow.

One implication of the downstream propagation of errors is that the
local grid resolution $\Delta/\delta$ is inappropriate to characterize
the errors in constant-size grids. Instead, an effective
upstream-averaged grid resolution of the form $\Delta/\bar{\delta}$
should be used, where $\bar{\delta}(x,y,z) = \int_{-\infty}^x G(x'-x)
\delta (x',y,z) \mathrm{d}x$ with $G$ the Gaussian kernel $G(x)
=1/(\sqrt{2\pi}\Delta x_e) \exp[-1/2x^2/(2\Delta x_e)^2]$ or any other
similar kernel. The new effective grid resolution is tested in
Figure~\ref{fig:Errors_all_custom}(b), which shows the errors in the
mean velocity as a function of $\Delta/\bar{\delta}$. In the fuselage,
the errors for constant-size grids collapse with the reference error
from turbulent channel flows, confirming that the diminished
performance of constant-size grids is merely a matter of downstream
error propagation. The collapse of the errors at the juncture and
trailing-edge regions is also significantly improved between
constant-size and BL-conforming grids using $\Delta/\bar{\delta}$,
suggesting that the same argument applies to these cases as
well. Errors in the fuselage for BL-conforming grids remain saturated
at roughly 2\%. In general, constant-size grids display a lower
effective resolution ($\Delta/\delta < \Delta/\bar{\delta}$) than
BL-conforming grids ($\Delta/\delta \approx \Delta/\bar{\delta}$ for
$Re_\Delta^\mathrm{min} < Re_x$), which explains the poorer
performance of the former. Hence, the long convective distance for
error propagation from the leading edge combined with the higher
upstream errors for constant-size grids accounts for the improved
performance of BL-conforming grids reported in
Figure~\ref{fig:Errors_all_custom}(a).

\subsection{Effect of leading-edge grid resolution}\label{subsec:LE}

\begin{figure}
\begin{center}
  \includegraphics[width=1.0\textwidth]{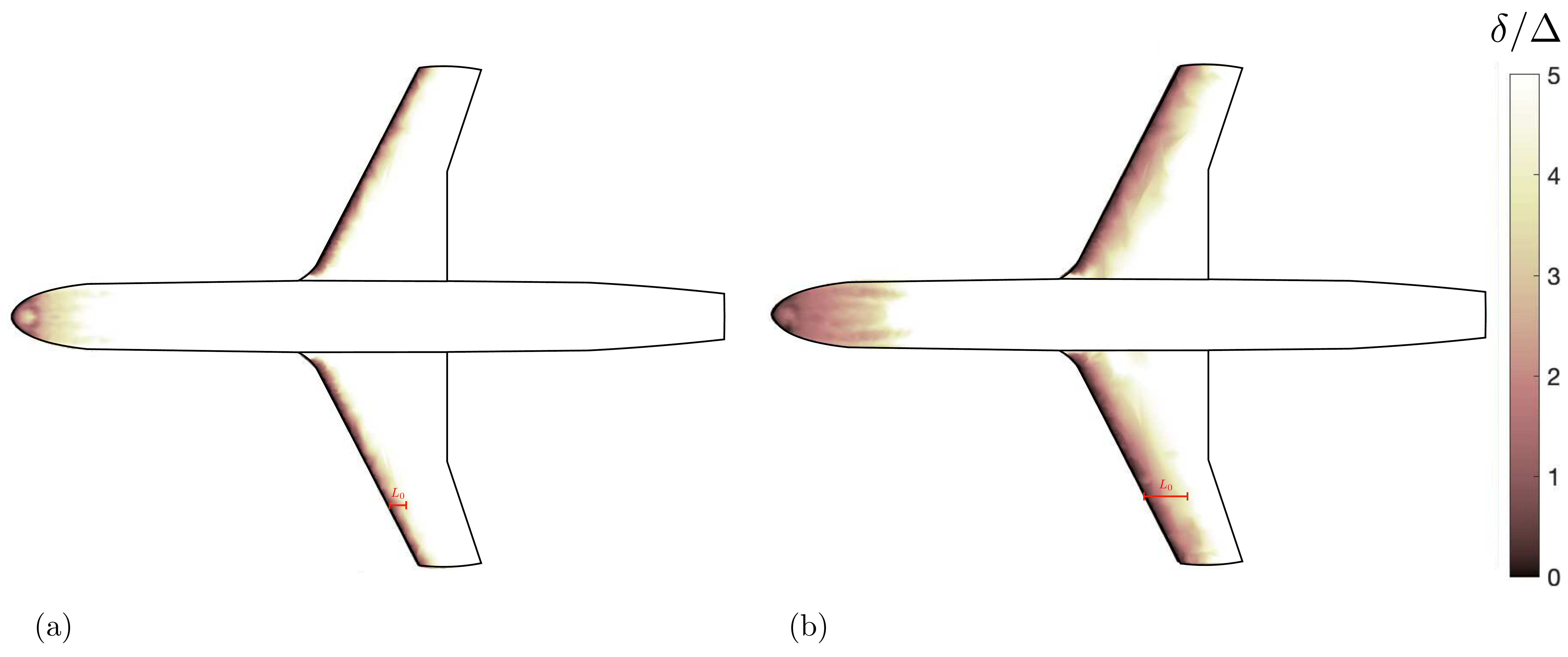}
\end{center}
\caption{ Resolution map to test the effect of leading-edge grid
  solution.  Points per boundary layer thickness ($\delta/\Delta$) for
  (a) case C-N5-R2e3 ($Re_\Delta^\mathrm{min}=2.8 \times 10^3$) and
  (b) case C-N5-R5e3 ($Re_\Delta^\mathrm{min}=5.6 \times 10^3$). $L_0$
  denotes the streamwise extent of the underresolved leading-edge,
  i.e., region with $\delta/\Delta<N_\mathrm{bl}$.\label{fig:LE}}
\end{figure}

The effect of leading-edge resolution on the accuracy of the mean
velocity profiles is evaluated by comparing C-N5-R2e3 and
C-N5-R5e3. Both cases share the same points per boundary-layer
thickness ($\delta/\Delta=5$) but differ on the value of
$Re_\Delta^\mathrm{min}$, namely $Re_\Delta^\mathrm{min}=2.6 \times
10^3$ and $Re_\Delta^\mathrm{min}=5.8 \times 10^3$ for C-N5-R2e3 and
C-N5-R5e3, respectively. It was argued that $Re_\Delta^\mathrm{min}$
can be understood as a quantification of the leading-edge grid
resolution such that, for constant $N_\mathrm{bl}$, decreasing
$Re_\Delta^\mathrm{min}$ implies finer resolution at the
leading-edge. Other characterizations of the leading-edge resolution
instead of $Re_\Delta^\mathrm{min}$ are possible and an alternative is
discussed in Appendix A.  A comparison of the resolution maps for
cases C-N5-R2e3 and C-N5-R5e3 is presented in Figure
\ref{fig:LE}. Increasing the value of $Re_\Delta^\mathrm{min}$ by a
factor of two lessens the portion of the wing area resolved with
$N_\mathrm{bl} = 5$. Following the assumptions introduced in \S
\ref{subsec:number} for ZPGTBL, the streamwise extent of the
underresolved leading-edge region ($L_0$, see
Figure~\ref{fig:grid_types}b) scales as $L_0/L \sim
(Re_\Delta^\mathrm{min} N_\mathrm{bl})^{7/6} Re^{-1}$ in first order
approximation. Thus, increasing $Re_\Delta^\mathrm{min}$ by a factor
of two will roughly double $L_0$, consistent with the growth of the
shading area observed in Figure \ref{fig:LE}.
%

\begin{figure}
\begin{center}
\subfloat[]{\includegraphics[width=0.47\textwidth]{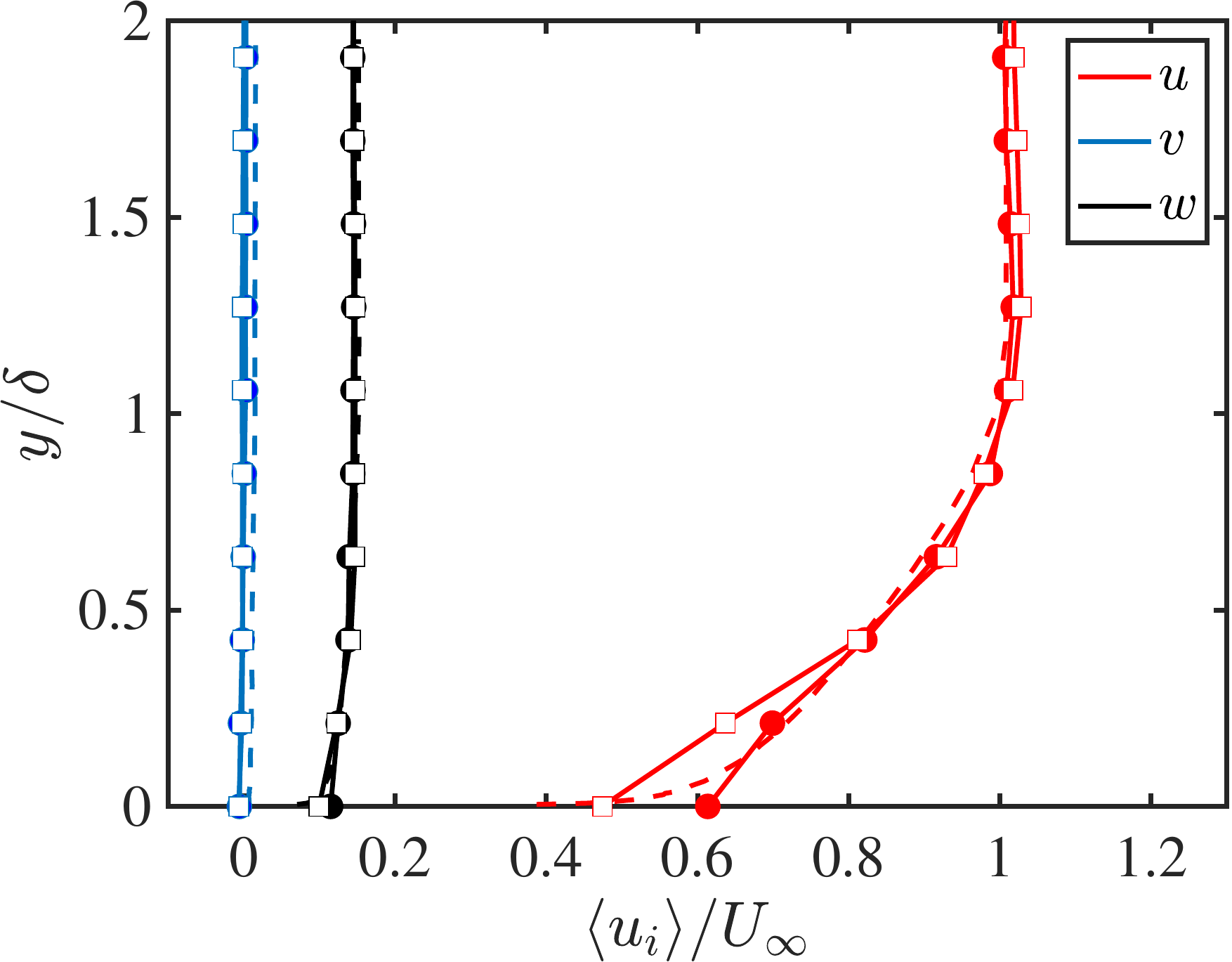}}
\hspace{0.5cm}
\subfloat[]{\includegraphics[width=0.47\textwidth]{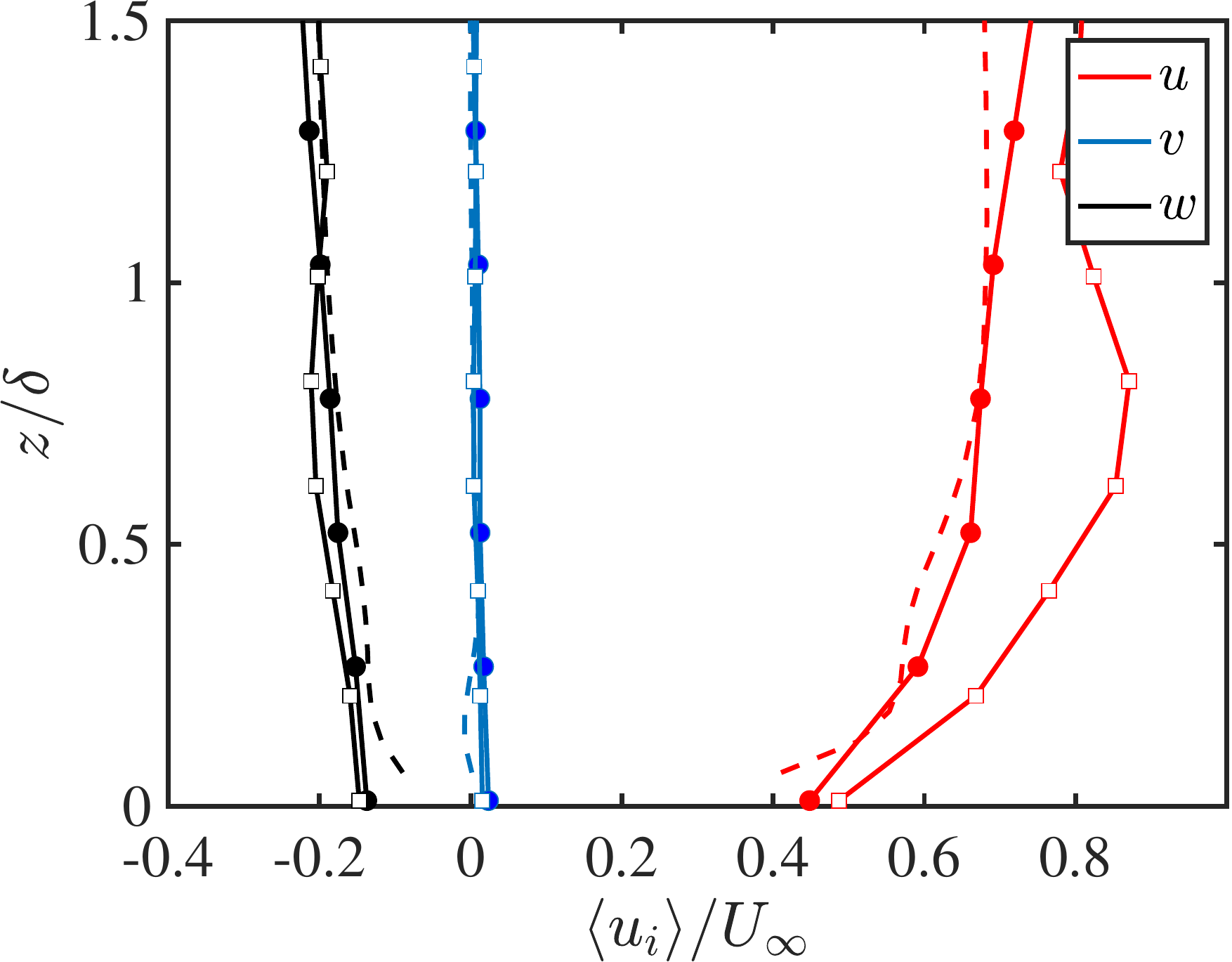}}
\end{center}
\begin{center}
\subfloat[]{\includegraphics[width=0.47\textwidth]{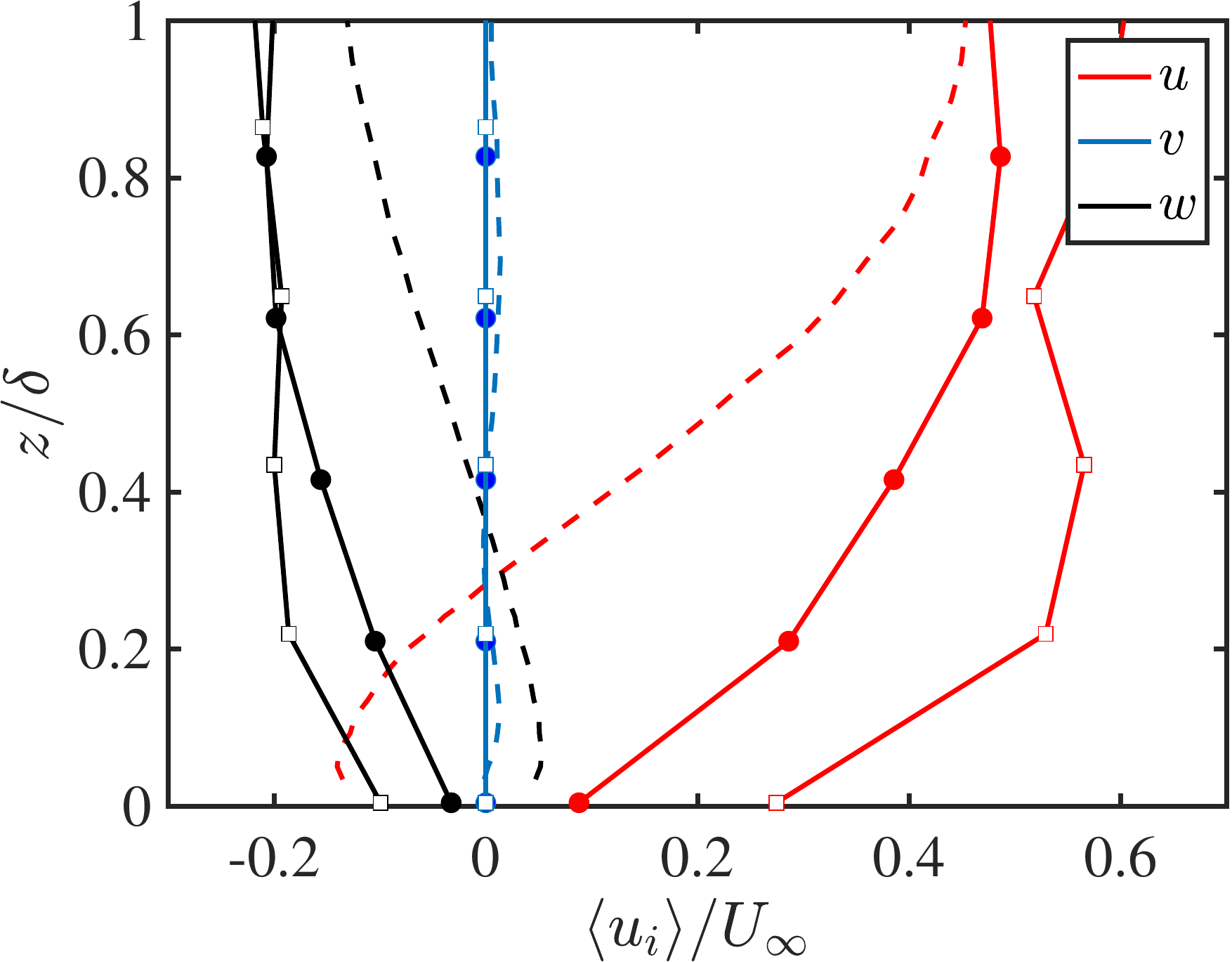}}
\hspace{0.5cm}
\subfloat[]{\includegraphics[width=0.45\textwidth]{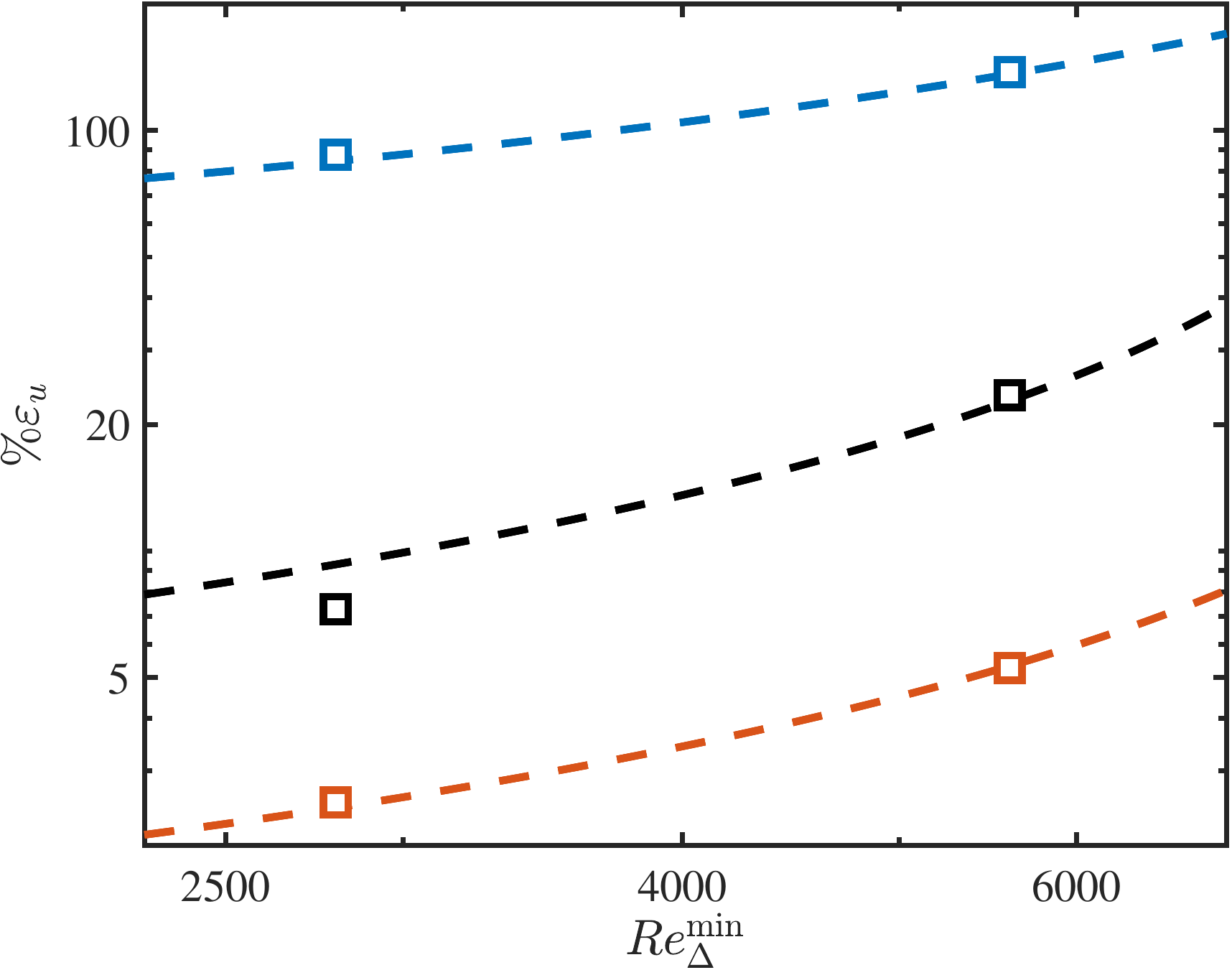}}
\end{center}
\caption{(a) Mean velocity profiles at (a) the fuselage region, (b)
  juncture region, and (c) separation zone for case C-N5-R2e3
  ($\circ$) and C-N5-R5e3 ($\square$). Colors denote different
  velocity components.  Experiments are denoted by dashed lines. The
  wall-normal distance is normalized by the local boundary-layer
  thickness $\delta$ at the respective location. (d) Errors in the
  mean velocity profile (symbols) as a function of
  $Re_\Delta^\mathrm{min}$ at the three locations shown in
  Figure~\ref{fig:locations} using the same color code: red
  (fuselage), black (juncture), and blue (separation). The dashed
  lines are proportional to $\varepsilon_u \sim 1/N_\mathrm{bl} [1 +
    \Gamma Re_\Delta^{\mathrm{min}} N_\mathrm{bl} /(Re_x -
    (Re_\Delta^{\mathrm{min}} N_\mathrm{bl}/K)^{7/6} ) ]$, with $Re_x$
  the leading-edge Reynolds number, $\Gamma = 300$, and
  $K=0.16$.\label{fig:results_LE}}
\end{figure}

The consequences of diminished leading-edge grid resolution are
appraised in Figure \ref{fig:results_LE}(a,b,c), which compares the
mean velocity profiles for C-N5-R2e3 and C-N5-R5e3 at the three
locations considered. Unsurprisingly, the mean velocity profiles
improve for decreasing $Re_\Delta^\mathrm{min}$. The relative error in
the mean velocity profile as a function of $Re_\Delta^\mathrm{min}$ is
shown in Figure \ref{fig:results_LE}(d), which quantifies more
rigorously the dependence of $\varepsilon_u$ with
$Re_\Delta^\mathrm{min}$. A simple model for the propagation of errors
from the underresolved leading-edge to the downstream location $x$ can
be constructed by assuming that $\varepsilon_u$ will decay linearly
with the streamwise distance to the leading-edge, $\varepsilon_u \sim
1/N_\mathrm{bl}( 1 + \Gamma \delta_\mathrm{min}/\Delta L)$, where
$\delta_\mathrm{min}$ is the minimum $\delta$ resolved with
$N_\mathrm{bl}$ points, $\Delta L=x-L_0$ is the distance to the
underresolved leading-edge, and $\Gamma$ controls the lasting effects
of errors propagation (as discussed in the previous section). For
$\Delta L \rightarrow \infty$ (i.e., $x$ far from the leading-edge),
the model recovers $\varepsilon_u \sim 1/N_\mathrm{bl} =
\Delta/\delta$ (i.e., no influence from leading-edge errors).
Expressing the errors in terms of the grid parameters
$(N_\mathrm{bl},Re_\Delta^{\mathrm{min}})$ under the assumption of
ZPGTBL yields $\varepsilon_u \sim 1/N_\mathrm{bl} [1 + \Gamma
  Re_\Delta^{\mathrm{min}} N_\mathrm{bl} /(Re_x -
  (Re_\Delta^{\mathrm{min}} N_\mathrm{bl}/K)^{7/6} ) ]$, where $Re_x$
is the Reynolds number based on the distance to the leading-edge and
$K$ is a constant. The model errors are included in Figure
\ref{fig:results_LE}(d) (dashed lines), and provide a reasonable
explanation of the trends obtained from WMLES.

Although not shown, a power law approximation of $\varepsilon_u$ in
the range of ${Re_\Delta^\mathrm{min}}$ considered in Figure
\ref{fig:results_LE}(d) shows that the WMLES predictions approach the
experimental solution as $(Re_\Delta^\mathrm{min})^{\alpha}$, with
$\alpha \approx 1-2$. These local convergence rates are steeper than
those for $N_\mathrm{bl}$ reported in the section above.  Considering
that the dependence of the total number of grid points is also milder
for $Re_\Delta^\mathrm{min}$ (i.e., $N_\mathrm{points} \sim
(Re_\Delta^\mathrm{min})^{-5/6}$) than for $N_\mathrm{bl}$
($N_\mathrm{points} \sim N_\mathrm{bl}^{13/6}$), the scaling of the
grid resolution requirements for the leading-edge region is less
demanding than the resolution to resolve the remaining turbulent
boundary layers.

An important caveat of the present setup is that both the experiment
and WMLES calculations were tripped at the leading-edge of the
wing. This prevent us from evaluating the grid limitations in the
presence of untripped laminar-to-turbulent transition. A second remark
is that our analysis was conducted at a low angle of attack, and
different conclusions might hold at higher values. In particular,
investigations on a NACA 0012~\citep{Goc2021} have shown that the flow
prediction along the whole chord of the airfoil could be dramatically
affected by the grid resolution details at the leading-edge. Our
results do not display this sensitivity and our model does not account
for the possibility of such drastic changes.
%

\subsection{Pressure coefficient}\label{subsec:pressure}

\begin{figure}
\begin{center}
\includegraphics[width=0.9\textwidth]{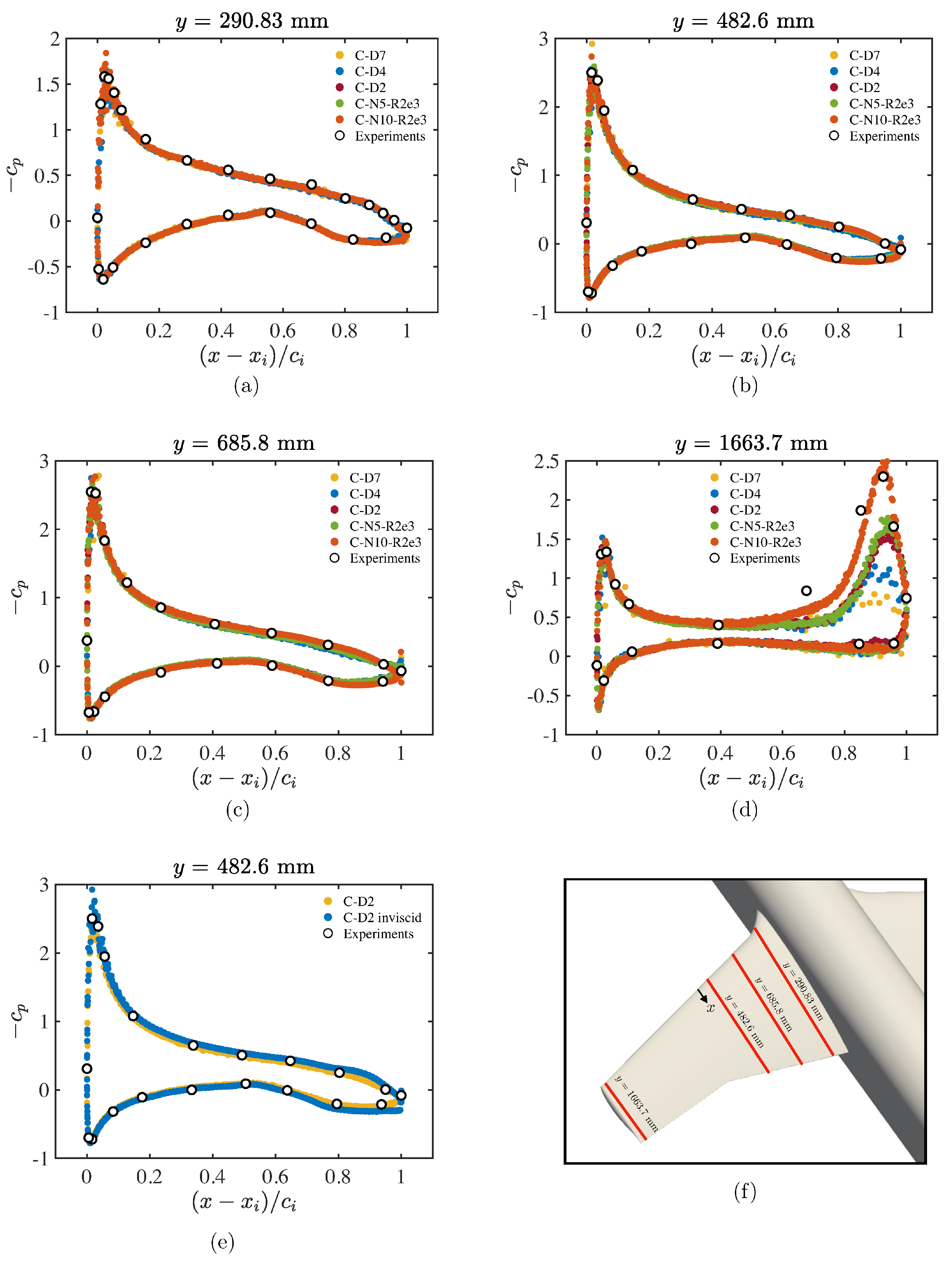}
\end{center}
\caption{ The surface pressure coefficient $C_p$ along the wing
  normalized by the local chord $c_i$ and with respect to the local
  leading-edge coordinate $x_i$. Panels (a), (b), (c) and (d) show
  $C_p$ for cases C-D7, C-D4, C-D2, C-N5-R2e3, and C-N5-R2e3 at
  different $y$ locations. Panel (e) shows $C_p$ for case C-D2 and a
  case identical to C-D2 but imposing free-slip boundary condition at
  the walls. (f) Locations over the wing selected to plot $C_p$ in
  panels (a), (b), (c), (d) and (e). \label{fig:Cp}}
\end{figure}

\begin{figure}
  \begin{center}
    \subfloat[]{\includegraphics[width=0.47\textwidth]{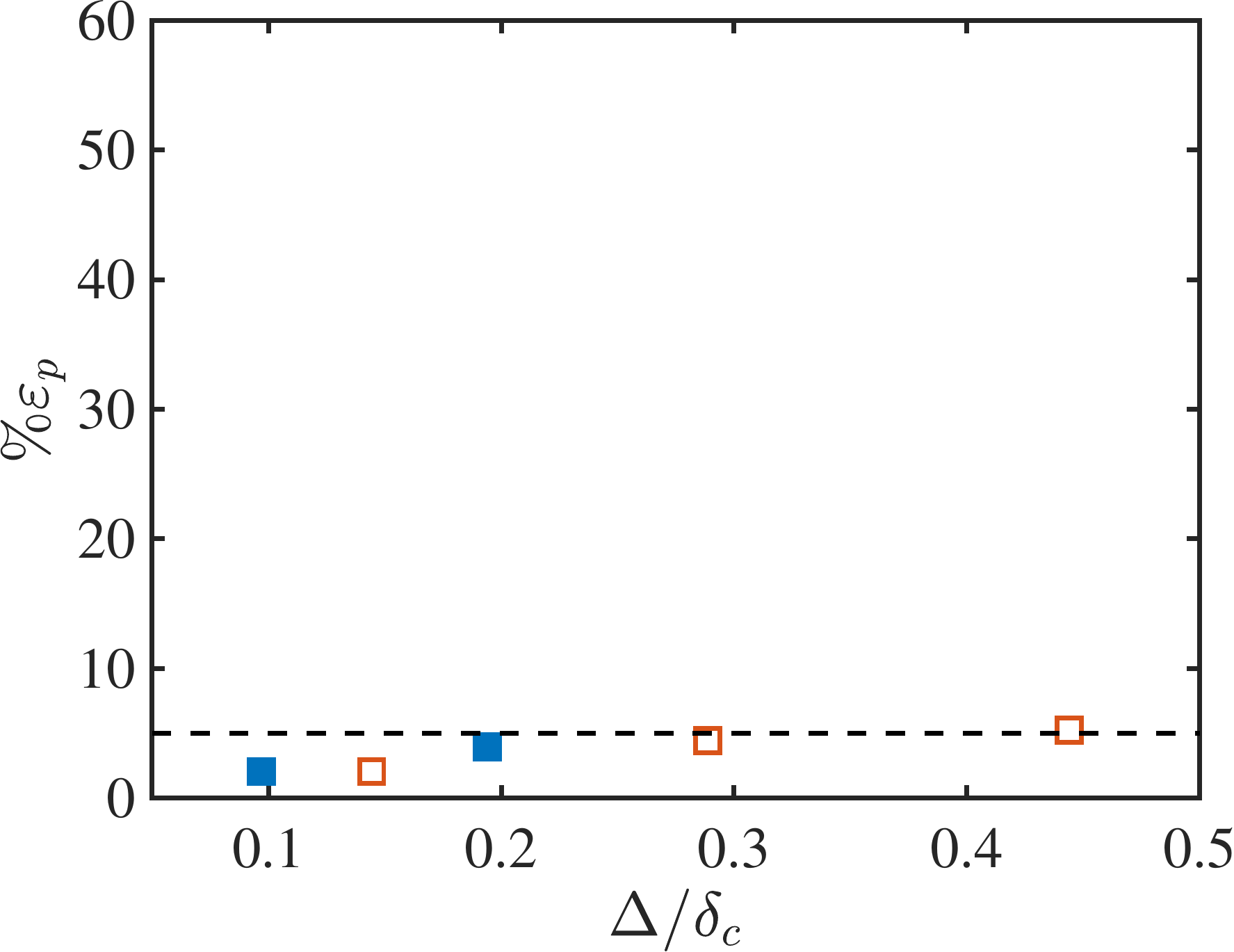}}
    \hspace{0.5cm}
    \subfloat[]{\includegraphics[width=0.47\textwidth]{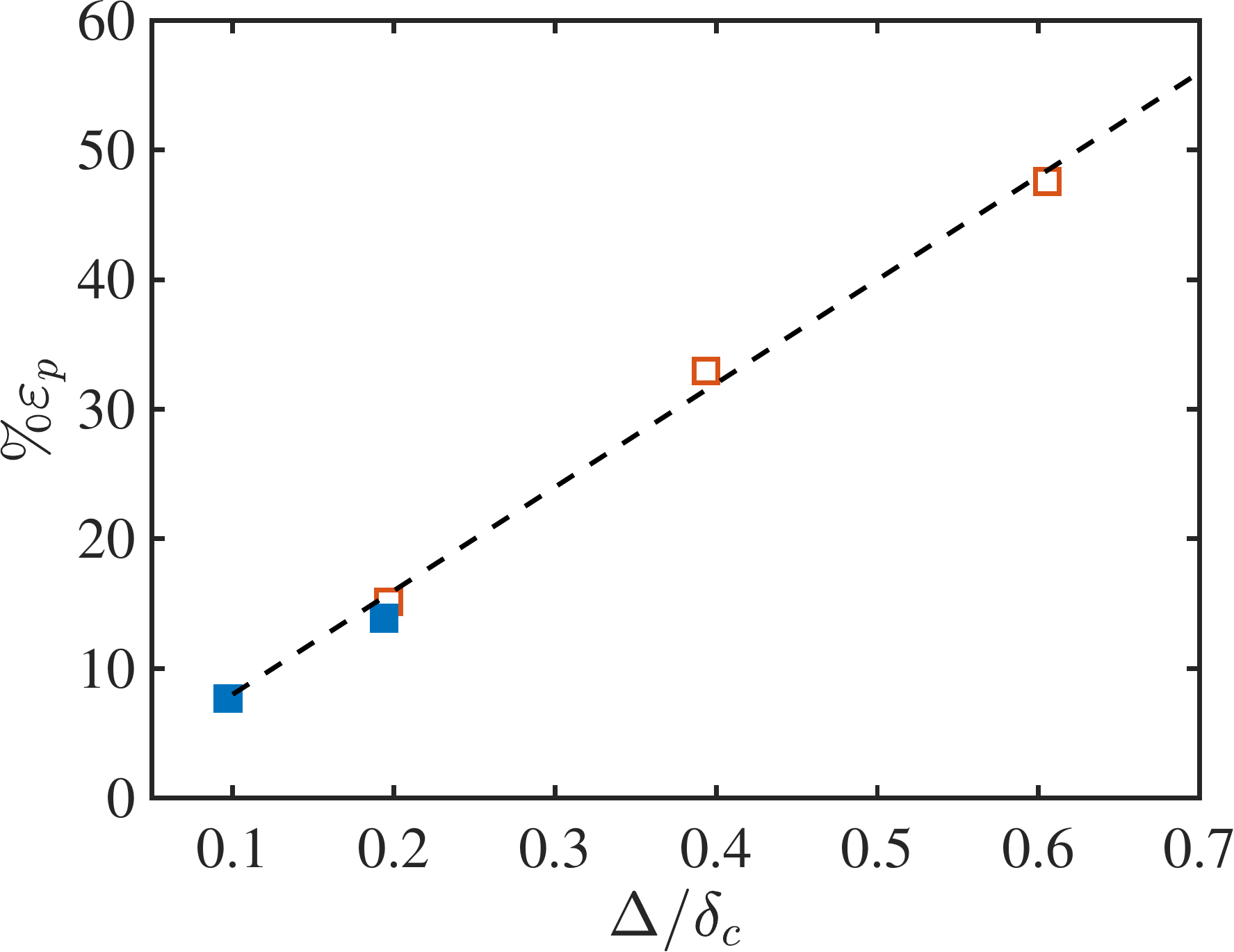}}
\end{center}
\caption{ Errors in the surface pressure coefficient at (a) $y=482.6$
  mm (close to the crank chord) and (b) $y=1663.7$ mm (wing tip) as a
  function of $\Delta/\delta_c$, where $\delta_c$ is the averaged
  $\delta$ along the streamwise direction of the station
  considered. The black dashed line in panel (a) is $\varepsilon_p =
  5\%$ and in panel (b) $\varepsilon_p \sim
  \Delta/\delta_c$.\label{fig:Cp_error}}
\end{figure}

The surface pressure coefficient along the chord of the wing is shown
in Figure \ref{fig:Cp}. The predictions are compared with experimental
data at four different $y$-locations denoted by the red lines in
Figure \ref{fig:Cp}(f).  The error in $C_p$ is quantified in Figure
\ref{fig:Cp_error} as function of $\Delta/\delta_c$, where $\delta_c$
is the averaged $\delta$ along the $x$ direction. Overall, WMLES
agrees with the experimental data to within 1--5\% error. The
predictions remain to within 5\% accuracy even when the boundary
layers are marginally resolved (i.e., 0--1 points per $\delta$).
The accurate prediction of $C_p$ along the main wing is a common
observation in CFD of external aerodynamic applications with attached
flows. As discussed in \S \ref{sec:errors}, the result can be
attributed to the inviscid nature of the mean pressure. Under the thin
boundary layer approximation, the inviscid outer-flow pressure is
directly imposed on the wall, which makes $C_p$ relatively insensitive
to the details of the near-wall turbulence.  This is demonstrated by
performing an additional inviscid calculation similar to C-D2 but
imposing the free-slip boundary condition at the wall such that
boundary layers are unable to develop (Figure \ref{fig:Cp}(e)). The
$C_p$ for the case with free-slip wall is strikingly similar to its
wall-modeled counterpart, confirming that $C_p$ is dominated by
inviscid contributions. The results also support the error analysis in
\S \ref{sec:errors}, where it was argued that errors in $C_p$ of the
form $\varepsilon_p = \beta_p (\Delta/\delta)^{\alpha_p}$ should have
$\beta_p\ll 1$ and $\alpha_q \approx 0$.
%
%

The main discrepancies with experiments are at the leading edge and
wing tip. The conditions for the thin boundary-layer approximation
require the wall radius of curvature to be much larger than the
boundary-layer thickness. This requirement might be satisfied to a
lower degree in the vicinity of the wing leading-edge, which explains
the slightly poorer predictions and higher sensitivity to $\Delta$ in
that region. The tripping methodology used in WMLES differs from the
experimental setup, which may also contribute to the discrepancies
observed with experiments at the leading edge. The inviscid assumption
for $C_p$ would not hold in the wing tip region due to the formation
of the wing tip vortex by viscous roll-up. This is corroborated in
Figure \ref{fig:Cp}(d), which shows that WMLES underpredicts the $C_p$
at the trailing edge of the wing tip with high sensitivity to the grid
resolution. Figure \ref{fig:Cp_error}(b) further reveals that the
errors at the wing tip follow a linear relation, $\varepsilon_p \sim
(\Delta/\delta_c)$.  The best $C_p$ prediction at the wing tip is
attained by case C-N10-R2e3, which supports the strongest vortex core
at the trailing edge as evidenced by the largest value of $-C_p$. By
construction, BL-conforming grids are designed to cluster points in
those regions where viscous effects are important (see Section
\ref{subsec:tbl}). In the case of the wing tip, the equalization of
the top and bottom pressure of the wing results in a crossflow
boundary-layer that is effectively captured by BL-conforming grids,
and this in turn permits the formation of stronger and more realistic
wing tip vortices.

Overall, the outer-flow nature of $C_p$ is encouraging for the
prediction of the pressure-induced components of the lift and drag
coefficients.  Our results, like previous studies in the literature,
suggest that $C_p$ might not be a particularly challenging quantity to
predict in the presence of wall-attached boundary layers (for example,
over the main body of the wing at low angles of attack).  Hence,
efforts should be devoted to improving the predictions at the leading
edge and wing tips, and to predict other viscous-dominated quantities
such as the skin-friction coefficient.

\subsection{Visualization of the separation bubble}\label{subsec:bubble}

\begin{figure}
\centering
\includegraphics[width=0.8\textwidth]{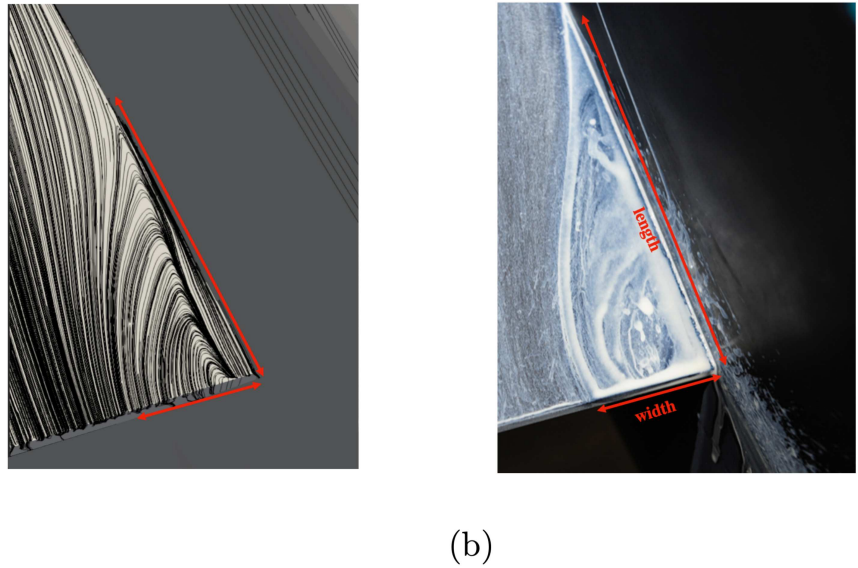}
\caption{(a) Streamlines of the average tangential wall-stress for
  C-D0.5. (b) Experimental oil-film
  visualization. \label{fig:bubble_tau}}
\end{figure}

The mean wall-stress streamlines for case C-D0.5 are shown in Figure
\ref{fig:bubble_tau}. The figure also contains a depiction of the
average length and width of the separation bubble, which are about
$100$ mm and $60$ mm, respectively, for case C-D0.5.  Direct
comparison of these dimensions with oil-film experimental results show
that the WMLES prediction is about 15\% lower than the experimental
measurements ($120 \times 80$ mm), consistent with previous WMLES
investigations~\citep{Lozano_AIAA_2020, Iyer2020,
  Ghate2020}. Nonetheless, it is important to note that the sizes of
the separation bubble from WMLES are calculated from the time-average
tangential wall-stress streamlines, whereas the experimental sizes are
obtained from the pattern resulting from the temporal evolution of the
oil film. Albeit both methodologies provide an average description of
the size of the separation zone, they do not allow for one to one
comparisons and we should not interpret the present differences as a
faithful quantification of the errors.

\section{Conclusions}\label{sec:conclusions}

We have investigated the cost and error scaling of WMLES for external
aerodynamic applications. The NASA Juncture Flow was selected as
representative of an aircraft with trailing-edge smooth-body
separation. The simulations were conducted using charLES with Voronoi
grids at an angle of attack of $5^\circ$ and $Re=2.4 \times 10^6$. Two
gridding strategies have been examined: i) constant-size grid, in
which the near-wall grid size has a constant value and ii)
boundary-layer-conforming grid (BL-conforming grid), in which the grid
size varies to accommodate the growth of the boundary-layer thickness
$\delta$.  In the latter, the boundary-layer thickness was estimated
by the difference between the solution of WMLES and an inviscid
calculation.  BL-conforming grids are characterized by $N_\mathrm{bl}$
(number of points per $\delta$) and $Re^\mathrm{min}_\Delta$ (minimum
grid-size Reynolds number), and their cost in terms of the number of
control volume scales roughly as $N_\mathrm{bl}^{13/6} Re \left(
Re^\mathrm{min}_\Delta \right)^{-5/6}$. This implies that the cost of
WMLES grows as $N_\mathrm{bl}^{13/6} \approx N_\mathrm{bl}^2$, and
from there the critical importance of assessing the minimum value of
$N_\mathrm{bl}$ required to attain the desired accuracy in the
quantities of interest.

We have focused on the prediction of the mean velocity profile and the
surface pressure coefficient $C_p$. Our analysis was accompanied by
theoretical estimations of the error in a zero-pressure gradient
turbulent boundary layer (ZPGTBL), which was used as the baseline to
assess the performance of WMLES in the NASA Juncture Flow. From a
theoretical viewpoint, the errors in the $C_p$ are expected to be low
and mostly grid independent at regions where the thin boundary layer
approximation holds. For the mean velocity profile, the errors should
scale linearly with the grid size for flows resembling a ZPGTBL.  We
have also distinguished two types of wall-modeling errors: internal
errors, due to the limitations of the wall model to represent the
near-wall turbulence (e.g., law of the wall); and external errors,
which originate from the outer LES solution.

Three different locations were considered to investigate the errors in
the mean velocity profile: the upstream region of the fuselage, the
wing-body juncture, and the wing-body juncture close to the
trailing-edge.  The last two locations are characterized by strong
mean-flow three-dimensionality and separation. A summary of the errors
incurred by WMLES in predicting mean velocity profiles is shown in
Figure \ref{fig:Errors_all_custom}(a).  The message conveyed by the
results is that WMLES performs as expected in regions where the flow
resembles a zero-pressure-gradient turbulent boundary layer,
consistent with our theoretical predictions. However, there is a
decline of the current models in the presence of wing-body junctions
and, more acutely, in separated zones. These errors are mitigated by
the use of BL-conforming grids, which enable a more efficient
distribution of grid points across the boundary layer. It was argued
that the improved accuracy provided by BL-conforming is related to the
reduced propagation of WMLES errors along the streamwise direction of
the flow. Nonetheless, the errors in the juncture and separated region
exhibit slow convergence rates regardless of the grid strategy,
rendering the brute-force grid-refinement approach too costly as a
pathway to improve the accuracy of the solution. The results reported
above for the mean velocity profile converge monotonically to the
experimental solution. However, we have shown that WMLES can exhibit a
non-monotonic convergence for some intermediate grid resolutions in
the range $\Delta \approx 0.03-0.05 \delta$, which might pose an
additional challenge to the robustness and reliability of WMLES.

The impact on the solution of the errors from the underresolved
leading-edge was also analyzed using BL-conforming grids. The
leading-edge grid resolution was controlled by the parameter
$Re_\Delta^{\mathrm{min}}$. In first order approximation, the
leading-edge errors decay linearly with the streamwise distance to the
underresolved leading-edge. However, these errors can still propagate
downstream for long distances, deteriorating the quality of the
solution even at the wing trailing-edge. In spite of that, the
convergence rate of the solution for decreasing
$Re_\Delta^{\mathrm{min}}$ suggests that the leading-edge grid
resolution might be less demanding than the grid resolution required
to resolve the subsequent turbulent boundary layers. Our study is
limited to low angles of attack and different conclusions might hold
at higher values, especially involving flow separation and stall. In
addition, the Juncture Flow Experiment and our simulations are both
tripped close to the leading-edge, which may also bias our
observations.

The errors in the mean pressure coefficient were assessed at four
spanwise locations ranging from the fuselage to the wing tip.  The
prediction of $C_p$ in the main wing is below 5\% error for all grid
sizes considered, even when boundary layers were marginally resolved.
This high accuracy can be attributed to the inviscid nature of the
mean pressure, which makes $C_p$ insensitive to flow details within
the turbulent boundary layer. The most challenging locations for WMLES
are the leading edge and wing tip, where the inviscid assumption
breaks down.  Similarly to the mean velocity profile, BL-conforming
grids deliver higher accuracy in the prediction of $C_p$ at the wing
tip with a lower computational cost.


Other relevant quantities of interest not investigated here include
the pointwise mean stress at the wall, and integrated quantities such
as lift, drag and moment coefficients, the last three being
particularly important for engineering applications. Unfortunately,
the pointwise friction coefficient is not available from the
experimental campaign of the NASA Juncture Flow, which
hinders our ability to assess the performance of the wall models more
thoroughly.  We also lack information on the lift, drag and moment
coefficients, and the impact of the modeling deficiencies identified
above on these quantities remains uncertain at this point.


The results presented here highlight the benefits of BL-conforming
grids is terms of accuracy and computational cost. Nevertheless, we
have also identified several outstanding issues that remain to be
solved. Among them, we can cite the decline in performance of current
modeling approaches in separated regions and corner flows, the
non-monotonic convergence of WMLES, and the necessity of acquiring
richer experimental measurements (such as pointwise skin friction
coefficient) to assess and aid the development of models.

\section*{Acknowledgments}

This work was supported by NASA under grant No. NNX15AU93A. The
authors also acknowledge the computational resources provided by the
NASA High-End Computing (HEC) Program through the NASA Advanced
Supercomputing (NAS) Division at Ames Research
Center. A.L.-D. acknowledges the MIT SuperCloud and Lincoln Laboratory
Supercomputing Center for providing HPC resources that have
contributed to the research results reported within this work.  We
thank Jane Bae and Konrad Goc for helpful comments. We are also
thankful to Michael Emory from Cascade Technologies, Inc. for his
guidance on grid generation.

\appendix

\section*{Appendix A: Grid point count of BL-conforming grids in ZPGTBL}

Let us consider a ZPGTBL over a flat plate of streamwise length $L_x$
spanwise width $L_z$, and leading edge at $x=0$. We aim at calculating
the number of grid points required for WMLES using BL-conforming grids
with $N_{\mathrm{bl}}$ and $Re_\Delta^\mathrm{min}$. An schematic of
the distribution of grid points was shown in Figure
\ref{fig:grid_types}(b). The grid resolution at a given $x$ is
$\Delta(x) = \delta(x)/ N_{\mathrm{bl}}$ for
$\delta>\delta_\mathrm{min} = \Delta_\mathrm{min}
N_{\mathrm{bl}}$. Once the boundary layer thickness is below
$\delta_\mathrm{min}$, the grid resolution is kept constant to
$\Delta_\mathrm{min}$.  The number of grid points can be divided into
points at the leading edge $N_\mathrm{LE}$ and the rest $N_R$.
Assuming a fully turbulent boundary layer growing as $\delta/x = K
Re_x^{-1/7}$ with $Re_x = x U_\infty/\nu$ and $K=0.16$, the number of
grid points is
\begin{equation}\label{eq:appendix_points}
N_\mathrm{points} = N_\mathrm{LE} +  N_\mathrm{R}=
\frac{L_z N_\mathrm{bl}^{13/6}}{L_x K^{7/6}}  \frac{Re}{\left(Re_\Delta^{\mathrm{min}} \right)^{5/6}}
+\frac{7}{5}\frac{L_z  N_\mathrm{bl}^{13/6} }{L_x K^{7/6}}
\frac{Re}{\left(Re_\Delta^{\mathrm{min}}\right)^{5/6}}\left(1-\frac{\left(N_\mathrm{bl}Re_\Delta^{\mathrm{min}}/K\right)^{5/6}}{Re^{5/7}}\right),
\end{equation}
where $Re= L_x U_\infty/\nu$. Equation (\ref{eq:appendix_points})
shows that the number of grid points for BL-conforming grids scales as
$N_\mathrm{bl}^{13/6}\left(Re_\Delta^{\mathrm{min}}\right)^{5/6} Re$,
which is exact for ZPGTBL with $\delta$ following the $1/7$-th growth
law. In the case of the NASA Juncture Flow, the value of
$N_\mathrm{points}$ differs from Eq. (\ref{eq:appendix_points}) (as
seen in figure \ref{fig:cost}) due to deviations in $\delta$ from a
ZPGTBL and the geometric complexities of the aircraft surface, which
makes $\Delta$ a function of the two wall-parallel
directions. Nonetheless, Eq. (\ref{eq:appendix_points}) offers a
simple model to rationalize the scaling of the cost of flow
simulations for an aircraft-like geometry dominated by attached
boundary layers.

We have characterized the quality of BL-conforming grids as a function
of $N_\mathrm{bl}$ and $Re_\Delta^{\mathrm{min}}$, where the latter is
a measure of the leading-edge resolution. An advantage of using
$Re_\Delta^{\mathrm{min}}$ is that it grants direct control over the
minimum grid size $\Delta_\mathrm{min}$.  However, the streamwise
extent of the underresolved leading-edge region ($L_0$ at which
$\delta=\delta_\mathrm{min}$, Figure~\ref{fig:grid_types}b) will
change with varying $N_\mathrm{bl}$ even if $Re_\Delta^{\mathrm{min}}$
is held constant. This can be seen from the relation $L_0/L_x =
(Re_\Delta^\mathrm{min} N_\mathrm{bl}/K)^{7/6} Re^{-1}$. To avoid
changes in $L_0$ with $N_\mathrm{bl}$, an alternative approach is to
replace $Re_\Delta^{\mathrm{min}}$ by the underresolved leading-edge
Reynolds number $Re_{L_0} = U_\infty L_0/\nu$. The price to pay for
using $Re_{L_0}$ instead of $Re_\Delta^{\mathrm{min}}$ is the lack of
direct control over $\Delta_\mathrm{min}$ and the complexities of
calculating $L_0$ in the actual geometry. The number of grid points to
resolve a flat plate as a function of $N_\mathrm{bl}$ and $Re_{L_0}$
is now given by
\begin{equation}\label{eq:appendix_points_v2}
N_\mathrm{points} = N_\mathrm{LE} +  N_\mathrm{R}=
\frac{L_z N_\mathrm{bl}^{3}}{L_x K^{2}} \frac{Re}{Re_{L_0}^{5/7}}
+\frac{7}{5}\frac{L_z  N_\mathrm{bl}^{3} }{L_x K^{2}}
\frac{Re}{Re_{L_0}^{5/7}}\left(1-\left(\frac{Re_{L_0}}{Re}\right)^{5/7}\right),
\end{equation}
which shows that the new expected cost scales as
$N_\mathrm{bl}^{3}Re_{L_0}^{-5/7} Re$. The modification of the cost
scaling law for WMLES with the selection of either
$Re_\Delta^{\mathrm{min}}$ or $Re_{L_0}$ should come as no surprise,
as each choice constitutes a different grid strategy.  In this work,
we have favored $Re_\Delta^{\mathrm{min}}$ due to its
simplicity. Nonetheless, other parametrizations, such as $Re_{L_0}$,
might also provide acceptable descriptions of the leading-edge grid
resolution.

\section*{Appendix B: Non-monotonic convergence of WMLES}

\begin{figure}
\begin{center}
\subfloat[]{\includegraphics[width=0.47\textwidth]{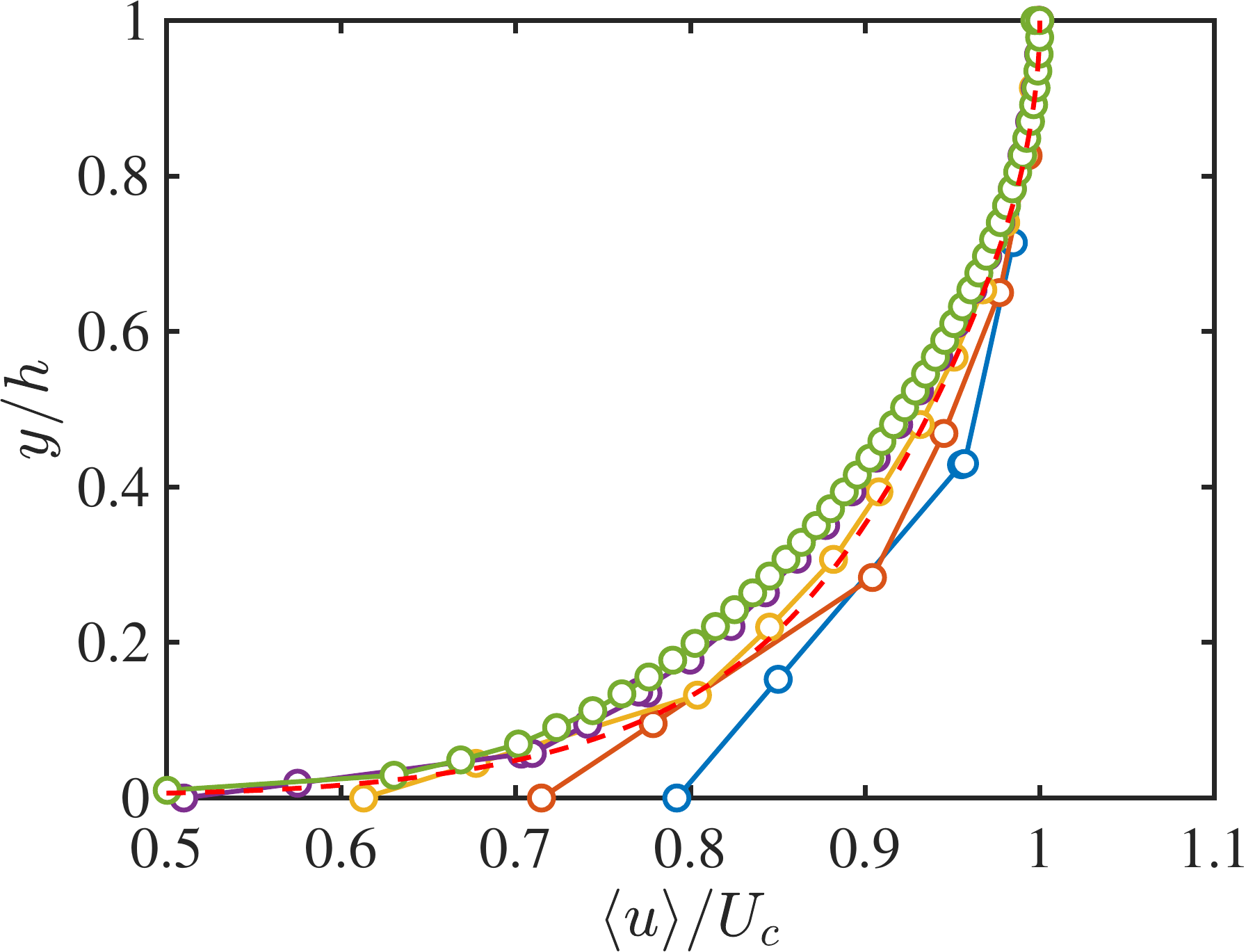}}
\hspace{0.5cm}
\subfloat[]{\includegraphics[width=0.47\textwidth]{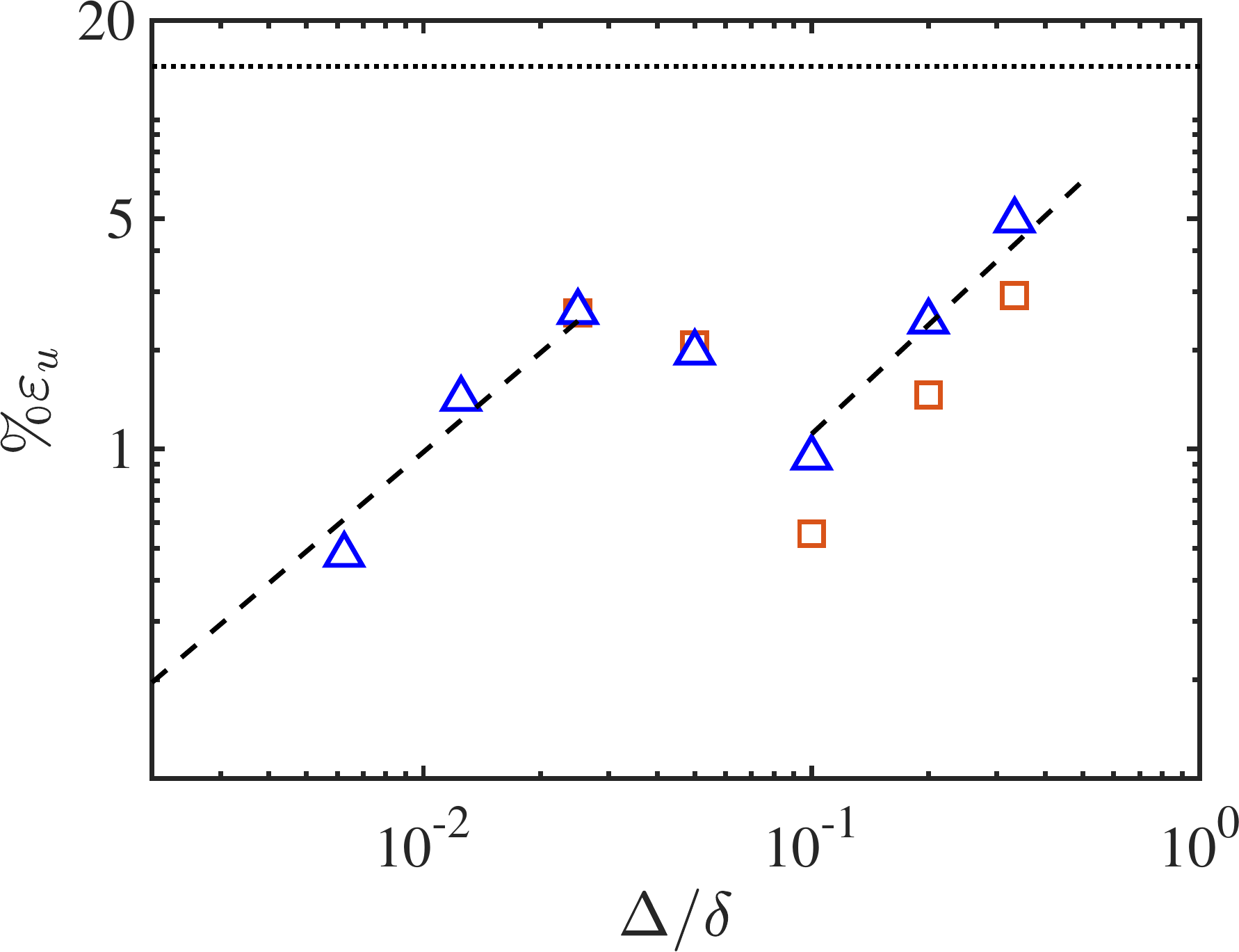}}
\end{center}
\caption{ (a) The mean velocity profile for WMLES with charLES of a
  turbulent channel flow. The colors denote different grid resolutions
  from coarser to finer: blue, red, yellow, purple, and green. (b)
  Error in the mean velocity profile $\varepsilon_u$ as a
  function of the grid resolution $\Delta$ for WMLES of turbulent
  channel flow.  The symbols denote simulations using charLES with
  Voronoi grids ($\square$, red) and finite-difference solver with
  staggered grids ($\triangle$, blue). The dashed line is
  $\varepsilon_u \sim \Delta/\delta$. The horizontal dotted line is
  the error from the inviscid solution. \label{fig:appendix_error}}
\end{figure}

The purpose of this appendix is to document more thoroughly the
convergence of WMLES in turbulent channel flows using charLES with
Voronoi isotropic grids. The friction Reynolds number is set to
$Re_\tau=4200$ and seven grid resolutions are considered:
$\Delta/\delta = 1/3, 1/5, 1/10, 1/20, 1/40, 1/80$ and $1/160$.  The
last four grid sizes are finer than ones we could afford for the
Juncture Flow Experiment.  To assess the effect of the numerical
scheme and gridding strategy, the results are compared with the error
obtained from WMLES of turbulent channel flows using the
finite-difference solver with staggered grid from \citet{Lozano2016,
  Lozano2018a}. All the simulations are conducted with dynamic
Smagorinsky model and the equilibrium wall model.  The channels are
driven by holding the centerline velocity $U_c$ to a constant
value. Figure \ref{fig:appendix_error}(a) contains the mean velocity
profiles obtained with charLES for the all the grid resolutions
considered. The error in the mean velocity profile is reported in
figure \ref{fig:appendix_error}(b) as a function of $\Delta$. The
results show a non-monotonic convergence of $\varepsilon_u$ and errors
from both solvers deviate from $\varepsilon_u \sim \Delta$ for
$\Delta/\delta \approx 0.03$--$0.05$.  The linear trend is recovered
again for $\Delta/\delta <0.03$.
%

\begin{figure}
\begin{center}
\includegraphics[width=1\textwidth]{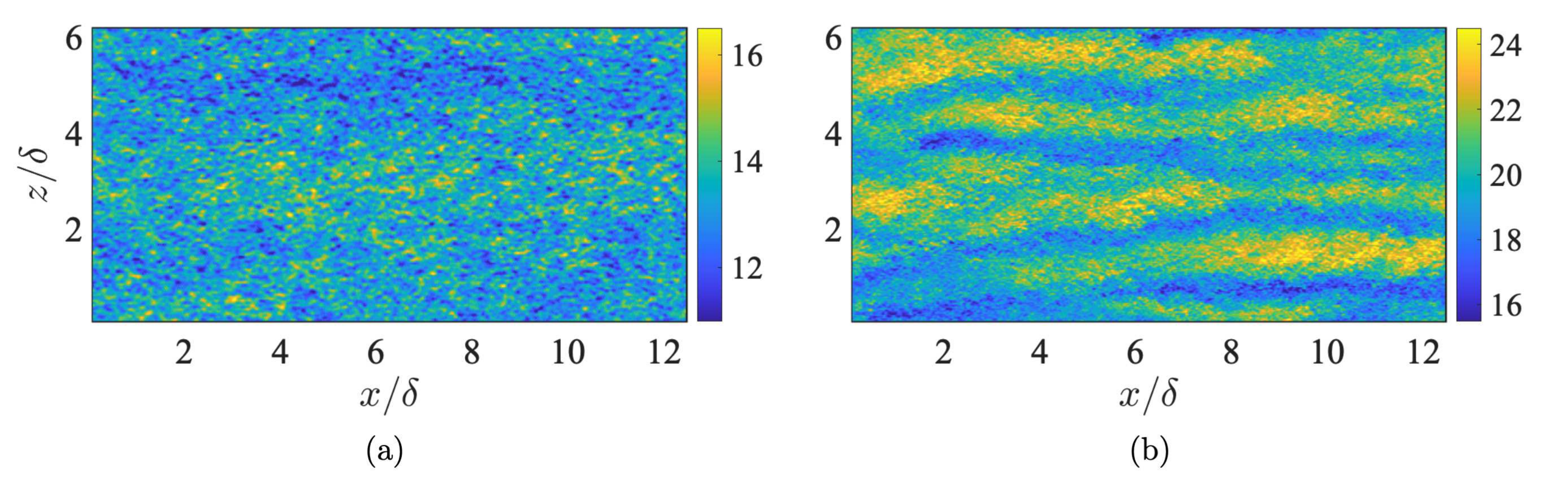}
\end{center}
Instantaneous streamwise velocity in a turbulent channel flow at the
wall-parallel plane $y=0.3\delta$. The simulations are performed using
charLES without SGS model with grid resolutions of (a) $\Delta =
0.05\delta$ and (b) $\Delta = 0.025\delta$. The latter case is able to
support streaky velocity structures even in the absence of explicit
SGS model.
\caption{ \label{fig:appendix_snapshots}}
\end{figure}

The non-monotonic convergence of the mean velocity profile was also
observed by \citet{Lozano2019a}.  They argued that this behavior can
be traced back to the ability of the LES grid to support streamwise
velocity streaks in the absence of SGS model. To test this idea, we
repeat the simulations in charLES without an explicit SGS model, i.e.,
the numerical truncation errors act as the SGS model. To avoid any
errors from the wall model, the equilibrium wall models is replaced by
an exact wall-stress boundary condition in which the mean wall stress
from DNS is directly imposed at the walls. Visual inspection of the
instantaneous streamwise velocity for cases without SGS model in
figure \ref{fig:appendix_snapshots} shows that there is a substantial
change in the flow structure at the critical grid size
$\Delta_c/\delta \approx 0.03$--$0.05$. For $\Delta> \Delta_c$, the
streamwise velocity lacks the characteristic features from wall
turbulence and exhibits instead a highly noisy structure (figure
\ref{fig:appendix_snapshots}(a)), as expected for LES in coarse grids
without an explicit SGS model. On the other hand, a clearly defined
streaky structure emerges for $\Delta < \Delta_c$ (figure
\ref{fig:appendix_snapshots}(b)), even in the absence of SGS model.
We can hypothesize that the transition observed at $\Delta_c$ will
also take place in the presence of an explicit SGS model, causing the
non-monotonic convergence of $\varepsilon_u$ reported in figure
\ref{fig:appendix_error}(b). Note that the flow transition observed at
$\Delta_c$ is independent of wall modeling errors, as the simulations
were performed by imposing the mean exact wall stress from DNS.  In
actual WMLES, the underperformance of the SGS model will propagate to
the wall model via external errors as discussed in \S
\ref{sec:errors}.  It is interesting that the critical grid resolution
$\Delta_c$ is roughly the same for the two solvers considered despite
the fact that they comprise different numerical schemes and grid
strategies. The latter observation points to a physical origin of
$\Delta_c$ in the sense that its value is dictated by physical
constraints rather than by the numerical details of the solver, at
least for low dissipation and energy preserving numerical schemes.
Indeed, \citet{Lozano2019a} showed that the grid resolution to resolve
90\% of the turbulent kinetic energy is $\Delta_{\mathrm{min}}
\approx0.04\delta$ at $y \approx 0.5\delta$, that is roughly equal to
$\Delta_c$. Although the results here provide some insight into the
origin of the non-monotonic convergence of WMLES, we still lack a
satisfactory explanation of the phenomenon and, more importantly, a
robust remedy.

\bibliography{references.bib}

\end{document}